\documentclass[a4paper,12pt,justification=centering]{article}
\usepackage{tikz, pgfplots}
\usepackage{tikz-cd} 
\usepackage{latexsym}
\usepackage{rotating}
\usepackage{amsmath,amssymb,amsxtra,amscd,amsthm}
\usepackage[mathscr]{eucal}
\usepackage{graphicx}
\usepackage[justification=centering]{caption}
\usepackage{subfig}
\usepackage{datetime}
\usepackage{pdfsync}
\usepackage{thumbpdf}
\usepackage{color}
\ifx\pdfoutput\undefined

\usepackage{amsmath}
\usepackage{amssymb}
\usepackage{pgfplots}
\usepackage{graphicx} 
\usepackage{caption}
\usepackage{ mathrsfs }
\usepackage{subcaption}
\usepackage{color}
\usepackage[nottoc,notlot,notlof]{tocbibind}

\usepackage{verbatim}
\usepackage{tensor}
\usepackage{xcolor}
\usepackage{titlesec}

\graphicspath {{images/}}
\usepackage{wrapfig}

\usetikzlibrary{shapes.geometric, arrows}

\usetikzlibrary{calc,decorations.markings}

\usetikzlibrary{arrows,decorations.markings}

\usetikzlibrary{calc,decorations.markings}

\tikzstyle{arrow} = [thick,->,>=stealth]

\usepackage{amsmath, amssymb, old-arrows}

\usepackage
[dvips,
 verbose=false,
 bookmarks=true,
 colorlinks=true,
 linktopage=true,
 linkcolor=webred,
 filecolor=webbrown,
 citecolor=webgreen,
 pagecolor=webblue,
 urlcolor=webblue,
 pdftitle={},
 pdfauthor={Murad Alim},
 pdfsubject={},
 pdfkeywords={},
 bookmarksopen=false,
 pdfpagemode=None,
 pdfview=FitH,
 pdfstartview=FitH,
 extension=pdf]{hyperref}
\else
\usepackage
[pdftex,
 verbose=false,
 bookmarks=true,
 colorlinks=true,
 linkcolor=webred,
 filecolor=webbrown,
 citecolor=webgreen,
 pagecolor=webblue,
 urlcolor=webblue,
 pdftitle={},
 pdfauthor={Murad Alim},
 pdfsubject={},
 pdfkeywords={},
 bookmarksopen=false,
 pdfpagemode=None,
 pdfview=FitH,
 pdfstartview=FitH,
 extension=pdf]{hyperref}
\fi
\definecolor{webred}{rgb}{.8,0,0}
\definecolor{webbrown}{rgb}{.6,0,0}
\definecolor{webgreen}{rgb}{0,0.5,0}
\definecolor{webdkgreen}{rgb}{0,0.3,0}
\definecolor{webblue}{rgb}{0,0,0.5}

\numberwithin{equation}{section}

\allowdisplaybreaks
\usepackage{bbm}
\usepackage{rotating}
\newcommand{\be}{\begin{eqnarray}}
\newcommand{\beq}{\begin{eqnarray}}
\newcommand{\ee}{\end{eqnarray}}

\def\e#1\e{\begin{equation}#1\end{equation}}
\def\ea#1\ea{\begin{align}#1\end{align}}

\theoremstyle{plain}
\newtheorem{thm}{Theorem}[section]
\newtheorem*{thm*}{Theorem}

\theoremstyle{definition}
\newtheorem{dfn}[thm]{Definition}
\newtheorem{ex}[thm]{Example}

\usepackage{hyperref}
\hypersetup{%
    linktocpage={true} 
}

\usepackage{enumitem}

\usepackage[a4paper,
            bindingoffset=0.2in,
            left=0.7in,
            right=0.9in,
            top=1.3in,
            bottom=1.3in,
            footskip=.25in]{geometry}

\begin{document}

\setlength{\abovedisplayskip}{20pt}
\setlength{\belowdisplayskip}{20pt}

\begin{titlepage}

\setlength{\parindent}{0cm}
\setlength{\baselineskip}{1.5em}
\title{Generating functions for $\mathcal{N}=2$ BPS structures}

\author{Murad Alim$^{1,2,3}$\footnote{\tt{m.alim@hw.ac.uk}}, Daniel  Bryan$^{3,4}$\footnote{\tt {daniel.bryan@fuw.edu.pl}}
\\
\\
\small $^1$ Maxwell Institute for Mathematical Sciences, Edinburgh EH14 4AS, UK\\
\small $^2$Department of Mathematics, Heriot-Watt University, Edinburgh EH14 4AP, UK\\
\small $^3$ Department of Mathematics, University of Hamburg, Bundesstr. 55, 20146, Hamburg, Germany\\
\small $^4$ Faculty of Physics, University of Warsaw, ul. Pasteura 5, 02-093 Warsaw, Poland\\
}

\date{}
\maketitle

\abstract{We propose generating functions which encode the degeneracies and wall-crossing phenomena of $\mathcal{N}=2$ BPS structures. The generating functions have a representation-theoretic origin and are the analogs of the 1/4-BPS dyon counting formula in $\mathcal{N}=4$ theories involving the Weyl denominator formula of a Borcherds-Kac-Moody Lie algebra. A general form of the generating function is suggested based on the Lie algebra associated to the adjacency matrix of the BPS quiver whenever the BPS spectrum of the $\mathcal{N}=2$ theory admits such a description. This proposal is tested for the BPS spectrum of Seiberg-Witten SU(2) theory as well as for the $D6$-$D2$-$D0$ BPS structure of the resolved conifold which are both captured by an affine $A_1$ Lie algebra and are obtained from limits of the $\mathcal{N}=4$ generating function.  The general proposal also reproduces the correct BPS spectra and wall-crossing structures for the Argyres-Douglas $A_2$ theory. We further discuss connections to scattering diagrams studied in the context of stability structures.

}

\nonumber

\thispagestyle{empty}

\end{titlepage}

\tableofcontents

\newpage

\section{Introduction} \label{section:chengverlindereview}

The study of BPS structures has been a rich source of insights into the underlying mathematical structure of quantum field theories, string theories and black hole microstate counting. BPS states can decay or form bound-states at walls of marginal stability in the moduli spaces of the underlying theories. Understanding these wall-crossing phenomena mathematically and physically has triggered a lot of interaction between mathematics and physics in the fields of enumerative geometry \cite{GopakumarVafaIhttps://doi.org/10.48550/arxiv.hep-th/9809187,gopakumar1998mtheory, D6D2D0JafferisMoorehttps://doi.org/10.48550/arxiv.0810.4909,BridgelandRH,bridgeland2020riemannhilbert,ASTThttps://doi.org/10.48550/arxiv.2109.06878}, black hole microstate counting \cite{K3T4dualN=4Sen_2007,Dabholkar_2008,Cheng_2007,Cheng_2008,Cheng:2008kt,Cheng:2009hm,Gaberdiel:2011qu,dabholkar2014quantum,kachru2017hiddenweyldenominator} as well as stability structures of categories \cite{kontsevich2008stability,Szendr_i_2008,joyce2010theory,BridgelandOriginalscatteringdiagram,Bridgeland}.

In this work we derive new generating functions for 4d $\mathcal{N}=2$ theories which count the BPS states in each chamber in the moduli space and determine the associated wall crossing phenomena. One of the models we are studying is the Argyres-Douglas $A_{2}$ theory \cite{Argyres_1995,shapere1999bps} used to describe the primitive wall crossing involving a dyon splitting up into its electric and magnetic constituents at the wall of marginal stability. We also study the models described by the affine $A_{1}$ BPS root system, denoted here as $\hat{A}_{1}$. The latter includes Seiberg-Witten SU(2) theory \cite{Seiberg_1994}, which has the full spectrum of the affine root system at weak coupling that jumps to just 2 basis states on the other side of the wall. The D6-D2-D0 brane system of the resolved conifold described by Jafferis-Moore \cite{D6D2D0JafferisMoorehttps://doi.org/10.48550/arxiv.0810.4909} is also described by this root system. This describes the appearance of bound states to a large core and has an interpretation in terms of non-commutative Donaldson-Thomas (NCDT) invariants that is further developed in \cite{Szendr_i_2008,joyce2010theory,ASTThttps://doi.org/10.48550/arxiv.2109.06878}. The new generating function outlined in this work describes the wall crossing for a particular set of BPS states in these cases but also has the potential for generalisation to other theories on CY-3 folds 
admitting general BPS quivers including for example those defined by ADE type singularities.

Our new $\mathcal{N}=2$ generating function is constructed as an analog of the $\mathcal{N}=4$ generating function that is known for counting $\frac{1}{4}$BPS dyonic black holes in 4d with a type II string theory description on $K3 \times T^{2}$. This was originally derived as a dyon counting function by Dijkgraaf, Verlinde, and Verlinde \cite{Dijkgraaf_1997} from fivebrane partition functions \cite{fivebraneBPSDijkgraaf_1997,fivebraneblackholeentropyDijkgraaf_1997}. This partly followed the computation of threshold corrections for BPS algebras defined by Harvey and Moore \cite{Harvey_1996,Harvey_1998} for perturbative BPS states and explicitly computed in \cite{Kawai_1996N=2}. The generating function itself is a weight 10 Siegel Modular form known as the Igusa cusp form, which is described in detail in the works of Gritsenko and Nikulin \cite{gritsenko1995siegel,Gritsenko_1996,classificationGritsenko_2002}. The first time this function arose in string theory was in the context of moduli dependent one loop gauge couplings or threshold functions of a heterotic model \cite{Mayr:1995rx}.
Subsequently, this generating function, and the process of extracting the degeneracies for particular charges has been studied extensively in e.g.  \cite{Dijkgraaf_1997,Cheng_2007, Cheng_2008,dabholkar2014quantum,N=4DegeneraciesDabholkar_2008,N=4DyonDegKKCMprodDavid_2006,KKoscfreeN=4DyonmodesDavid_2006,K3T4dualN=4Sen_2007}. The degeneracies are extracted as Fourier coefficients using a particular contour prescription in a complex 3 dimensional space. Cheng and Verlinde \cite{Cheng_2008} determined that the charge invariants of the BPS black holes can be written as positive roots $\alpha_{i} \in \Delta^{+}_{(A,S)}$ \footnote{Here we write $A$ as the Cartan matrix for the BKM algebra and (following \cite{gritsenko1995siegel}) $S$ is a subset of the indicies of the generators of the algebra.} of a Borcherds-Kac-Moody superalgebra $\mathfrak{g}(A,S)$ which are a linear combination of 3 basis roots. We will also use this formulation here. The generating function itself is a square of the Weyl denominator of this Borcherds-Kac-Moody algebra.   
This can be written as
\begin{align} \label{eq:dyoncountingformula}
(-1)^{P \cdot Q+1} D(P,Q) = \oint_{\mathcal{C}} d^{3}\Omega \Bigg{(}\frac{e^{i \pi( \frac{1}{2}\Lambda_{P,Q}+ \rho,\Omega)}}{\prod_{\alpha \in \Delta^{+}}(1-e^{-i\pi (\alpha, \Omega)})^{\text{mult}(\alpha)}} \Bigg{)}^{2},
\end{align}
where, $D(P,Q)$ is the degeneracy of a dyon with electric and magnetic charges $(P,Q)$, $d^{3}\Omega$ is a complex measure form in the three variables corresponding to the entries of the symmetric $2 \times 2$ matrix $\Omega$, $\Lambda_{P,Q}$ is the charge vector, $\rho$ is the Weyl vector, $\text{mult}(\alpha)$ is the multiplicity of the root $\alpha$ and $\alpha(\Omega)=(\alpha, \Omega)$
\footnote{This can be written as $(X,Y)= -\det(Y)\text{Tr}(XY^{-1})$.}.
The integral is over a complex 3d contour $\mathcal{C}$ where the real parts of the 3 complex variables is taken to be range [0,1] and the imaginary parts are large.

There are wall crossing phenomena that have been studied in this model which involve the 1/4-BPS dyons splitting into their electric and magnetic 1/2-BPS constituents. This can be understood in terms of a moduli dependent contour prescription \cite{Cheng_2007}, whereby when the contour crosses the poles $(\alpha_{i}, \Omega) = 0$, where $i$ labels the positive roots, there is a jump in the degeneracies. This configuration also has an M-theory description in terms of M5 branes wrapping a Riemann surface which degenerates at the walls and has periods corresponding to the arguments of the generating function \cite{ChengHollands:2009hm}. Alternatively, one can interpret this, as done in \cite{Cheng_2008}, as a jump in the Fourier coefficients for the different possible expansions of the Weyl denominator in the different Weyl chambers. These have boundaries at $\text{Im}[(\alpha_{i}, \Omega)] = 0$. The number of BPS dyons that exist in a particular chamber can be counted in the highest weight of the Verma module in that chamber. 

In this paper we first restrict to particular subalgebras of the Kac-Moody algebra that retain this pattern of wall crossing, although now with a subset of BPS states that can exist in the full algebra. We find that in these subalgebras, for example $\hat{A}_{1}$, the BPS states and walls can be matched with analogs in 4d $\mathcal{N}=2$ theories such as D6-D2-D0 boundstates in type IIA string theory on the resolved conifold or the BPS spectrum of pure $SU(2)$ Seiberg-Witten theory. We then generalise this principle to the root system in Argyres-Douglas $A_{2}$ by considering the Weyl denominator of the associated Lie algebra as the analog generating function. The wall crossing in these models has been well understood physically in terms of 4d $\mathcal{N}=2$ theories \cite{Gaiotto_2010,FramedBPSGMNhttps://doi.org/10.48550/arxiv.1006.0146} and encapsulated mathematically for BPS structures \cite{bridgeland2020riemannhilbert, BridgelandRH} in terms of stability conditions on Donaldson-Thomas (DT) invariants.  This generating function in the form of the Weyl denominator reproduces the count of the BPS states in all chambers. We do this by choosing a prescription for the complex arguments of the generating function in terms of the moduli. For the known $\mathcal{N}=4$ dyon counting function the central charge is used in the literature by letting the imaginary part of the contours be proportional to the normalized central charges $X = \frac{Z_{P,Q}}{|Z_{P,Q}|}$. Using the same principle, we also choose, for the 4d $\mathcal{N} = 2$ theories, the central charges as the analog for the complex parameters present in the original generating function.

Framed BPS states can be described as bound states of the BPS states to a large core object (or defect) and are described by Gaiotto, Moore, Neitzke (GMN) in \cite{FramedBPSGMNhttps://doi.org/10.48550/arxiv.1006.0146} and in \cite{BPSgalaxiesAndriyash_2012} from a supergravity perspective. The process in which constituent BPS states interact to form boundstates can be formulated in terms of a scattering algebra in which the S-matrices of the interaction form the structure constants of the algebra \cite{Harvey_1998}. 
An algebra of this form also describes these framed states and was constructed by Kontsevich and Soibelman \cite{CoHAKS}. This is known as the Cohomological Hall Algebra (CoHA) and  has recently been matched \cite{Galakhov_2019BPSalgebrasandframedBPSstates} with the original Harvey-Moore \cite{Harvey_1998} BPS  scattering algebras.
When one uses the central charges in the generating function then the boundaries of the Weyl chambers become the BPS walls for the existence of these boundstates.  This also means that the Verma modules, described by the generating function with a particular highest weight, count a special set of boundstates or “framed halo” BPS states. However, this generating function is distinct from that of framed BPS states in \cite{FramedBPSGMNhttps://doi.org/10.48550/arxiv.1006.0146} which also tracks the core charges.
The chamber with the maximal number of framed BPS states that can enter the spectrum should also match the total number of BPS states present in the theory, where one must continually interpret the framed BPS states as the boundstates of the known vanilla  states, defined as the original BPS states in the theory before insertion of the defect \cite{FramedBPSGMNhttps://doi.org/10.48550/arxiv.1006.0146}, with the large core. If there is a wall of marginal stability in the theory, this maximum number is distinct on the different sides of the wall, which matches the known number of BPS states in both chambers. We determine this maximum number by crossing a different number of BPS walls in each chamber. From this we also reproduce the scattering diagrams developed by Bridgeland \cite{BridgelandOriginalscatteringdiagram} for the $\hat{A}_{1}$ (Seiberg-Witten) and $A_{2}$ examples, using Kontsevich-Soibelman operators \cite{kontsevich2008stability}.

This chamber structure and the BPS content of each chamber is determined by mapping onto the attractor flow existence conditions for the central charges developed in \cite{MooreArithmeticAttractorshttps://doi.org/10.48550/arxiv.hep-th/9807087, DenefGreeneRaugas_2001, DenefSeibergWittenemptyhole_2000,Bousseau:2022snmnewattractorflowscatteringdiagram,Alim:2023doi}, where existing BPS states must not have regular endpoints for the flow where the central charges vanish. The split attractor flow can be embedded into the scattering diagrams \cite{Bousseau:2022snmnewattractorflowscatteringdiagram}. We find that one can write the central charges of the states that are excluded after crossing the wall of marginal stability in terms of the others where they align on the wall. One can then continue this through the wall in a way that avoids the termination of flow lines at regular points where the central charges vanish. Now the term represented by the non-existing BPS state is no longer in the lattice of positive roots and must be dropped from the spectrum that is counted in the highest weight. Therefore, in models with a wall of marginal stability, the wall crossing phenomena (of the vanilla BPS states) can be clearly understood in terms of this generating function. 

We propose that this generalises to $\mathcal{N}=2$ BPS structures which can be described by a BPS quiver as in \cite{cecotti2010rtwisting, cecottivafacompleteclassificationhttps://doi.org/10.48550/arxiv.1103.5832,Alim:2011ae,N=2BPSquivershttps://doi.org/10.48550/arxiv.1112.3984}, which in particular includes all ADE type Argyes-Douglas theories. The Lie algebra in these cases is associated to the adjacency matrix of the BPS quiver and the Weyl denominator of the associated Lie algebra becomes the generating function. In this last generalisation we expect the generating function in each chamber bound by walls of marginal stability to no longer be the Weyl denominator of a Lie algebra but a more general denominator formula in which the factors are determined by the sequence of quiver mutations (see Cecotti and Vafa \cite{cecottivafacompleteclassificationhttps://doi.org/10.48550/arxiv.1103.5832} and also \cite{N=2BPSquivershttps://doi.org/10.48550/arxiv.1112.3984}) generating the BPS spectrum in that chamber.

\section[$\mathcal{N}=2$ subalgebras of the Borcherds-Kac-Moody algebra]{$\mathcal{N}=2$ subalgebras of the Borcherds-Kac-Moody \\ algebra} \label{section:startonprojectongeneratingfunction}

We now start by deriving generating functions for subalgebras from the $\mathcal{N} =4$, 1/4-BPS dyon counting function, derived in \cite{Dijkgraaf_1997}, for type II string theory on $K3 \times T^{2}$ that can also give us information about wall crossing in $\mathcal{N} =2$ theories. The generating functions for the $\mathcal{N}=2$ subalgebras can then be written in terms of a special subset of the root lattice of the full Borcherds-Kac-Moody algebra that one can choose. This can done by looking at the wall crossing for 1/4-BPS dyons that have charges corresponding to those that exist in the BPS spectrum of 4d $\mathcal{N} =2$ theories. For example the BPS spectra described by an affine $A_{1}$ (also denoted by $\hat{A}_{1}$) root lattice, such as Seiberg-Witten theory or the D6-D2-D0 bound states of the resolved conifold described by Jafferis-Moore \cite{D6D2D0JafferisMoorehttps://doi.org/10.48550/arxiv.0810.4909}. Then we can use such a generating function as an analog counting function to determine existence of such BPS states in these theories.

We first review the $\mathcal{N}=4$ interpretation. One can start by extracting BPS degeneracies from the generating function (\ref{eq:dyoncountingformula}) in the full $\mathcal{N} = 4$ case. They  are extracted in the Siegel upper half plane $\mathbb{H}^{+}$. Once this is done one can then look at their wall crossing phenomena. As discussed above in sec.~\ref{section:chengverlindereview}, it is understood that this generating function describes the wall crossing phenomenon of the splitting of multicentered BPS black hole dyons. This has been shown to work from 2 perspectives: the first is directly from the generating function \cite{Cheng_2008} and the second is by a contour prescription \cite{Cheng_2007}. Starting with the generating function (\ref{eq:dyoncountingformula}), for particular values of the moduli the function can be expanded as a Fourier series expansion of the form \footnote{Here we have redefined $\alpha(\Omega)$ by absorbing the factor $i\pi$ into the complex variables in $\Omega$.}
\begin{align} \label{eq:expansionFouriercoefficients}
\frac{1}{\Phi_{10}(\Omega)} = \frac{e^{( 2\rho,\Omega)}}{\prod_{\alpha \in \Delta^{+}}(1-e^{- (\alpha, \Omega)})^{2\text{mult}(\alpha)}} = \sum_{p,q,r=-1}^{\infty}  F_{KM}(p,q,r)e^{-p\alpha_{1}(\Omega)-q\alpha_{2}(\Omega)-r \alpha_{3}(\Omega)},
\end{align}
where $\alpha_{1}, \alpha_{2}, \alpha_{3}$ are a basis of positive roots $ \Delta^{+}_{(A,S)} $ of the Borcherds-Kac-Moody (BKM) superalgebra $\mathfrak{g}(A,S)$, where we are using the notation of Gritsenko and Nikulin \cite{gritsenko1995siegel} such that $A$ represents the Cartan matrix of the BKM algebra and $S$ is a subset of the indicies labelling  the generators. The $F_{KM}(p,q,r)$ are the Fourier coefficients of the square Weyl denominator that are obtained by series expansion. However, recalling the condition for the expansion of a geometric series (also used in \cite{Cheng_2008}), we note that this Fourier series expansion is only well defined for $\text{Im}[i\alpha_{j}(\Omega) ] > 0 $ \footnote{The factor $i$ is inserted here, following the convention of \cite{Cheng_2007,Cheng_2008}, as $\text{Re}[\alpha_{j}(\Omega) ] > 0$ for the sum to converge. } for all roots $\alpha_{j}$ that occur within the product in the Weyl denominator. Only in this case do we know how to extract the Fourier coefficients. This means that at a point in the moduli space where one has a finite number of roots with negative imaginary part
\begin{align}
\text{Im}[i\alpha_{j}(\Omega)] < 0,   \   \   \forall j \in \{ 1,\dots,n \},  \   \   \  \text{Im}[i\alpha_{k}(\Omega)] > 0, \   \  \forall k \neq j\,,
\end{align}
one must rewrite the generating function such that it has only factors in the denominator with exponents all in the form $\text{Im}[i\alpha'_{j}(\Omega)] > 0, \   \  \forall  j \in  \{ 1,\dots,n \} $. It can then be expanded again as a Fourier series where coefficients can be extracted. However, in this case the coefficients will jump. This is because, in rewriting the equation, one moves the exponential factors into the numerator from which one obtains a shift in the degeneracies, meaning that the degeneracies of a particular electric and magnetic charge  before wall crossing become those of a different charge on the other side of the wall. In this case one can also use S-duality to change the basis used in the denominator back to one of positive roots, such that the denominator is again written in terms of these roots, as it was before the jump occurred.

Alternatively, according to the second prescription \cite{Cheng_2007}, this can be understood in terms of the contour $\mathcal{C}$ in (\ref{eq:dyoncountingformula}) and the poles. The walls of marginal stability occurring at the points $\text{Im}[i\alpha_{j}(\Omega)] = 0$  are passed as the contour crosses the poles at $\alpha_{j}(\Omega) = 0$. For every wall crossed, the BPS state associated to the root $\alpha_{i}$ disappears from the spectrum.


Our aim is to gain information about wall crossing of BPS structures in 4d $\mathcal{N} =2$ theories. We therefore start looking at multi-centered BPS dyons that decay in this theory that are analogous to the BPS states with corresponding electric and magnetic charges existing in 4d $\mathcal{N} =2$ theories. These charges should exist as subsets of the root lattice of the Borcherds-Kac-Moody superalgebra $\mathfrak{g}(A,S)$ discussed in \cite{Cheng_2008}, with an S-duality group that leaves this sublattice invariant. Examples include the algebras $\mathfrak{g}_{A_{1}} \subset \mathfrak{g}_{\hat{A}_{1}} \subset \mathfrak{g}(A,S)$ with root systems $\Delta_{A_{1}} \subset \Delta_{\hat{A}_{1}} \subset \Delta_{(A,S)}$. In particular one can look at the simplest case of the $A_{1}$ Lie algebra which just describes a single BPS state. For this case a discussion was started in \cite{N=4DegeneraciesDabholkar_2008} involving an S-duality transformation for a set of degeneracies controlled by a single factor representing the Weyl denominator of just $A_{1}$. In this algebra, the only change of basis of the roots that can be done corresponds to permuting the positive and negative root $\pm \alpha \rightarrow \mp \alpha$.
Furthermore, there is also the case of the affine Lie algebra denoted by $\hat{A}_{1}$ which has been shown to be a subalgebra in \cite{subalgebra1Govindarajan:2021pkk,subalgebra2Govindarajan:2022jmo}. This has an S-duality and modular group given by $PSL(2, \mathbb{Z})$ \cite{Cheng_2007,Cheng_2008}. Both the Weyl denominators of $A_{1}$ and $\hat{A}_{1}$ exist as factors within the Weyl denominator of the Borcherds-Kac-Moody algebra. It should therefore be possible to proceed by extracting suitable degeneracies which correspond to the BPS states formed from the expansions of the roots that occur in this subalgebra, and determine the jumps of these degeneracies as the BPS states decay. This then acts as an analog for $\mathcal{N}=2$ BPS states with a matching set of charges described by the same roots and where particular roots can disappear across a wall.

\section{$\mathcal{N}=2$ BPS structures and generating functions} \label{section:analogsforcountingfunctionSWandA2}

We subsequently use and then generalise this concept of the generating function to construct analog generating functions in a class of 4d $\mathcal{N}=2$ theories \cite{Seiberg_1994,Argyres_1995,shapere1999bps} from which one can observe the wall crossing phenomena \cite{N=2BPSquivershttps://doi.org/10.48550/arxiv.1112.3984,RefinedMotivicQuantumDimofte_2009,kontsevich2008stability} in the same way and that can hence also act as counting functions for BPS states. This is done for the $\hat{A}_{1}$ Lie algebra involving Seiberg-Witten SU(2) in section \ref{workforanalogs} as well as the $A_{2}$ system in the context of Argyres-Douglas theories in section \ref{A2subsection}.
We first derive the generating function for $\hat{A}_{1}$ using the
$\hat{A}_{1}$ subalgebra of the Borcherds-Kac-Moody algebra where the root system describes $\mathcal{N}=4$, 1/4-BPS states. This is presented in section \ref{extractingaffineA1fromKacMoody}. In this case, as with the full $\mathcal{N}=4$ dyon counting formula, the BPS degeneracies of states with particular charges can be read off as Fourier coefficients or extracted using a contour integration. We found that formulae of this form encode wall crossing through either a jump in the Fourier coefficients  or in the highest weight of a Verma module (see Eq. \ref{eqverma}). The $\hat{A}_{1}$ example is an ideal testing case that can in future be generalised to further theories described by BPS quivers including those described by ADE type root systems. We also conjecture a generalisation to different BPS quivers e.g. those with added matter such that one can write down a generating function from the sequence of quiver mutations encoded in the denominator. We conjecture that, in the ADE examples, this generating function is always the Weyl denominator associated to the root system of the particular quiver describing the BPS states in the theory. We explicitly work out the examples of Seiberg-Witten SU(2) theory and the Argyres-Douglas $A_{2}$ theory as good models to start with. We look for a different spectrum of framed BPS states existing in each Weyl chamber and also see that the wall of marginal stability is encoded in the generating function. To do this we first review (in the next subsections \ref{sec:BPSstructures}-\ref{section:rootsystemBPSliealgebras}) the definition of BPS structures, their quiver descriptions and the concept of root systems in Lie algebras.

\subsection{BPS structures}
\label{sec:BPSstructures}
In this subsection we introduce the relevant notions characterizing $\mathcal{N}=2$ BPS structures, following \cite{Alim:2023doi,Gaiotto_2010,BridgelandRH} in introducing the necessary data for studying the BPS problem. 

Let $\mathcal{B}$ be a complex manifold of complex dimension $r$, this corresponds to the Coulomb branch or moduli space of vacua of a physical theory. We denote a local coordinate on $\mathcal{B}$ by $(w_1, \dots, w_r)$, where
 $r$ the dimension of the moduli space. This is also the rank of the gauge group.  $\mathcal{B}$ carries a local system $\Gamma$ with fiber $\Gamma_w \cong \mathbb{Z}^{2r}$ at each $w \in \mathcal{B}$.
$\Gamma_w$ is the charge lattice. 

$\Gamma$ further carries a symplectic pairing
\begin{equation}\label{pairing}
\langle -, - \rangle \colon \Gamma \times \Gamma \to \underline{\mathbb{Z}},
\end{equation}
where $\underline{\mathbb{Z}}$ is the constant sheaf with fiber $\mathbb{Z}$.
On $\mathcal{B}$ one can locally split $\Gamma = \Gamma^e \oplus \Gamma^m$ into Lagrangian sub-lattices, which correspond to the \emph{electric} and \emph{magnetic charge lattices} respectively. 
One can construct a basis $\{ \alpha_1,\dots,\alpha_r\}$ for $\Gamma^m$ and $\{ \beta^1,\dots,\beta^r\}$ for $\Gamma^e$ such that
\begin{equation}
\langle \alpha_I , \beta^J \rangle =\delta_I^J \,, \quad  \langle \alpha_I , \alpha_J \rangle = 0 = \langle \beta^I , \beta^J \rangle
\end{equation}
for $I, J = 1, \dots, r$.
This form of basis of $\Gamma$ is also known as an electric-magnetic basis. 
This means that every $\gamma \in \Gamma$ can be written as
\begin{equation}
\gamma=\sum_{I=1}^r p^I \alpha_I + q_I \beta^I\,,
\end{equation}
where $(q,p)$ are the electric and magnetic charges respectively. 
One can now write down a pairing between $\gamma_{i},\gamma_j \in \Gamma$  which corresponds to the Dirac pairing of the charges $(q_{i}, p_{i})$:
\begin{equation}
\langle \gamma_{i}, \gamma_{j}\rangle = \sum_{I=1}^{r} (p_{i})^I (q_{j})_I - (p_{j})^I (q_{i})_I\,.
\end{equation}
We assume there exists a holomorphic map associated to $\Gamma$: 
\begin{equation}
Z \colon \mathcal{B} \to \mathrm{Hom}(\Gamma, \mathbb{C})\,.
\end{equation}

From this map we can write down a holomorphic function
$
Z_\gamma(w) := Z(w) \cdot \gamma
$
which is known as the \emph{central charge of $\gamma$}. 
One can also write down a \emph{mass function} as the map 
\begin{equation}
M \colon \mathcal{B} \to \mathrm{Map}(\Gamma, \mathbb{R})\,,
\end{equation}
such that for every $\gamma$ and $w\in \mathcal{B}$ the BPS bound, defined by 
\begin{equation}\label{eq:bpsbound}
M_\gamma(w):= M(w)\cdot \gamma \geq |Z_\gamma (w)|\,,
\end{equation}
is satisfied.

The BPS states can now be defined as the states with charge $\gamma \in \Gamma$ that saturate the BPS bound in equation \eqref{eq:bpsbound}.
We denote by $\mathcal{S}_w$ the BPS spectrum, defined as all the existing BPS states in $\Gamma_w$, at the point $w \in \mathcal{B}$.

\subsection{BPS walls and walls of marginal stability}\label{framedbpsstates}

Framed BPS states were introduced by Gaiotto, Moore, Neitzke (GMN) \cite{FramedBPSGMNhttps://doi.org/10.48550/arxiv.1006.0146} to describe BPS states in 4d $\mathcal{N}=2$ theories, now referred to as vanilla BPS states, bound to a line operator $L_{\zeta}$. 
These were re-interpreted in \cite{BPSgalaxiesAndriyash_2012} from a supergravity perspective, in which case the BPS states are bound to a large core charge in a multicentered supergravity solution. There is a charge located at the position of the line operator known as the core and a charge that can bind to it known as a halo. The walls for this halo to bind to the core are called BPS walls. The wall crossing we are familiar with for 4d $\mathcal{N}=2$ theories, in which the vanilla BPS states bind with each other, occurs when the second type of wall, the wall of marginal stability, is crossed. The two types of walls are distinct and are labelled by: 

\begin{align}
 \text{$MS_{\gamma_{1}, \gamma_{2}}$}: & \   \   \  \text{Wall of marginal stability} \   \   \   \   \  \  \   \   \  \   \   \   \    \    \   \ \text{Im}[ Z_{\gamma_{1}}(w)\bar{Z}_{\gamma_{2}}(w)] = 0, \\ \nonumber
 \\
 \text{$W_{\gamma_{i}}$}: & \   \   \  \text{BPS wall}
 \   \   \   \  \   \  \  \  \   \   \   \   \   \   \  \   \   \   \   \  \   \   Z_{\gamma_{i}}(w)/\zeta \in \mathbb{R}_{-}\   \  \   \  \text{Im}[Z_{\gamma_{i}}(w)/\zeta] = 0, \   \   \   \zeta \in e^{i\theta}\,,\label{BPSwall}
\end{align}
where the BPS wall also depends on the complex phase $\zeta \in \mathbb{C}$. Therefore if we write the walls of the $\mathcal{N}=2$ theories purely in terms of the central charges of the theory (and the phase), then the walls become not the walls of marginal stability $MS_{\gamma_{1}, \gamma_{2}}$ for the BPS invariants of the theory itself, but those of framed halo BPS states represented by BPS walls $W_{\gamma_{i}}$. These represent boundstates of the $\gamma_{i}$ to a large core charge. However, the intersection of these BPS walls $W_{\gamma_{i}}$ occurs on the wall of marginal stability. This is because if the imaginary part of all the central charges is vanishing, or in general $\text{Im}[Z_{\gamma_{i}}(w)/ \zeta] =0, \ \forall \gamma_{i},$ then the ratio of all the $Z_{\gamma_{i}}(w) /\zeta$ must be real. Hence the condition on the wall of marginal stability $\text{Im}[ Z_{\gamma_{i}}(w)\bar{Z}_{\gamma_{j}}(w)] = 0$ is satisfied. It should be noted that the generating functions derived here in the 4d $\mathcal{N}=2$ examples only count particular framed halo states and do not consider the core charge which is not included when one considers BPS walls in terms of central charges.

\subsection{Attractor flow and scattering diagrams} \label{attractorflowexistence}

The attractor mechanism has been derived in the context of supergravity \cite{Ferrara_1995,Strominger_1996Attractor,Ferrara_1997,Ferrara_2008} and has been developed to determine the wall crossing phenomena by using existence conditions on the endpoints of the flow lines \cite{MooreArithmeticAttractorshttps://doi.org/10.48550/arxiv.hep-th/9807087,DenefSeibergWittenemptyhole_2000,DenefGreeneRaugas_2001}, which were recently applied in \cite{Bousseau:2022snmnewattractorflowscatteringdiagram}. We used this in a previous work \cite{Alim:2023doi} applying these methods outlined in \cite{DenefSeibergWittenemptyhole_2000,DenefGreeneRaugas_2001} to Argyres-Douglas $A_{2}$ and Seiberg-Witten theory. Now to give an explanation of the wall crossing at $MS$ we state the conditions:  

\begin{enumerate}[label=(\roman*)]

\item 
We recall that for a BPS state to exist the endpoint of the flow must terminate at a singular point.

\item 
If the flow terminates at a regular point in Seiberg-Witten and Argyres-Douglas theories the central charges vanish. This can be interpreted as contradictory, as the central charges vanish at a point where the cycles in the elliptic curve do not pinch. Hence the BPS state does not exist. 

\end{enumerate}

One way of looking at the flow lines is as lines in $\mathcal{B}$ where the central charges have constant phase, and indeed the BPS walls $W_{\gamma_{i}}$ satisfy these conditions because $Z_{\gamma_{i}}(w)/\zeta \in \mathbb{R}$ on the $W_{\gamma_{i}}$. Therefore, one can use the attractor flow existence conditions on the BPS walls $W_{\gamma_{i}}$ by excluding walls that pass through regular attractor points at which the central charge vanishes. We will apply them in this way for the 4d $\mathcal{N}=2$ theories that have a wall of marginal stability $MS$ in sections \ref{SeibergWittenexamplesubsection} and \ref{A2subsection} for Seiberg-Witten and Argyres-Douglas $A_{2}$ theory respectively. When a flow line only exists on one side of the wall of marginal stability $MS$ it can be drawn as a split flow line where it is one line splitting into the flow lines of its everywhere existing constituents at $MS$.

Scattering diagrams were initially developed by Gross and Siebert \cite{https://doi.org/10.48550/arxiv.math/0703822grosssiebertscattering} and Kontsevich-Soibelman \cite{kontsevich2006affine} in their mirror symmetry programme. A good overview is given by \cite{Fantini2023ScatteringDI}. These diagrams consist of lines corresponding to wall crossing automorphisms of BPS walls. Given that, in theories with an additional wall of marginal stability, one can look at the wall crossing product identities in \cite{kontsevich2008stability} 
\begin{align}
 &\text{Argyres-Douglas $A_{2}$}: \  \   \   \  \   \   \   \   \  \  \    \    \   \   \   \ \  \  \  \  \  \   \mathcal{K}_{\gamma_{1}}\mathcal{K}_{\gamma_{2}} =  \mathcal{K}_{\gamma_{2}} \mathcal{K}_{\gamma_{1}+\gamma_{2}} \mathcal{K}_{\gamma_{1}}   \\ \nonumber
 \\ 
&\text{Seiberg-Witten theory}: \  \   \    \mathcal{K}_{\gamma_{1}}\mathcal{K}_{\gamma_{2}}=  \mathcal{K}_{\gamma_{2}}\mathcal{K}_{\gamma_{1}+2\gamma_{2}}\mathcal{K}_{2\gamma_{1}+3\gamma_{2}}...\mathcal{K}_{\gamma_{1}+\gamma_{2}}^{-2}... \mathcal{K}_{3\gamma_{1}+2\gamma_{2}}\mathcal{K}_{2\gamma_{1}+\gamma_{2}}\mathcal{K}_{\gamma_{1}}
\end{align}
where $\mathcal{K}$ are operators that act as symplectomorphisms and $\gamma_{1}, \gamma_{2}$ are the basis charges for each theory respectively.

These relations determine a different number of rays on either side of the wall of marginal stability. These rays all intersect and those that only exist on one side of the wall of marginal stability only appear as half lines on the side in which they exist. The scattering diagrams for Seiberg-Witten and Argyres-Douglas $A_{2}$ theory are described by Bridgeland \cite{BridgelandOriginalscatteringdiagram} and again in \cite{Bousseau:2022snmnewattractorflowscatteringdiagram}. The split flow lines for states existing only on one side of $MS$, that then become the seperate flow lines of the constituents on the other side of $MS$, can therefore be embedded into the scattering diagram \cite{Bousseau:2022snmnewattractorflowscatteringdiagram}. We show this explicitly for the Argyres-Douglas $A_{2}$ theory in section \ref{subsec:A2attractorflow} (see figures \ref{A2attractorlinechamber1} and \ref{A2attractorlinechamber2}).

\subsection{BPS quivers}
One can write down a directed graph called a quiver that describes the BPS spectrum of $4d$ $\mathcal{N}=2$ theories. These quiver descriptions were originally introduced for D-brane systems in string theory \cite{QuiverBackgroundDouglasMoorehttps://doi.org/10.48550/arxiv.hep-th/9603167,Diaconescu_1998BPSquivers,BPSorbifoldquiverFiol_2000,QuiversnoncompactCYDouglas_2005} and later used by Cecotti and Vafa \cite{cecottivafacompleteclassificationhttps://doi.org/10.48550/arxiv.1103.5832} in the classification of 4d $\mathcal{N}=2$ theories. A good review of quivers in this context is given in \cite{Lecturenotescecotti2010trieste}. Here we define and introduce the concept of a BPS quiver, following \cite{N=2BPSquivershttps://doi.org/10.48550/arxiv.1112.3984}, in which the BPS spectrum $\mathcal{S}_w$ of these theories was determined using the method of quiver mutations as well as the representation theory of quivers. To define a quiver we first require information about the BPS states.   
We start by fixing a point $w \in \mathcal{B}$ in the moduli space where we know $\mathcal{S}_w$. We then proceed by splitting the BPS spectrum into two sets, otherwise known as the \emph{particles} and the \emph{antiparticles}. The particles are defined to be the BPS states with central charges taking values in the upper half of the complex $Z$ plane, conversely the antiparticles have their central charges in the lower half-plane. The theories we are studying have CPT invariance which means that every BPS particle of charge $\gamma$ has a corresponding antiparticle with charge $-\gamma$. This means that the full BPS spectrum consists of the set of all the BPS particles combined with their corresponding CPT conjugate antiparticles. 

At first we just consider the particles, and then choose a minimal basis set of hypermultiplets. The charge lattice $\Gamma$ has rank $2r+f$, which means that the basis of BPS states we have chosen will consist of $2r+f$ BPS hypermultiplets, and their charges can be labelled as $\gamma_{i}$. The basis set of particles can be thought of as elementary building blocks of the complete BPS spectrum, which means that these particles must form a positive integral basis for every BPS particle that is occupied within the charge lattice $\Gamma$.  This implies that all the charges $\gamma$ of the BPS particles can now be written as
\begin{equation}
\gamma =\sum_{i=1}^{2r+f}n_{i}\gamma_{i}. \hspace{.2in}n_{i}\in \mathbb{Z}^{+} \label{nsum}
\end{equation}

One can encode such a basis of hypermultiplets with charges $\{\gamma_{i}\}$ on a natural diagram, called a \emph{quiver}.  This quiver can be constructed in the following steps \cite{N=2BPSquivershttps://doi.org/10.48550/arxiv.1112.3984}:
\begin{itemize}
\item For every element $\gamma_{i}$ in the basis, one must draw a node of the quiver.
\item For every pair of constituent charges in the basis one must compute the electric-magnetic inner product $\gamma_{i}\circ \gamma_{j}$.  If $\gamma_{i}\circ \gamma_{j}>0,$ one draws $\gamma_{i}\circ \gamma_{j}$ arrows connecting the corresponding nodes $\gamma_{i}$ and $\gamma_{j}$. Each of these arrows points from node $j$ to node $i$.
\end{itemize}

\begin{dfn}
\cite{Jankowski:2022qdpnew,browne_2021} For a quiver with a set of nodes $\{1, ... , n\}$ one can define the adjacency matrix $R$. The matrix $R_{ij}$ is then defined by the number of arrows connecting the nodes from $i$ to $j$.

One can also define the adjacency matrix $R_{\Gamma}$ of the undirected graph $\Gamma$ underlying the quiver. Now $R_{ij}$ is the number of arrows connecting nodes $i$ and $j$ without considering their direction. 
\end{dfn}

BPS quivers are defined by choosing a basis of particles in a half plane. As described by \cite{cecottivafacompleteclassificationhttps://doi.org/10.48550/arxiv.1103.5832} and also in \cite{N=2BPSquivershttps://doi.org/10.48550/arxiv.1112.3984}, a quiver can be mutated by rotating that half plane such that some particles become antiparticles. Then one can change the basis of positive roots so that one again has particles in the new half plane. These mutations occur at the walls of the second kind \cite{kontsevich2008stability,N=2BPSquivershttps://doi.org/10.48550/arxiv.1112.3984}. These can be considered for $\gamma_{i}$ when $\zeta =1$ from (\ref{BPSwall}) but for the whole real axis such that $Z_{\gamma_{i}}(w) \in \mathbb{R}$.

For example one can look at the central charge of a $\gamma_{1}$ changing sign such that the mutation one must take is:
\begin{align} \label{eq:antimutation}
\gamma'_{1} = - \gamma_{1}. 
\end{align}
The action of the mutation on the other charges then takes the form
 \cite{N=2BPSquivershttps://doi.org/10.48550/arxiv.1112.3984, Lecturenotescecotti2010trieste,quivermutationsCachazo_2002}:

\begin{align} \label{quivermutation}
\gamma'_{i} =
\begin{cases}
	\gamma_{i}+ \langle \gamma_{i},\gamma_{1} \rangle \gamma_{1},  & \text{for}  \   \ \langle \gamma_{i},\gamma_{1} \rangle > 0,\\
		\gamma_{i},  \     \    \  \  \   \   \   \  \  \ &  $for$ \  \  \langle \gamma_{i},\gamma_{1} \rangle  \leq 0.
\end{cases}
\end{align}

These mutations can therefore involve charges becoming their negative or Weyl reflections and are applied later in this work for example in subsection \ref{sec:expansiondifferentchambers}.

\subsection{Root systems and Lie algebras} \label{section:rootsystemBPSliealgebras}

For a $4d$ $\mathcal{N}=2$ ADE type Argyres-Douglas theory the BPS states can be represented by a specific Lie algebra such that the roots represent all the charge vectors in the theory and a generating function can be written down in the form of the Weyl denominator. This is constructed as a generalisation of the generating function for degeneracies of 1/4-BPS dyons studied in \cite{Cheng_2007,Cheng_2008, dabholkar2014quantum}. This being a function of the Cartan subalgebra $\mathfrak{h}(A,S)$ of the Borcherds-Kac-Moody Lie superalgebra $\mathfrak{g}(A,S)$ associated with the  $\mathcal{N}=4$ supersymmetric type II string theory on $K3 \times T^{2}$ that describes the 4d black holes. In this example the positive roots of the Lie algebra represent the possible BPS states that can exist in the theory. The Weyl chambers represent the regions in which these states are stable - the boundaries can be connected back to the moduli space of the theory. This is also the case for the Lie algebras representing analogous BPS states in $\mathcal{N}=2$ theories. For example as described in \cite{cecottivafacompleteclassificationhttps://doi.org/10.48550/arxiv.1103.5832} the $\hat{A_{1}}$ root system describes the BPS spectrum of Seiberg-Witten SU(2) theory at weak coupling. A quiver can be constructed from the Cartan matrix, and mutations in the quiver can correspond to Weyl transformations of the 
roots or the exchange of positive and negative roots depending on which side of the wall of marginal stability one does the analysis \cite{cecottivafacompleteclassificationhttps://doi.org/10.48550/arxiv.1103.5832,N=2BPSquivershttps://doi.org/10.48550/arxiv.1112.3984}. 

The general construction of a Lie algebra and its root system $\Delta$ is well known and reviewed for example in \cite{Liealgebrasrepresentations2nla.cat-vn1972748, fulton1991representation,humphreys1992reflection,Liealgebras1Hall2000Elementary, Serre2001,vermamodulenotes,WeylCharacterWalton_2013}. A good introduction to infinite dimensional examples such as affine and Kac-Moody algebras is given by \cite{KacbookMR1104219}. A Lie algebra \footnote{If it is semisimple or a generalisation thereof.} can be decomposed as $\mathfrak{g} = \mathfrak{h} \bigoplus_{\alpha \in \Delta} \mathfrak{g}_{\alpha}$  where $\mathfrak{h}\subset \mathfrak{g}$ is the Cartan subalgebra. The root subspaces are defined as $\mathfrak{g}_{\alpha} = \{ x, \  \  [h, x] = \alpha(h)x, \  \ \forall h \in \mathfrak{h}  \}$. 
A root therefore corresponds to an eigenvalue of the action of the linear adjoint operators on a vector in the eigenspace $\mathfrak{g}_{\alpha}$.  The explicit action on the coordinates can be written as $Ad_{h}(x_{b}) = [h, x_{b}] = \alpha_{b}(h)x_{b}$. These roots live in the dual linear vector space $ \Delta \subset \mathfrak{h}^{*}/\{0 \}$. One can also consider for a representation $V$ of $\mathfrak{g}_{\mathbb{C}}$ the weight space which is given by $V_{\lambda} : = \{ v \in V: \forall h \in \mathfrak{h}, \   h \cdot v = \lambda (h) v \} $. One can then generate all the weights in a representation from the root system using: $h \cdot (x \cdot v) = [(\lambda+\alpha)(h) ](x \cdot v)$. If a root is simple it cannot be written as a linear combination of other basis roots. If not it can be written in such a linear combination.

\begin{dfn}

The Killing form is an inner product represented by the trace normalization for adjoint representation generators. Consider $a,b \in \mathfrak{g}_{\mathbb{C}}$. Their Killing form
is represented by
\begin{align}
(a,b) = \text{Tr}[Ad_{a}, Ad_{b}],   \      \     \     \    \mathfrak{g} \times \mathfrak{g} \rightarrow        \mathbb{C}.
\end{align}
This can be computed by finding matrices representing adjoint operators for a particular basis and then calculating matrix products.

\end{dfn}

Alternatively this can also be calculated by writing the form in terms of brackets as \newline
$Ad_{a_{i}}Ad_{b_{j}}(x_{k}) = [x_{i},[x_{j},x_{k}]]$. We recall that the root space is in the dual $\mathfrak{h}^{*}$ of the Cartan subalgebra $\mathfrak{h}$ which can be understood in terms of a Killing form. For every root $\alpha \in \mathfrak{h}^{*}$, there
exists an isomorphism $\mathfrak{h}^{*} \rightarrow \mathfrak{h}$ such that we have $u, h_{\alpha} \in \mathfrak{h}$ such that $\alpha(u) = (\alpha,u) = (h_{\alpha}, u)$. One can now write the central charges with the phase $\zeta$ for the Line operator (subsection \ref{framedbpsstates}) for the Framed BPS states in terms of this product as
\begin{align} \label{eq:rootscharges}
\alpha_{i}(u) = Z_{\gamma_{i}}(w)/ \mu = Z_{\alpha_{i}}(w)/ \mu, 
\end{align}
where $\mu = \epsilon \zeta$ such that $\epsilon \in \mathbb{R}$ is a small parameter. In an $\mathcal{N}=2$ context of framed BPS states it can be associated to the large core charge. For the $\mathcal{N}=4$ black holes it is used to define the contour used to extract the degeneracies \cite{Cheng_2007}. In section \ref{subsection:affineA1contourprescription} we will show how this contour prescription can be restricted to the $\hat{A}_{1}$ subalgebra. Therefore the analogous $\mathcal{N}=2$ contour one can take to obtain BPS degeneracies can also be parameterised in the Cartan subalgebra $u \in \mathfrak{h}$.  

One can also define another inner product between real roots and real linear combinations of them
\begin{align}
\langle \alpha, \beta \rangle   \  = \  (h_{\alpha}, h_{\beta}),   \      \     \     \ \mathfrak{h}^{*} \times \mathfrak{h}^{*} \rightarrow   \mathbb{R}. 
\end{align}
Now we can construct the Cartan matrix for a Lie algebra from its root system and the inner product shown here. We show the examples for $A_{1}, \hat{A}_{1} $ and $A_{2}$. The matrix contains entries of the form 
\begin{align} \label{cartanmatrix}
A_{i,j} =    2\frac{\langle \alpha_{i}, \alpha_{j} \rangle}{\langle \alpha_{i}, \alpha_{i} \rangle}.    
\end{align} 

\begin{center}

\begin{align*}
\hspace{-4cm}
A_{1}:   \    \     \    \ (2) 
\hspace{5cm}
\begin{tikzcd} [execute at end picture={
\node[circle,draw=black,fill=red, line width=0.5mm,minimum size=17pt,inner sep = 0pt] (1) at(1.1,0) {};}]
\text{\Huge{$\alpha$}}  \ \ 
\end{tikzcd}	
\end{align*}
\end{center}
\vspace{0.65cm}
$\   \    \    \   \      \    \  \   \     \    \    \    \hat{A}_{1}:   \    \     \     \
\begin{pmatrix}
2 &  -2\\
-2 & 2
\end{pmatrix}$ \   \   \   \   \   \   \   \  \   \   \    \   \    \   \     \begin{Huge}
		\begin{tikzcd} [execute at end picture={
\node[circle,draw=black,fill=yellow, line width=0.5mm,minimum size=17pt,inner sep = 0pt] (1) at(1.6,0.11) {};
\node[circle,draw=black,fill=green, line width=0.5mm,minimum size=17pt,inner sep = 0pt] (1) at(-1.55,0.11) {};
}] \alpha_{\text{\small{0}}}\ \ \circ \arrow[r, shift left, "  \scalebox{2.7}{\tiny{b}}_{\text{\small{1}}} \    \scalebox{2.7}{\tiny{b}}_{\text{\small{2}}} \ \ "] 
		  \arrow[r,  shift right]  
		& \circ \ \ \alpha_{\text{\small{1}}}    
\end{tikzcd}
\end{Huge}
\\
\\
\\
\\
\\
\hspace{7cm}$\ \ \phantom{aaaaaa}   A_{2}:   \  \  \   \ 
\begin{pmatrix}
2 &  -1\\
-1 & 2
\end{pmatrix}$ \   \   \    \    \  \   \     \   \ \phantom{.'''''}  \ \begin{Huge}
		\begin{tikzcd}[execute at end picture={
\node[circle,draw=black,fill=cyan, line width=0.5mm,minimum size=17pt,inner sep = 0pt] (1) at(1.6,0.1) {};
\node[circle,draw=black,fill=cyan, line width=0.5mm,minimum size=17pt,inner sep = 0pt] (1) at(-1.55,0.1) {};
}]	\alpha_{\text{\small{1}}} \ \ \circ \arrow[r,  "\scalebox{2.7}{\tiny{b}}_{\text{\small{1}}} "]  
		& \circ \ \  \alpha_{\text{\small{2}}} 
		\end{tikzcd}
	\end{Huge}
\\
\\
\\
\\
The quiver can be constructed from this by plotting the roots and the arrows $\mathbf{b}_{i}$ between them. The Cartan matrix is related to the adjacency matrix $R_{\Gamma}$ 
for the undirected graph associated to the quiver
\begin{align}
 A = 2I - R_{\Gamma}.
\end{align}
We give the examples for the undirected graph for the  $\hat{A}_{1}$ and $A_{2}$ quivers:

\begin{align*} 
&R_{ij}: \\
\\
&\hat{A}_{1}:   \    \     \     \
\begin{pmatrix}
0 &  2\\
2 &  0
\end{pmatrix} \    \     \     \  \  \  \  \  \  \  \  \  \          
\\
\\
&A_{2}:   \    \     \     \
\begin{pmatrix}
0 &  1\\
1 & 0
\end{pmatrix}  \    \     \     \  \  \  \  \  \  \  \  \           
\\
\end{align*}

These are shown above to the right of the Cartan matrices. 
Finally, the Weyl group is a subgroup of the isometry group of the root system which is generated by reflections through the spaces 
perpendicular to the roots. These represent quiver mutations in the quiver encoding the roots $\alpha_{i}$ representing the BPS states. The transformations acting on the element of the Cartan subalgebra $u$ are then generated by the maps 
\begin{align}
\mathbf{w}_{\alpha_{i}}(u) = u - 2 \frac{(u,\alpha_{i})}{(\alpha_{i}, \alpha_{i})} \alpha_{i}.
\end{align}

\subsection{Weyl denominator and generating function} \label{subsectionADEWeyldenominator}

From a general root system associated with a Lie algebra one can define the Weyl denominator formula that was introduced for the Borcherds-Kac-Moody algebra (\ref{eq:dyoncountingformula}) and its subalgebras (\ref{Weyldenominatorsubrepresentations}) as the inverse of a product of factors involving all the positive roots of the Lie algebra. We aim to find a generating function for BPS degeneracies corresponding to a particular root for the $\mathcal{N}=2$ examples following the approach previously carried out for the $\mathcal{N}=4$ cases. In general, as in the literature \cite{Cheng_2007, dabholkar2014quantum, N=4DyonDegKKCMprodDavid_2006, K3T4dualN=4Sen_2007} we
expect this to take the form of a contour integral over the Weyl denominator, although it should be possible to read off the BPS state count from the denominator itself in terms of a highest weight. We conjecture the following general result for this formula, that an integral of the Weyl denominator over the Cartan subalgebra is related to a degeneracy of a particular root or combination of roots in the Lie algebra.

Now to proceed, one must write the inverse Weyl denominator and multiply it by an additional charge factor $e^{\Lambda(u)}$ to obtain the generating function:
\begin{align} \label{eq:originalweyldenominatordefinition}
 g(u) = \frac{e^{\Lambda(u)-\rho(u)}}{\prod_{\alpha \in \Delta^{+}}(1-e^{-\alpha(u)})},
\end{align}
where $r$ is the rank of the Lie algebra and $\Lambda$ is analogous to the charge vector in \cite{Cheng_2008}. The product is over positive roots $\alpha \in \Delta^{+}$. The Weyl vector $\rho$ is the half sum of positive roots $\rho = \frac{1}{2}\sum_{\alpha \in \Delta_{+}} \alpha_{i}$.  All roots $\alpha_{i}$ and weights $\lambda_{i}$ are contained in a charge lattice $\alpha_{i}, \lambda_{i} \in \Gamma$.

There is another object one can define in general from a semi-simple  Lie algebra $\mathfrak{g}$ and its Borel (maximally solvable) subalgebra $\mathfrak{b}$. This is the Verma module. A detailed review of Verma modules is given in \cite{Vermamodulesbookalma9949384756902959}. 

\begin{dfn}

One first considers the Cartan subalgebra $\mathfrak{h}$ again. Then one can define a Verma module $M_{\lambda}$ with highest weight $\lambda \in \mathfrak{h}^{*}$ in the adjoint by
\begin{align} \label{eqverma}
   \text{Hom}_{\mathfrak{g}}(M_{\lambda},V) = \text{Hom}_{\mathfrak{b}}(\mathbb{C}_{\lambda},V),
\end{align}
where $V$ is a representation of $\mathfrak{g}$, and $\mathbb{C}_{\lambda}$ is the one-dimensional module on which elements of $\mathfrak{h}$ act on with $\lambda$. 

\end{dfn}

Verma modules are infinite dimensional, but one can take quotient modules with highest weights that correspond to those of representations of finite semi-simple Lie algebras. The character of a Verma module is (up to a shift by the Weyl vector $\Lambda(u) \rightarrow \lambda(u)$ \footnote{One can redefine $\lambda(u) = \Lambda(u)-\rho(u) $ to absorb this shift if one is working with an inverse product to move between these definitions.}) the inverse of the Weyl denominator. The exponent in the numerator of (\ref{eq:originalweyldenominatordefinition}) should extract Fourier coefficients from the denominator and can also be assigned a representation of the Lie algebra or module to which it is the highest weight. Hence, we expect the stability of BPS states to coincide with the stability of representations.

\begin{dfn}
 The character of the Verma module is defined as
\begin{align} \label{eq:verma }
 v_{\lambda}(u) = \frac{e^{\lambda(u)}}{\prod_{\alpha \in \Delta^{+}}(1-e^{-\alpha(u)})}.
\end{align}
\end{dfn}

\begin{dfn}

A Lie algebra also has a character function which includes a numerator summing over the possible Weyl transformations of the representation in question. The full Weyl character formula takes the form
\begin{equation} \label{eq:firstfullcharacter}
ch_{\lambda}(u) = \frac{\sum_{\mathbf{w}\in W} (\det{\mathbf{w}}) e^{\mathbf{w}(\lambda+ \rho)(u)}}{e^{\rho(u)}\prod_{\alpha \in \Delta^{+}}(1-e^{-\alpha(u)})},
\end{equation}
where $\mathbf{w}$ are elements of the Weyl group and $\lambda$ is the highest weight of the representation chosen. 

\end{dfn}

Our work presented in section \ref{workforanalogs} determines how this generating function counts BPS states in simple Argyres-Douglas examples. In the table below we list what we expect the generating function to become when we substitute the Weyl denominators of the respective Lie algebras into the general formulation. This is expected to be the function which determines what BPS states exist in each chamber of the moduli space and will be verified as such in the subsections of chapter \ref{workforanalogs}. This should reproduce the counts that have been obtained from other methods such as quiver representations. In the next sections we look explicitly at each example to reproduce the BPS state counts described for these theories in the literature and describe the wall crossing in the new language involving root systems and Weyl chambers.

\begin{center}
	\begin{tabular}{ | l | l | p{5cm} |}
		\hline
		Generating \vspace{-0.3cm} & \vphantom{\Huge{HH}} $ \frac{\sum_{\mathbf{w} \in W} (\det{\mathbf{w}}) e^{\mathbf{w}(\lambda+\rho)(u)}}{e^{\rho(u)}\prod_{\alpha \in \Delta^{+}}(1-e^{-\alpha(u)})}   \ \  \  \   \  \  \    \text{The inverse denominators are shown below}$   \\[8pt] \  \ function  &  \\[0pt] \hline
		$A_{1}$ & \vphantom{\Huge{HH}} $   \frac{e^{ \Lambda(u) }}{e^{\rho(u)} (1-e^{-\alpha(u)} ) }  $  \\[8pt] \hline
        $\hat{A}_{1}$  & \vphantom{\Huge{HH}} $  \frac{e^{\Lambda(u)}}{e^{\rho(u)} \prod_{m=1}^{\infty}(1- e^{-m (\alpha_{0}(u)+\alpha_{1}(u))})(1- e^{-m(\alpha_{0}(u)+\alpha_{1}(u))-\alpha_{1}(u)})(1- e^{-(m-1) (\alpha_{0}(u)+\alpha_{1}(u))+\alpha_{1}(u)})}   $  \\[8pt]  \hline
	$A_{2}$ & \vphantom{\Huge{HH}} $  \frac{e^{\Lambda(u)}}{e^{\rho(u)}(1-e^{-\alpha_{1}(u)})(1-e^{-\alpha_{2}(u)})(1-e^{-\alpha_{3}(u)})}  $  \\ [8pt]
	\hline 
\end{tabular}
\end{center}

\subsubsection{Wall crossing as a Fourier expansion}

Here we show the simplest example of how wall crossing is encoded in this generating function for a single root appearing or disappearing at a wall. This can be taken as the single root $\alpha$ of the $A_{1}$ Lie algebra before we consider the affine version. For every such factor in such a generating function there are 2 possible ways of expanding it in 2 distinct regions of the moduli space bounded by a wall:  
\begin{align} \label{eq:A1exp}
 &f_{A_{1}}(u) = \\ \nonumber
 \\ \nonumber
 &\begin{cases}
 \frac{e^{\lambda(u)}}{(1-e^{-\alpha(u)})} = e^{\lambda(u)}\left(1+e^{-\alpha(u)}+e^{-2\alpha(u)}+e^{-3\alpha(u)}+... \  \right) \  & |e^{-\alpha(u)}| < 1, \  \text{Im}[i\alpha(u)]>0,  \\ 
 \\
 -\frac{e^{(\lambda+\alpha)(u)}}{(1-e^{\alpha(u)})} =-e^{(\lambda+\alpha)(u)}\left( 1+e^{\alpha(u)}+e^{2\alpha(u)}+e^{3\alpha(u)}+... \right) \   & |e^{-\alpha(u)}| > 1, \  \text{Im}[i\alpha(u)]<0.
\end{cases}
\end{align}

One can also change the basis of the root to re-write it in its original form. In this example this involves just the antiparticle $\alpha' = -\alpha$. For the more complicated examples of the $\hat{A}_{1}$ and $A_{2}$ discussed later in this work this will also involve Weyl reflections. In this case the generating function becomes:
\begin{align*}
-\frac{e^{(\lambda'-\alpha')(u)}}{(1-e^{-\alpha'(u)})}. 
\end{align*}

Here we can see that in the new basis the highest weight of the Verma module has shifted to become $\lambda'-\alpha'$. This represents a discontinuous jump in the number of BPS states. In general a theory will have many factors of this form in the denominator. The total number of vanilla BPS states existing in a particular theory can be counted by considering the highest weight in the Weyl chamber where the maximal number of roots are added to the highest weight.  

\subsubsection{Physical interpretation and types of  wall crossing}

The wall crossing described here in (\ref{eq:A1exp}) represents, in an $\mathcal{N}=4$ setup, the decay of the $\frac{1}{4}$BPS dyons into electric and magnetic $\frac{1}{2}$BPS states. We construct an $\mathcal{N}=2$ analog of this wall crossing where this jumping at Weyl chamber boundaries represents the decay of the framed halo BPS states, bound to a large core charge, described by Gaiotto, Moore, Neitzke \cite{FramedBPSGMNhttps://doi.org/10.48550/arxiv.1006.0146} and Jafferis-Moore \cite{D6D2D0JafferisMoorehttps://doi.org/10.48550/arxiv.0810.4909}. Hence these are BPS walls or rays. There is another type of wall crossing at the walls of marginal stability in the $\mathcal{N}=2$ theories.
This is where the vanilla BPS states decay into constituents. In the generating function this is represented by the exclusion of the factor in the denominator containing the root that no longer exists on the other side of the wall of marginal stability. This exclusion can be read off from the attractor flow existence conditions where a flow line representing a BPS wall is excluded by a regular termination point. This is described in sections \ref{SeibergWittenexamplesubsection} and \ref{A2subsection} for $\hat{A}_{1}$ and $A_{2}$ respectively. The 2 types of walls are as follows:

\begin{enumerate}[label=(\roman*)]

\item 
The BPS walls for $\mathcal{N}=2$ framed BPS states are encoded in the jump of the Fourier series due to the expansions distinguished by positive or negative imaginary part of $i \alpha(u)$ for a given root.

\item 
The wall of marginal stability for a root is encoded in the generating function by a loss of a factor in the denominator associated to the root. 

\end{enumerate}

\subsubsection{ADE generalisation}

All the above generating functions and wall crossing phenomena should be generalisable to ADE type Lie algebras when we input the data of the root system, Weyl denominator formula, Verma module character, Cartan matrix $A_{i,j}$ and the undirected adjacency matrix of the associated quiver $R_{\Gamma}$. We should then also expect the quiver mutation sequences in the different chambers in the moduli space seperated by $MS$ to match the factors in the denominator in the different chambers. Furthermore we can also conjecture a generalisation to all 4d $\mathcal{N}=2$ quivers in which the factors in the denominator are associated to the mutation sequence for that particular quiver in the chambers (seperated by the walls of marginal stability $MS$) that we are considering. However, because the mutations would in this general case no longer be Weyl reflections, we do not expect the resulting generating function to be a Weyl denomoninator formula but a generalisation of this.  

\newpage

\section{Generating function and wall crossing for $\hat{A}_{1}$ theories} \label{sec:affineA1extraction}

In the following, we derive the generating function for the theories with BPS states described by the BPS root system of $\hat{A}_{1}$. We first consider this root system as the $\mathcal{N}=4$ $\hat{A}_{1}$ subalgebra of the Borcherds-Kac-Moody algebra. To do this we will extract it in the form of a Weyl denominator of the subalgebra (see \cite{subalgebra1Govindarajan:2021pkk,subalgebra2Govindarajan:2022jmo} for the construction of the subalgebra) from the full Weyl denominator describing the Borcherds-Kac-Moody algebra in the dyon counting function. Then we take the restrictions for the degeneracies and wall crossing phenomena for the original dyon counting formula.

\subsection{Weyl denominator for the $\mathcal{N}=4$ subalgebras} \label{Weyldenominatorsubrepresentations} 

First we find the Weyl denominators for the subalgebras within the Weyl denominator for the Borcherds-Kac-Moody algebra. To do this we can write the partition function (\ref{eq:dyoncountingformula}-\ref{eq:expansionFouriercoefficients}) more explicitly in the form of
\begin{align} \label{eq:partitionfunctionfirstline}
&\frac{1}{\Phi_{10}(\Omega)} =
\frac{e^{( 2\rho,\Omega)}}{\prod_{\alpha \in \Delta^{+}}(1-e^{- (\alpha, \Omega)})^{2\text{mult}(\alpha)}} \\ \nonumber
\\ \nonumber
& = e^{-  2\pi i \tau} \prod_{n=1}^{\infty}(1-e^{ 2\pi i n \tau})^{-18}  \prod^{\infty}_{l=1} \frac{\mathcal{Z}(\Omega)}{(1-e^{2\pi i l \tau})^{2} (1-e^{ 2\pi i l \tau+ 2\pi i v})^{2} (1-e^{ 2\pi i l \tau- 2\pi i v})^{2}} \frac{e^{ -2\pi i v}}{(1-e^{ -2\pi i  v})^{2}}, \nonumber
\end{align}
where here the partition function from (\ref{eq:dyoncountingformula}-\ref{eq:expansionFouriercoefficients}) is split into the various components of the center of mass motion of the black hole, and a factor that contains the partition function for a rotating D1-D5 Strominger-Vafa system, as discussed in \cite{dabholkar2014quantum}. This contains the third complex variable $\sigma$. We call this $\mathcal{Z}(\Omega)$. This can be written as follows \footnote{$c(4kl-m^{2})$ are Fourier coefficients of the $K3$ elliptic genus.} 
\begin{align}
 \mathcal{Z}(\Omega) = e^{-  2\pi i \sigma} \prod^{\infty}_{l \geq 0,  k >0, m} \frac{1}{(1-e^{ 2\pi i l \tau+ 2\pi i mv+2\pi i k \sigma})^{c(4kl-m^{2})} }.
\end{align}
Now that we know the partition function, one can extract the Weyl denominators for the subalgebras $A_{1}$ and its affinisation $\hat{A}_{1}$. These should be present within the partition function (\ref{eq:partitionfunctionfirstline}) on the first line. Specifically these can be read off from the partition function as
\begin{align} \label{eq:differentdenominatorbrackets}
\frac{1}{\Phi_{10}(\Omega)} =
 &  \Bigg(\prod^{\infty}_{l=1} \frac{e^{-  2\pi i \tau} }{(1-e^{2\pi i l \tau})^{2} (1-e^{ 2\pi i l \tau+ 2\pi i v})^{2} (1-e^{ 2\pi i l \tau- 2\pi i v})^{2}} \Bigg( \frac{e^{ -2\pi i v}}{(1-e^{ -2 \pi i  v})^{2}  } \Bigg)_{A_{1}} \Bigg)_{\hat{A}_{1}} \nonumber \\ & \times \prod_{n=1}^{\infty}(1-e^{ 2\pi i n \tau})^{-18} \mathcal{Z}(\Omega).
\end{align}
The factors shown here in brackets, denoted by $A_{1}$ and $\hat{A}_{1}$, are the Weyl denominators of the subalgebras. Hence, we can recalling from section \ref{section:chengverlindereview}, that the (real) roots existing within such a Weyl denominator correspond to  multicentered dyons and undergo wall crossing across  the Weyl chamber boundary associated to each root. Next one can extract particular degeneracies that know about the wall crossing for the subalgebras. This is greatly simplified if one chooses particular limits. For example if we take the limit of large $\sigma$ 
such that we have $\sigma \rightarrow +i \infty $, this ensures that none of the walls in the $ \mathcal{Z}(\Omega)$ partition function are crossed, because the imaginary parts of the products with the roots are well above the walls of marginal stability. This also means the contour $\mathcal{C}$ from \cite{Cheng_2007}, used to extract the degeneracies of the roots is always satisfied. This is because, for: $ \text{Im}[\tau] \text{Im}[\sigma] >> (\text{Im}[v])^{2}+1$, one can let $\text{Im}[\tau]$ and $\text{Im}[v]$ be small and still be well within the bound because of the large value for $\text{Im}[\sigma]$.

\subsection{Degeneracies} \label{degeneracies}

Considering the fact that the partition function determines black hole degeneracies \cite{Dijkgraaf_1997,Cheng_2008,dabholkar2014quantum} and their jumps during wall crossing \cite{Cheng_2008},  we look to extract the particular degeneracies that contain the Weyl denominators in (\ref{eq:differentdenominatorbrackets}). These  will thereby exhibit wall crossing of the roots in the subalgebra. We look at particular sets of Fourier coefficients. These Fourier coefficients can be extracted from the dyon degeneracy formula (\ref{eq:dyoncountingformula}) for the Borcherds-Kac-Moody algebra. This generating function can be written more explicitly in terms of the electric and magnetic charge invariants, and in general takes the form
\footnote{$c(|\alpha|^{2})$ are again Fourier coefficients of the $K3$ elliptic genus defined by the value $|\alpha|^{2}$ \cite{Cheng_2008}.}
\begin{equation}
(-1)^{P \cdot Q+1}D(P,Q) = \oint_{\mathcal{C}}d \rho dv d\sigma \   e^{i \pi \Lambda_{P,Q}(\Omega)} \Bigg( \frac{e^{-(\pi iv+\pi i\tau+\pi i\sigma)}}{\prod_{\alpha \in \Delta^{+}}(1-e^{ -i \pi (\alpha, \Omega)})^{\frac{1}{2}c(|\alpha|^{2})}} \Bigg)^{2}. 
\end{equation}

where (see e.g. \cite{Cheng_2008,dabholkar2014quantum}) the contour $\mathcal{C}$ for charges $(P,Q)$ is given by: $\text{Re}(\Omega) \in T^{3}, \ \text{Im}(\Omega) =  \frac{1}{\epsilon}\frac{Z_{P,Q}}{|Z_{P,Q}|}$,

and we also recall that 
\begin{align}
\Lambda_{P,Q} = 	
\begin{pmatrix}
P \cdot P &  P \cdot Q  \\
P \cdot Q   &  Q \cdot Q \\
\end{pmatrix},
\  \  \  \   \   \   \   \   \  \  \   \  \Omega = 	
\begin{pmatrix}
\sigma &  v  \\
v   &  \tau \\
\end{pmatrix}. 
\end{align}

The action of the charge vector on the moduli in $\mathcal{N} =4$ can in general be written as: \footnote{One can also let $P\cdot Q \rightarrow -P \cdot Q$ with a different basis of positive roots.} $\Lambda_{P,Q}(\Omega) = -P \cdot P  \tau -Q \cdot Q   \sigma - 2 P \cdot Q  v$. One can define integers \footnote{The inner product is invariant under $SO(22,6)$ and is defined on the lattice $\Gamma^{22,6}$ \cite{Dijkgraaf_1997,SEN_1994dualitygroup, Harveydualitygrouphttps://doi.org/10.48550/arxiv.hep-th/9407111}.}
\begin{equation}
l =  \frac{1}{2}P \cdot P , \  \ k = \frac{1}{2}Q \cdot Q, \ \ m = P \cdot Q,
\end{equation}
and use these to relabel the degeneracies in terms of the Fourier coefficients introduced in (\ref{eq:expansionFouriercoefficients}) for these integers called
\begin{equation} \label{degeneraciesFKM}
(-1)^{P \cdot Q+1}D(P,Q) = F_{KM}(k,l,m).
\end{equation}
One can extract degeneracies in such a way that the coefficients describe the wall crossing associated with a particular Weyl denominator of a chosen subalgebra. This can be done by choosing a suitable sublattice of charges. One does this by fixing some of the above invariants in the lattice, while letting others vary. For example one could fix $k  =1$. From this, one can define further degeneracies as
\begin{equation} \label{eq:degeneraciesflm}
F_{KM}(1,l,m) = f(l,m) = \oint_{\mathcal{C}}e^{-2\pi i mv-2\pi i l \tau} d \rho d v d \sigma \Bigg( \frac{e^{-(\pi iv+\pi i\tau)}}{\prod_{\alpha \in \Delta^{+}}(1-e^{ -i \pi (\alpha, \Omega)})^{\frac{1}{2}c(|\alpha|^{2})}} \Bigg)^{2}. 
\end{equation}
If we start with the 3 dimensional complex integral over $\rho, v$ and $\sigma$ we can integrate out $\sigma$, so that what is left is a complex 2 dimensional integral over $\rho$ and $\tau$. The degeneracies we are left with depend just on 2 charges $l,m$. These degeneracies then produce an integer valued count of the roots in the algebra also defined as in \cite{Cheng_2008} as a “second quantised multiplicity” of the roots. This is a count of how many combinations of roots can add to produce the root associated to the charge vector in the degeneracy, where one can choose 2 sets of positive roots that one can use to form the combinations.

\subsection{$\hat{A}_{1}$ as a subalgebra} \label{extractingaffineA1fromKacMoody}

We can now go back to the original equation (\ref{eq:differentdenominatorbrackets}) and extract a different set of coefficients. This time we look for the Weyl denominator of $\hat{A}_{1}$, which is a form of the Jacobi-theta function or Jacobi triple product. This affine Lie algebra is another subalgebra \cite{subalgebra1Govindarajan:2021pkk,subalgebra2Govindarajan:2022jmo} of the Borcherds-Kac-Moody algebra which the generating function for the $\mathcal{N} =4$, $1/4$-BPS states describes. This can be seen clearly when looking at the (real part of the generalised) Cartan matrix for the Borcherds-Kac-Moody algebra used in \cite{Dijkgraaf_1997,Cheng_2008}. In this case the inner product between the roots $\alpha_{i}, \alpha_{j} \   \   i,j  = 1,2,3$ of the form $\tilde{A}_{i,j} = (\alpha_{i}, \alpha_{j})$. The construction of this algebra was already classified and related to the Igusa cusp form by Gritsenko and Nikulin in \cite{gritsenko1995siegel,Gritsenko_1996,classificationGritsenko_2002}. For the affine Lie algebra $\hat{A}_{1}$ the basis only has 2 elements such that $\alpha_{i}, \alpha_{j} \   \   i,j  = 1,2$. Here we again consider $A_{ij} = (\alpha_{i}, \alpha_{j})$ the 2 matrices become 
\begin{align}
\tilde{A}_{i,j} = 	
\begin{pmatrix}
2 &  -2 & -2 \\
-2   &  2 & -2  \\
-2   &  -2 & 2   \\
\end{pmatrix}, 
\   \     \   \ 
A_{i,j} = 	
\begin{pmatrix}
2 &  -2  \\
-2   &  2  \\
\end{pmatrix}, 
\end{align}
where we can see that we can embed the matrix  $c_{i,j}$ within the larger matrix and hence one can see how the $\hat{A}_{1}$ Lie algebra is contained within that for the Borcherds-Kac-Moody algebra. In fact when one can see that there are now 2 different ways that the Cartan matrix $\hat{A}_{1}$, $c_{i,j}$ is contained in the larger Cartan of the Bocherds-Kac-Moody algebra $C_{i,j}$. This information can be used to 
find the Weyl chambers, degeneracies and wall crossing phenomena of $\hat{A}_{1}$ within the construction for the Kac-Moody algebra.

This Weyl denominator can again be extracted as a generating function of particular degeneracies of the full Borcherds-Kac-Moody algebra. 
As discussed in sec. \ref{degeneracies} the degeneracies of the sum of roots (\ref{eq:expansionFouriercoefficients}-\ref{degeneraciesFKM}) in the Kac-Moody algebra, given by $F_{KM}(p,q,r)$ \footnote{Here we re-write $F_{KM}(k,l,m)$ as $F_{KM}(p,q,r)$ by using a different basis of roots.}, are then a product of those of the subalgebras including that of the $\hat{A}_{1}$. This can be seen by decomposing the generating function from (\ref{eq:expansionFouriercoefficients}) as
\begin{align} \label{splittingdegeneracies}
& F_{KM}(p,q,r) = \oint d^{3}\Omega \frac{e^{p \alpha_{1}(\Omega)+q\alpha_{2}(\Omega)+r\alpha_{3}(\Omega) }}{\Phi_{10}(\alpha_{1}, \alpha_{2},\alpha_{3})(\Omega)} \    = \oint d^{3}\Omega \frac{e^{p \alpha_{1}(\Omega)+q\alpha_{2}(\Omega)+r\alpha_{3}(\Omega) }}{\theta(\alpha_{1},\alpha_{2})(\Omega)^{2} \tilde{\Phi}_{10}(\alpha_{1}, \alpha_{2}, \alpha_{3})(\Omega)}\\ \nonumber
\\ \nonumber
& = \oint d^{3}\Omega \  e^{p \alpha_{1}(\Omega)+q\alpha_{2}(\Omega)+r\alpha_{3}(\Omega)} \\ \nonumber \\ \nonumber  & \Big(\sum_{k,h}f_{\hat{A}_{1}}(k,h)e^{-k \alpha_{1}(\Omega)-h\alpha_{2}(\Omega)}\Big)\Big(\sum_{c,d,e}d_{\tilde{\Phi}}(c,d,e)e^{-c \alpha_{1}(\Omega)-d\alpha_{2}(\Omega)-e\alpha_{3}(\Omega)}\Big),
\end{align}
where we have decomposed the denominator into the product of the degeneracies of the affine Lie algebra $f_{\hat{A}_{1}}(k,h)$ and that of the partition function left over $d_{\tilde{\Phi}}(c,d,e)$ after the factor corresponding to the affine Lie algebra is extracted. In this case the degeneracies of the Kac-Moody algebra can be written as the product 
\begin{align} \label{eq:affineA1degeneracies}
F_{KM}(k+c,h+d,e) = \sum_{k,h,c,d} f_{\hat{A}_{1}}(k,h)d_{\tilde{\Phi}}(c,d,e).
\end{align}
The next step for us is to look at the degeneracies $f_{\hat{A}_{1}}(k,h)$ of the affine Lie algebra $\hat{A}_{1}$. The degeneracies $d_{\tilde{\Phi}}(c,d,e)$ then come from the rest of the product including the factor $\eta(\tau)^{-18}$ from (\ref{eq:differentdenominatorbrackets}).

\subsubsection{Extracting generating function for $\hat{A}_{1}$}

The factor $f_{\hat{A}_{1}}(k,h)$ itself in (\ref{eq:affineA1degeneracies}) is not directly a degeneracy (of the form $f(l,m)$ in (\ref{eq:degeneraciesflm})) of the Borcherds-Kac-Moody Weyl denominator because of the removal of the factor of $\eta(\tau)^{-18}$ . However, it contains all the information about the wall crossing. This is because this is a factor containing only imaginary roots such that it does not change under the Weyl reflections associated to the permutations of the other roots. Furthermore, the wall $\text{Im}[\tau] =0$ is not crossed in the $\mathcal{N}=4$ theory because the moduli stay within the Siegel upper half-plane. If one considers only functions with 2 complex variables, this can be interpreted as preserving the modularity of the eta and theta functions, with $\tau$ being defined on the complex upper half plane.  If we look back to the product of the different partition functions (\ref{eq:differentdenominatorbrackets}) and look at what is contained in the bracket, we can extract Fourier coefficients of this generating function, now denoted by $g_{\hat{A}_{1}}(\tau,v)$: 
\begin{align} \label{equation:thetafunction}
g_{\hat{A}_{1}}(\tau,v) =\prod^{\infty}_{  l = 1 }  \frac{e^{ -2\pi i v-  2\pi i \tau} }{(1-e^{2\pi i l \tau})^{2} (1-e^{ 2\pi i l \tau+ 2\pi i v})^{2} (1-e^{ 2\pi i l \tau- 2 \pi i v})^{2}(1-e^{ -2\pi i  v})^{2} }.
\end{align}
Now one can write this in terms of the roots of the affine Lie algebra: 
\begin{align} \label{variablesintermsofroots1}
\alpha_{0}(u) = -2\pi i v,  \     v \in \mathbb{C} \  \  \text{and}  \  \  \delta(u) = -2\pi i \tau, \   \  \tau \in \mathbb{H}^{+}.
\end{align}
Then we can write the function in the form
\begin{align}
g_{\hat{A}_{1}}(u) = \prod^{\infty}_{  m = 1 }  \frac{e^{ \alpha_{0}(u)+ \delta(u)} }{(1-e^{ -m \delta(u)})^{2} (1-e^{  -m \delta(u)+  \alpha_{0}(u)})^{2} (1-e^{-m \delta(u)- \alpha_{0}(u) })^{2}(1-e^{ \alpha_{0}(u)})^{2} }  = \\ \nonumber
\prod^{\infty}_{  m = 1 }  \frac{e^{ \alpha_{0}(u)+ \delta(u)} }{(1-e^{ -m \delta(u)})^{2} (1-e^{-(m-1) \delta(u)+ \alpha_{0}(u)})^{2} (1-e^{  -m \delta(u)- \alpha_{0}(u) })^{2} }. 
\end{align}
Now the coefficients we want to extract become 
\begin{align} \label{eq:affinecharge1}
f_{\hat{A}_{1}}(k,h) = & \oint_{\gamma} d\alpha_{0}(u) d \delta(u)  e^{k\alpha_{0}(u)+h\delta(u)}  \\ \nonumber  &  \times \frac{e^{ \alpha_{0}(u)+ \delta(u)} }{\prod^{\infty}_{  m = 1 }(1-e^{ -m \delta(u)})^{2} (1-e^{-(m-1) \delta(u)+ \alpha_{0}(u)})^{2} (1-e^{  -m \delta(u)- \alpha_{0}(u) })^{2} }, 
\end{align}
where the contour $\gamma \subset \mathcal{C}$ is again that which was previously defined in the Siegel upper half-plane, but this time restricted to the $\tau$ and $v$ planes. We define the charge vector restricted to the $\hat{A}_{1}$ subalgebra as
\begin{align} 
\lambda_{k,h} = k \alpha_{0}+h\delta \  \  k,h \in \mathbb{Z}.
\end{align}
We assume that the imaginary part of $\sigma$ is large so that it satisfies all the conditions in the contour $\mathcal{C}$:
\begin{align}
\       \  \   \   \ & 0 \leq \text{Re}[\tau], \text{Re}[v] < 1,  \    \     \text{Im}[\tau] > 0, \    \     \   \text{Im}[\tau] (\text{Im}[\sigma] \rightarrow + \infty) >> (\text{Im}[v])^{2}, \\ \nonumber
\\ \nonumber
& \text{so the generating function now becomes}
\end{align}
\begin{align} \label{eq:affinecharge2}
& f_{\hat{A}_{1}}(k,h) =  \oint_{\gamma} d\alpha_{0}(u) d \delta(u) e^{\lambda_{k,h}(u)} \\ \nonumber  & \frac{e^{ \alpha_{0}(u)+ \delta(u)} }{\prod^{\infty}_{  m = 1 }(1-e^{ -m \delta(u)})^{2} (1-e^{-(m-1) \delta(u)+ \alpha_{0}(u)})^{2} (1-e^{  -m \delta(u)- \alpha_{0}(u) })^{2} }. 
\end{align}
  Now we see that we have a function that acts in the same way (extracting the multiplicity) of particular roots just for the affine Lie algebra $\hat{A}_{1}$ as it does for the Borcherds-Kac-Moody Lie algebra in (\ref{eq:dyoncountingformula}). However, now we are just looking at the $PSL(2, \mathbb{Z}) $ part of the full modular group in the Igusa cusp form.

 \subsubsection*{S-duality} 
  
 This generating function and coefficients $f_{\hat{A_{1}}}(k,h)$ from (\ref{eq:affineA1degeneracies}) can be rewritten using an S-duality transformation. These are $PSL(2, \mathbb{Z})$ transformations that
 act as Weyl reflections, or automorphisms of the fundemental Weyl chamber, of the $\hat{A}_{1}$ Lie algebra at each point in the moduli space \cite{Cheng_2008}. However, the S-duality condition of constant $f_{\hat{A_{1}}}(k,h)$ at $u$ and $u'$ does not hold across the walls where roots can enter or leave the spectrum depending on which direction the wall is crossed. The imaginary roots are not transformed under S-duality.
 
\subsubsection{Contour prescription for $\hat{A}_{1}$} \label{subsection:affineA1contourprescription}

To apply the contour prescription $\mathcal{C}$ discussed in \cite{Cheng_2007} for the full Borcherds-Kac-Moody algebra to the $\hat{A}_{1}$ part $\gamma \subset \mathcal{C}$ one must again separate the degeneracies associated to $\hat{A}_{1}$ Lie algebra (\ref{splittingdegeneracies}-\ref{eq:affineA1degeneracies}) from that associated to the other roots of the Kac-Moody algebra and then restrict the possible contours one can take to those extracting the degeneracies of $\hat{A}_{1}$ roots. 

\begin{figure}[h!]
	
\begin{center}

\begin{tikzpicture}
\draw (-7,-1) -- (9,-1);  
\draw (-7,-7) -- (-7,7.5);   

\node at (6.7,-0.5){$x$};
\node at (-6.3,6.7){$iy$};
\node at (-0.8,6.5){$\gamma_{R_{1}}$};
\node at (-0.8, -5.5){$\gamma_{R_{2}}$};

\node at (8.3, -1.5 ){$\text{Im}[v] = 0$};

\node at (-6, -7.5 ){$\text{Re}[v] = 0$};

\node at (2.0, 7){$\text{Im}[v] = N_{2}$};

\node at (2.0, -7){$\text{Im}[v] = -N_{2}$};

\node at (-0.3, 4.5){$\Big\downarrow$};

\node at (-6.63, 4.5){$\Big\downarrow$};

\node at (-0.3, -3.5){$\Big\downarrow$};

\node at (-6.63, -3.5){$\Big\downarrow$};

\node at (-6.8, -1.5 ){$0$};
\node at (-0.1, -1.5 ){$1$};
\node at (6.9, -1.5 ){$2$};

\draw[dashed,red,xshift=4pt,
]
(-7.16,-6) --  (0,-6);
\draw[thick,red,xshift=4pt,
]
(0,-6)  -- (0,7);

\draw[thick,red,xshift=4pt,
]
(0,7)  -- (-7.16,7);

\draw[thick,red,xshift=4pt,
]
(-7.16,7)  -- (-7.16,-6);

\draw[thin,blue,xshift=4pt,]
(0,7.5)  -- (0,-7);
\draw[thin,blue,xshift=4pt,]

(-7.17,7.5)  -- (-7.17,-7); 

\draw[thin,blue,xshift=4pt,]
(0,7.5)  -- (0,-7);

\draw[thin,blue,xshift=4pt,]
(7,7.5)  -- (7,-7);

\draw[thick,black,xshift=4pt,
]
(-7.16,-1) --  (0,-1);

\draw[thick,black,xshift=4pt,
]
(-7.16,-1.01) --  (0,-1.01);

\end{tikzpicture}

\end{center}

\caption{This diagram shows the part of the contour $\gamma_{R_{1}}$ in the $v$ plane used to extract the Fourier coefficients. This is depicted as the solid red line. The pole at $v=0$ is crossed as the contour jumps to $\gamma_{R_{2}}$ from positive to negative infinity. In this case a wall is crossed.}

	\label{fig:Fouriercontour}
\end{figure}
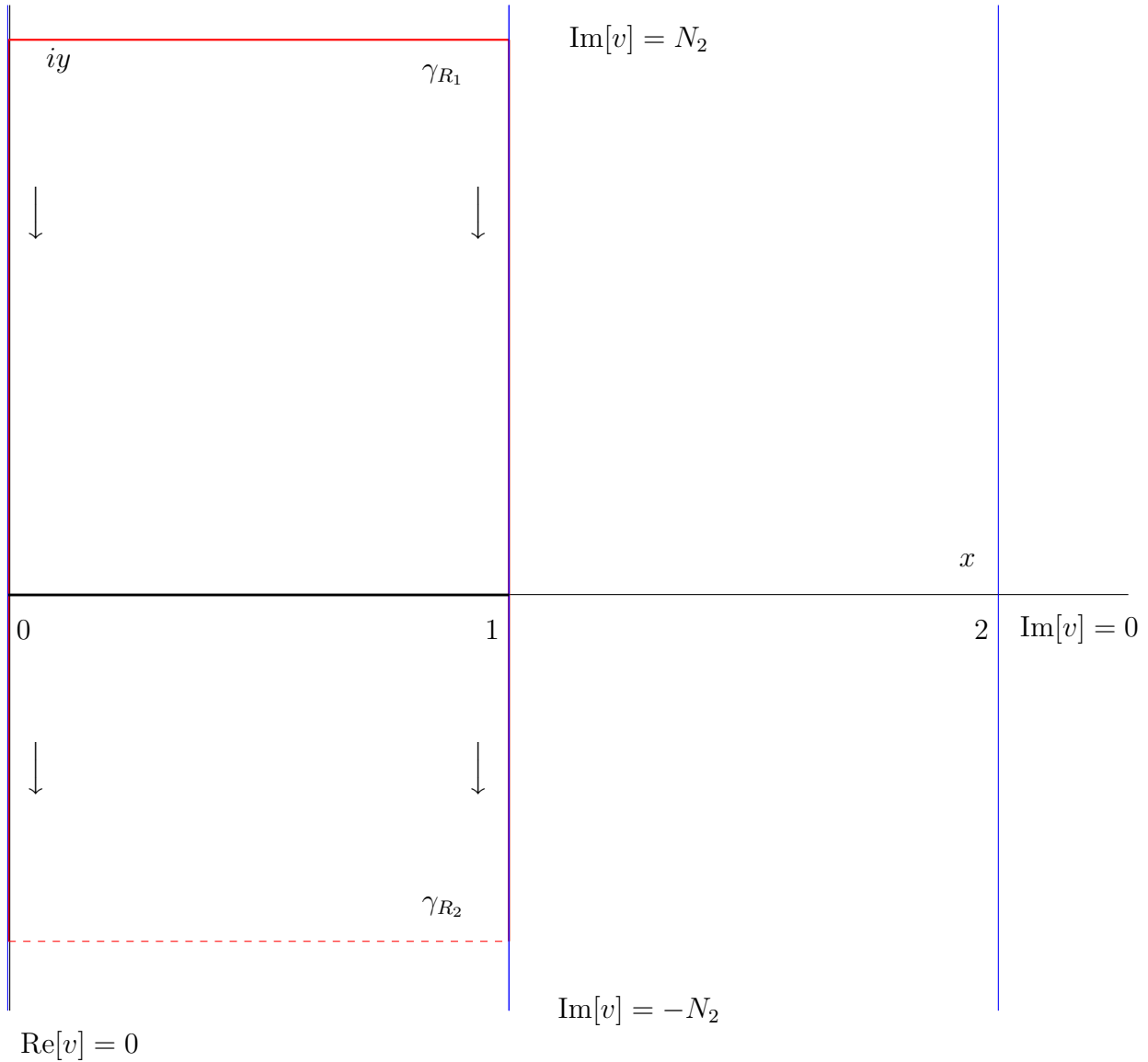

For this we look at the contour prescription again but this time only over the 2 variables $v$ and $\tau$. We define the contour in a modular invariant way. This time it must remain invariant under $SL(2, \mathbb{Z})$. Here we are again left with the contour:
\begin{align}
\text{Re}[\tau] \in [0,1], \   \    \   \text{Re}[v] \in [0,1].
\end{align}
The imaginary part of this contour is a large positive constant such that we have:
\begin{align}
\text{Im}[\tau] = N_{1},  \   \     \    \text{Im}[v] = N_{2},
\end{align}
where $N_{1}, N_{2} \rightarrow +\infty$. This contour is taken to enclose part of the fundamental domain of the $SL(2, \mathbb{Z})$ group. This means that every element in this region is contained here only once for the orbit of this group. In the Figure \ref{fig:Fouriercontour} below we show 2 domains and enclose the first one by a contour. For this contour, following \cite{Cheng_2007}, we are now looking at the pole at $v=0$ on the real axis but in general there are for $\hat{A}_{1}$ also poles at $v+n\tau,v-n\tau  = 0, \ \  n \in \mathbb{Z}$ that can be mapped to the $v=0$ pole under the modular transformations. The imaginary part $\text{Im}[v]$ is 
defined in such a way that the contour crosses the real axis
as it jumps from a large positive $+N_{2}$ to a large negative number $-N_{2}$ as the modulus changes sign 
\begin{align}
\text{Im}[v] > 0 \longrightarrow \text{Im}[v] < 0.
\end{align}
As the contour must remain invariant as this happens one must deform it back to its original position. In doing so one picks up the contribution of the pole as the contour crosses the real axis. This is the cause of wall crossing of the generating function. The residue from this pole creates the jump in the degeneracies. 
This should also give a consistent description for the other poles. These are crossed when the contour for the particular charge one wants to extract the degeneracy of crosses $\text{Im}[v \pm n \tau] = 0$. As long as the the contour is taken to be at a  sufficiently large imaginary value, then only one pole (and one wall) is crossed at a time. This is because only that particular combination of moduli change sign.

\subsubsection{Wall crossing for $\hat{A}_{1}$ in $\mathcal{N}=4$} \label{subsection1N=4affinewallcrossing}

Now we understand how the degeneracies of $\frac{1}{4}$BPS dyons can be extracted from the Igusa cusp form as Fourier coefficients using a contour prescription. These degeneracies include both single and multicentered black holes. We have also seen how wall crossing represents the split of a multicenterd black hole into 2 electric and magnetic $\frac{1}{2}$BPS states and its effect on the dyon degeneracies as the multicentered contribution to the wall is removed. We have seen in (\ref{eq:expansionFouriercoefficients}) that this wall crossing can be encoded in a jump in the Fourier coefficients representing the degeneracies. We know from \cite{Cheng_2008} that this wall crossing for the full Borcherds-Kac-Moody algebra can be encoded in a change in the highest weight of a Verma module \footnote{See section \ref{subsectionADEWeyldenominator} equation (\ref{eqverma}) for a definition.} which counts the BPS states. We will now proceed to show that this also holds in the case of the $\hat{A}_{1}$ subalgebra. So far the Weyl denominator of $\hat{A}_{1}$ has been found as a factor in the generating function and the contour prescription found in a limit of the 3 complex variables in the full Igusa cusp form. What remains is the Lie algebraic formulation of the wall crossing in terms of Verma modules. In the section \ref{subsection2N=4affinewallcrossing} below we explicitly calculate the change in the highest weight in the different chambers of $\hat{A}_{1}$.

\subsubsection{Wall crossing} \label{subsection2N=4affinewallcrossing} 
  
 The function we have above in (\ref{equation:thetafunction}) exhibits wall crossing behavior as the roots disappear from the spectrum. This happens when the imaginary part of a particular root shown in the Weyl denominator vanishes and changes sign. As we remember from (\ref{eq:expansionFouriercoefficients}) the exponentials must have modulus less than 1 for a well defined series expansion which is convergent and allows one to extract the Fourier coefficients. This means that in our case, for the Weyl denominator written in terms of the roots of the affine Lie algebra, shown in (\ref{eq:affinecharge2}), we must have: 
  \begin{align}
| e^{ -m \delta(u)} | < 1,  \    \   \   | e^{-(m-1) \delta(u)+ \alpha_{0}(u)} | < 1,  \    \   | e^{  -m \delta(u)- \alpha_{0}(u) } | < 1,   \   m \geq 1. 
  \end{align} 
  Or equivalently this means that the imaginary parts  $\text{Im}[ i  m \delta(u)+ i  \alpha_{0}(u)], \  \text{Im}[ i  (m-1) \delta(u)- i  \alpha_{0}(u) ]    > 0$. If we, as in the $\mathcal{N}=4$ black hole literature \cite{N=4DyonDegKKCMprodDavid_2006,KKoscfreeN=4DyonmodesDavid_2006,Cheng_2007,Cheng_2008,ellipticgenusandintegralDavid_2006,dabholkar2014quantum}, stay in the Siegel upper half plane, or in the case of the Jacobi theta function just the modular upper half plane, we must also constrain $\text{Im}[ im \delta(u)] > 0$. 
  So we can now look at what happens when a particular root crosses a wall: 
   \begin{align} \label{exampleroot}
  \text{Im}[ i  (m-1) \delta(u)- i  \alpha_{0}(u) ]   > 0 \longrightarrow \text{Im}[ i  (m-1) \delta(u)- i  \alpha_{0}(u) ]  < 0.
  \end{align}
 We assume all the other roots retain a positive imaginary part. In this case the particular root  $- (m-1) \delta(u)+ \alpha_{0}(u)$ disappears or appears (depending on the direction) from the spectrum while all other roots remain in the spectrum. This is an example of wall crossing. In this case the denominator formula must be rewritten such that all the roots in the denominator again satisfy the conditions on the imaginary part.  This can be done by moving a root, for example  the $-(m-1)\delta(u)+\alpha_{0}(u)$ root from (\ref{exampleroot}), into the numerator.
 
 \subsubsection{Jumping between chambers}
 
 We can start with a simple example: we start in a chamber in which no root exists - the fundamental Weyl chamber $W_{F}$ and we move in a direction in the moduli space such that at the first wall of marginal stability a root appears. But only this root and no other roots. For example if we let a root $-  \delta(u)- \alpha_{0}(u)$ enter the spectrum. The generating function from (\ref{eq:affinecharge2}) is rewritten as:
  \begin{align}
  f_{\hat{A}_{1}}(k,h) = &  \oint_{\gamma} d\alpha_{0}(u) d \delta(u) e^{\lambda_{k,h}(u)}  \\ \nonumber  & \frac{e^{ \alpha_{0}(u)+ \delta(u)} }{\prod\limits^{\infty}_{l=1}(1-e^{ -l \delta(u)})^{2} (1-e^{-(l-1) \delta(u)+ \alpha_{0}(u)})^{2}\prod\limits^{\infty}_{m =1, m \neq 1} (1-e^{  -m \delta(u)- \alpha_{0}(u) })^{2} } \nonumber \\ \nonumber & \times  \frac{1}{(1-e^{- \delta(u)-\alpha_{0}(u)})^{2}} \\ \nonumber   \longrightarrow 
 \oint_{\gamma} d\alpha_{0}(u) d \delta(u) e^{\lambda_{k,h}(u)}  &  \frac{e^{ \alpha_{0}(u)+ \delta(u)} }{\prod\limits^{\infty}_{l =1 }(1-e^{ -l \delta(u)})^{2} (1-e^{-(l-1) \delta(u)+ \alpha_{0}(u)})^{2} \prod\limits^{\infty}_{m =1,  m \neq 1}(1-e^{  -m \delta(u)- \alpha_{0}(u) })^{2} } \\  & \times \frac{e^{2 \delta(u)+2 \alpha_{0}(u)}}{(1-e^{ \delta(u)+ \alpha_{0}(u)})^{2}}. 
  \end{align}
  Now it can again be expanded in terms of the positive roots within the denominator, but with an additional factor in the numerator. This factor causes a jump in the degeneracies of the various roots. Depending on which region of the moduli space one is in, or in which Weyl chamber $W$, there can be many of these factors which one must move into the numerator. In general we can keep moving in that direction such that we cross $k$ walls of marginal stability. If we start within the fundamental Weyl chamber we now have $n \in  1,\dots, k$ additional roots in the spectrum. This is a finite set of roots for which one must modify the generating function as:
  \begin{align}
  f_{\hat{A}_{1}}(k,h) = & \oint_{\gamma} d\alpha_{0}(u) d \delta(u) e^{\lambda_{k,h}(u)} \\  & \times \frac{e^{ \alpha_{0}(u)+ \delta(u)} }{\prod^{\infty}_{l = 1 }(1-e^{ -l \delta(u)})^{2} (1-e^{-(l-1) \delta(u)+ \alpha_{0}(u)})^{2}  \prod^{\infty}_{m = k+1 }(1-e^{  -m \delta(u)- \alpha_{0}(u) })^{2} } \nonumber \\ \nonumber & \times \frac{1}{\prod^{k}_{m = 1 }(1-e^{-m \delta(u)- \alpha_{0}(u)})^{2} },  \\ \nonumber
  \\ \nonumber
= &  \oint_{\gamma} d\alpha_{0}(u) d \delta(u) e^{\lambda_{k,h}(u)} \\ \nonumber  & \times \frac{e^{ \alpha_{0}(u)+ \delta(u)} }{\prod^{\infty}_{l =1 }(1-e^{ -l \delta(u)})^{2} (1-e^{-(l-1) \delta(u)+ \alpha_{0}(u)})^{2} \prod^{\infty}_{m =k+1 }(1-e^{  -m \delta(u)- \alpha_{0}(u) })^{2} } \\  & \times \frac{\prod^{k} _{m = 1 }e^{2m\delta(u)+ 2\alpha_{0}(u)} }{\prod^{k}_{m = 1 }(1-e^{m \delta(u)+ \alpha_{0}(u)})^{2} }.    
  \end{align}
 Again this is rewritten in a way such that the denominator is now expandable in terms of Fourier coefficients which can then be extracted by the charge vector $\lambda_{k,h}$ introduced in (\ref{eq:affinecharge1}-\ref{eq:affinecharge2}). There is an alternative way to cross the walls. If one moves in the other direction starting in $W_{F}$, where no roots initially exist, a different set of roots start to appear. In this case the product formula after rewriting in a way that can be expanded as a Fourier series, then becomes:
 \begin{align}
 f_{\hat{A}_{1}}(k,h)  = & \oint_{\gamma} d\alpha_{0}(u) d \delta(u) e^{\lambda_{k,h}(u)} \\ \nonumber & \times \frac{e^{ \alpha_{0}(u)+ \delta(u)} }{\prod^{\infty}_{l =1 }(1-e^{ -l \delta(u)})^{2}  (1-e^{  -l \delta(u)- \alpha_{0}(u) })^{2}\prod^{\infty}_{m =k+1}(1-e^{-(m-1) \delta(u)+ \alpha_{0}(u)})^{2} } \nonumber  \\ \nonumber  & \times \frac{\prod^{k} _{m = 1} e^{2(m-1) \delta(u)- 2\alpha_{0}(u)}}{\prod^{k}_{m = 1 }(1-e^{(m-1) \delta(u)- \alpha_{0}(u)})^{2} }.    
 \end{align}
 This time it is the other set of roots that appear in the spectrum and cause a shift in the degeneracies.

 \subsubsection{Wall crossing in terms of highest weights}
 
 As done in \cite{Cheng_2008} we can again write this in terms of a Verma module associated to the affine Lie algebra with highest weight 
 \begin{align}
 \lambda = \frac{1}{2}(\lambda_{k,h} +\alpha_{0}(u)+ \delta).
 \end{align}
 In this case the formula, as with the Bocherds-Kac-Moody algebra in \cite{Cheng_2008}, can be written as a square: 
 \begin{align} \label{eq:squareofaffineweyldenominator}
 f_{\hat{A}_{1}}(k,h) =  \oint_{\gamma} d\alpha_{0}(u) d \delta(u) \Bigg(    \frac{e^{ \lambda(u)} }{\prod^{\infty}_{  m = 1 }(1-e^{ -m \delta(u)})(1-e^{-(m-1) \delta(u)+ \alpha_{0}(u)}) (1-e^{  -m \delta(u)- \alpha_{0}(u) }) } \Bigg)^{2}.
 \end{align} 
This can be used to see that each Weyl chamber $W$ can be associated to a highest weight for a particular representation, $W_{\lambda}$ as when  one moves in the moduli space, for example starting in the fundamental Weyl chamber and moving in a particular direction, this highest weight picks up a certain number of roots with each root corresponding to a particular wall that has been crossed. So if one starts in the fundamental chamber and moves, the highest weight of the representation is modified in a way such that 
\begin{align} \label{changeinN=4affinehighestweight1}
\lambda'_{1}= \lambda + \sum_{m =1}^{k} ((m-1) \delta- \alpha_{0}),
\end{align} 
in one direction, and 
\begin{align} \label{changeinN=4affinehighestweight2}
\lambda'_{2} = \lambda + \sum_{m =1}^{k} (m \delta+ \alpha_{0}),
\end{align}
in the other direction. One could use the S-duality transformations to map the denominator that is now expandable in terms of a new Fourier series into one containing only positive roots. This would involve a suitable change of basis of positive roots. However, we keep the basis constant here such that we can directly compare all the highest weights. Here we can see that in every chamber the highest weight is different - representing a different Fourier series and different black hole degeneracies. 

\subsubsection*{Highest weight in the different chambers}

The Weyl chambers of $\hat{A}_{1}$ are shown in Fig. \ref{fig:A1chambersN=4} below. Each region in the moduli space has a unique combination of roots. The Figure shows the combinations of roots that exist within each chamber.
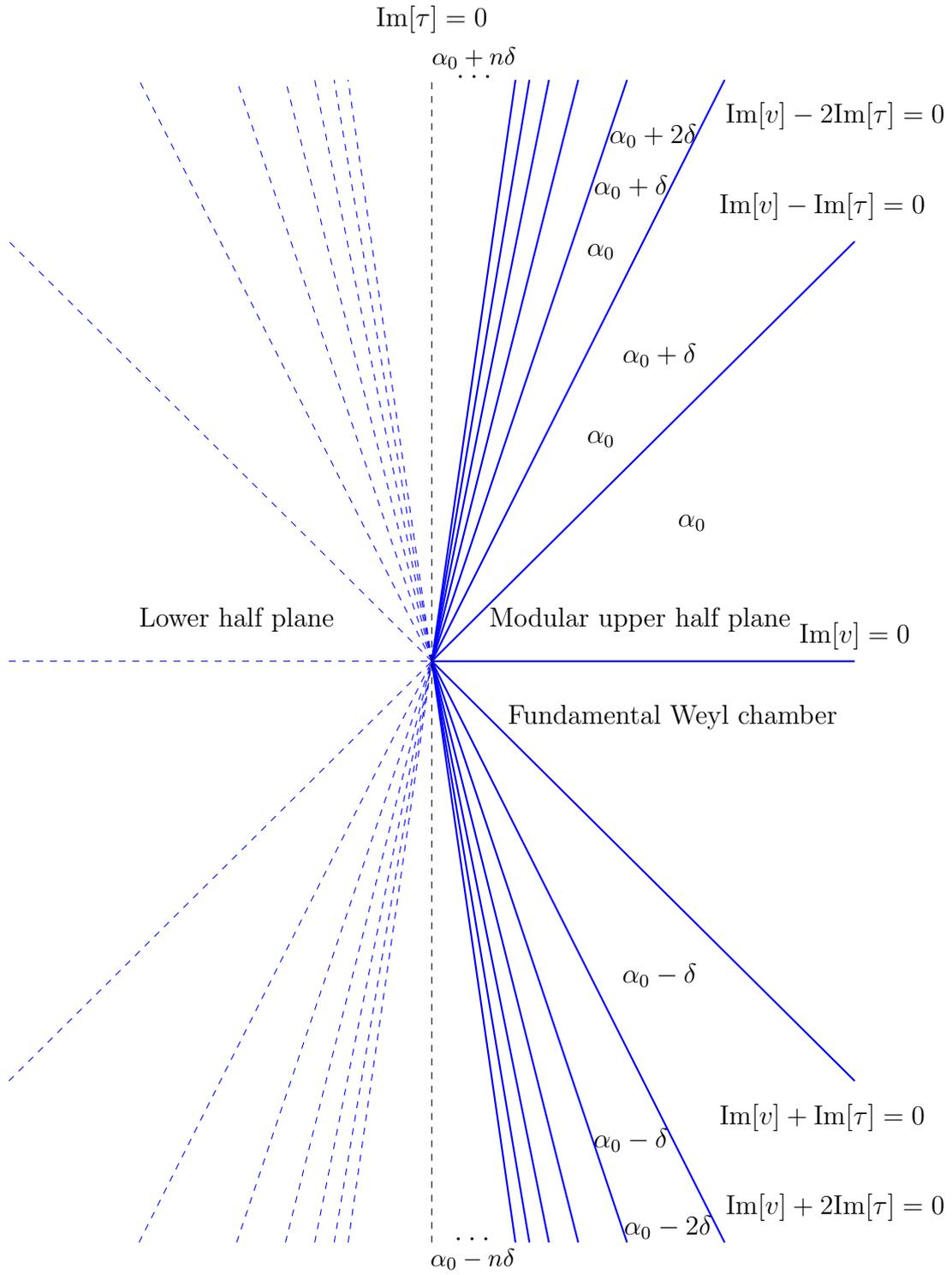
\begin{figure}

\begin{center}

\vspace*{-3cm}%

\begin{tikzpicture}

\node at (3.2, 0.5) {$\text{Modular upper half plane}$};

\node at (3.7, -1) {$\text{Fundamental Weyl chamber}$};

\node at (-3, 0.5) {$\text{Lower half plane}$};

\node at (6.5, 0.25) {$\text{Im}[v] = 0$};

\node at (0, 9.8) {$\text{Im}[\tau] = 0$};

\node at (6, 6.9) {$\text{Im}[v]-\text{Im}[\tau] = 0$};

\node at (6, -7.2) {$\text{Im}[v]+\text{Im}[\tau] = 0$};

\node at (6.2, -8.6) {$\text{Im}[v]+2\text{Im}[\tau] = 0$};

\node at (6.2, 8.3) {$\text{Im}[v]-2\text{Im}[\tau] = 0$};

\node at (0.65, 8.9) {$\cdot  \cdot  \cdot$};

\node at (0.62, -9.1) {$\cdot  \cdot  \cdot$};

\node at (3.4, 8) {$\alpha_{0}+2\delta$};

\node at (3.06, 7.2) {$\alpha_{0}+\delta$};

\node at (2.6, 6.2) {$\alpha_{0}$};

\node at (3.06, -7.55) {$\alpha_{0}-\delta$};

\node at (3.63, -8.9) {$\alpha_{0}-2\delta$};

\node at (3.5, -5) {$\alpha_{0}-\delta$};

\node at (3.5, 4.6) {$\alpha_{0}+\delta$};

\node at (2.6, 3.3) {$\alpha_{0}$};

\node at (4, 2) {$\alpha_{0}$};

\node at (0.65, 9.2) {\small$\alpha_{0}+n\delta$};

\node at (0.63, -9.4) {\small$\alpha_{0}-n\delta$};

\draw[thick,blue,yshift=-4pt,
]
(0,0) -- (6.5,6.5) ;

\draw[dashed,blue,yshift=-4pt,
]
(-6.5,-6.5) -- (0,0) ;

\draw[dashed,blue,yshift=-4pt,
]
(0,0)  -- (-6.5,6.5);

\draw[thick,blue,yshift=-4pt,
]
(6.5,-6.5)  -- (0,0);

\draw[thick,blue,yshift=-4pt,
]
(0,0)  -- (6.5,0);

\draw[dashed,blue,yshift=-4pt,
]
(-6.5,0)  -- (0,0);

\draw[thick,blue,yshift=-4pt,
]
(0,0)  -- (4.5,9);

\draw[dashed,blue,yshift=-4pt,
]
(-4.5,-9)  -- (0,0);

\draw[dashed,blue,yshift=-4pt,
]
(0,0)  -- (-4.5,9);

\draw[thick,blue,yshift=-4pt,
]
(4.5,-9)  -- (0,0);

\draw[thick,blue,yshift=-4pt,
]
(0,0)  -- (3,9);

\draw[dashed,blue,yshift=-4pt,
]
(-3,-9)  -- (0,0);

\draw[dashed,blue,yshift=-4pt,
]
(0,0)  -- (-3,9);

\draw[thick,blue,yshift=-4pt,
]
(3,-9)  -- (0,0);

\draw[thick,blue,yshift=-4pt,
]
(0,0)  -- (9/4,9);

\draw[dashed,blue,yshift=-4pt,
]
(-9/4,-9)  -- (0,0);

\draw[dashed,blue,yshift=-4pt,
]
(0,0)  -- (-9/4,9);

\draw[thick,blue,yshift=-4pt,
]
(9/4,-9)  -- (0,0);

\draw[thick,blue,yshift=-4pt,
]
(0,0)  -- (9/5,9);

\draw[dashed,blue,yshift=-4pt,
]
(-9/5,-9)  -- (0,0);

\draw[dashed,blue,yshift=-4pt,
]
(0,0)  -- (-9/5,9);

\draw[thick,blue,yshift=-4pt,
]
(9/5,-9)  -- (0,0);

\draw[thick,blue,yshift=-4pt,
]
(0,0)  -- (1.5,9);

\draw[dashed,blue,yshift=-4pt,
]
(-1.5,-9)  -- (0,0);

\draw[dashed,blue,yshift=-4pt,
]
(0,0)  -- (-1.5,9);

\draw[thick,blue,yshift=-4pt,
]
(1.5,-9)  -- (0,0);

\draw[thick,blue,yshift=-4pt,
]
(0,0)  -- (9/7,9);

\draw[dashed,blue,yshift=-4pt,
]
(-9/7,-9)  -- (0,0);

\draw[dashed,blue,yshift=-4pt,
]
(0,0)  -- (-9/7,9);

\draw[thick,blue,yshift=-4pt,
]
(9/7,-9)  -- (0,0);

\draw[dashed,black,yshift=-4pt,
]
(0,0)  -- (0,9);

\draw[dashed,black,yshift=-4pt,
]
(0,-9)  -- (0,0);

\end{tikzpicture}

\caption{This figure shows the walls associated with the $\hat{A}_{1}$ roots. These are the blue lines and are found in the modular upper half of the $\tau$-plane. The dashed blue lines shown are the continuation into the lower half plane.  }

\label{fig:A1chambersN=4}

\end{center}

\end{figure}
\begin{table}[h!]
    
\begin{center}

\begin{tabular}{ |l|l|}
	\hline
	\multicolumn{2}{|c|}{ \vphantom{\Huge{H}} All possible highest weights} \\[7pt]
	\hline
   \vphantom{\Huge{H}}	Highest weight in one direction & \vphantom{\Huge{H}} In other direction \\[7pt]
	\hline
	\vphantom{\Huge{H}}	
 $\lambda + \sum_{n=1}^{\infty}\Big(n \delta+ \alpha_{0}  \Big)$ & $\lambda + \sum_{m=1}^{\infty} \Big((m-1) \delta- \alpha_{0}  \Big) $  \\[7pt]
	\hline
 \vphantom{\Huge{H}}	
 $\lambda + \sum_{n=1}^{k}\Big(n \delta+ \alpha_{0}  \Big)$ & $\lambda + \sum_{m=1}^{h} \Big((m-1) \delta- \alpha_{0}  \Big) $ \\[7pt]
	
	\hline
	\multicolumn{2}{|c|}{\vphantom{\Huge{H}} $\lambda$} \\[7pt]
	\hline
\end{tabular}

\end{center}

    \caption{Highest weight in modular upper half plane representing existing BPS boundstates. This is also shown in Fig. \ref{fig:A1chambersN=4}. }
    \label{tab:highestweightmodularhalfplane}
\end{table}
This shows how one can start in a chamber with none of the roots existing and move to a chamber, where the roots that exist appear when crossing a finite number of walls in either direction. In the $\mathcal{N} = 4$ theory one obtains discrete attractor flow as one moves towards the fundamental Weyl chamber from either side.

\section{$\hat{A}_{1}$ root system in $\mathcal{N}=2$ theories}\label{workforanalogs}

The generating function for the $\hat{A}_{1}$ Lie algebra has now been derived in the previous section by taking it as a subalgebra of the $\mathcal{N} = 4$ example and is the Weyl denominator formula. Specifically, this was established in subsection \ref{extractingaffineA1fromKacMoody} by extracting particular degeneracies from the Borcherds-Kac-Moody algebra Weyl denominator in \cite{Cheng_2007, Cheng_2008}. 
Furthermore, the wall crossing for this generating function has now been derived in subsections \ref{subsection1N=4affinewallcrossing} and \ref{subsection2N=4affinewallcrossing} above. This is depicted in Fig. \ref{fig:A1chambersN=4} and is encoded in a change in highest weight of a Verma module.

Now we can construct an analog generating function by extending this pattern to $\mathcal{N}=2$ theories where the BPS states are described by the $\hat{A}_{1}$ root system. This root system is important in an $\mathcal{N}=2$ context because it describes several interesting examples of wall crossing in $\mathcal{N}=2$ theories. This includes both the D6-D2-D0 brane system of type IIA string theory on the resolved conifold, described by Jafferis and Moore \cite{D6D2D0JafferisMoorehttps://doi.org/10.48550/arxiv.0810.4909} as well as Seiberg-Witten SU(2) theory.

It is important to note that in the $\mathcal{N}=4$ example, the generating function is only defined in the Siegel upper half plane of the complex moduli $ v, \sigma, \tau$. This is because the Jacobi-theta function and the full Igusa cusp form are modular and Siegel modular forms respectively. This means that to preserve modularity of the Siegel modular form, or of the Jacobi theta and eta functions, the analysis in subsection \ref{subsection1N=4affinewallcrossing} as well as that in the literature \cite{Dijkgraaf_1997,Cheng_2007,Cheng_2008, dabholkar2014quantum} of which we have taken a limit here must be confined to the values $\text{Im}[\tau] > 0$. Here the variable $\tau$, from (\ref{variablesintermsofroots1}),  can be parameterised as the $\tau$ parameter of an elliptic curve, which always has a positive imaginary part. 

\begin{dfn}
 We can define the imaginary or “affine” wall as the wall $\text{Im}[\tau] = 0$ separating the 2 half-planes corresponding to the 2 domains of definition of the modular form and the Fourier coefficients.    
\end{dfn}

Hence, the imaginary or affine wall is never crossed in the $\mathcal{N}=4$ theory. This is interpreted in \cite{Cheng_2008} as imaginary roots always existing in the spectrum which is not the case in $\mathcal{N}=2$ theories we are looking at. If one is to use this function as an analog counting function for theories in $\mathcal{N} = 2$, to obtain the full spectrum for example of Seiberg-Witten theory, one must analytically continue to a region where the full spectrum of $\hat{A}_{1}$ exists. Now values of $\text{Im}[\tau] < 0$ are allowed. This then gives 2 domains of definition of the expansion coefficients.  
 
One can start by considering that the theta function is defined for $\tau$ in the upper half plane as $\theta(\tau, v)$. However, in the $\mathcal{N}=2$ examples the parameter $\tau$ has an imaginary part that changes sign at the affine wall. For this one must define an analytic continuation of the generating function into the lower half plane and hereby define a region in which all the roots can exist including the imaginary root. 

\begin{dfn}

In this case the actions of the roots on the Cartan subalgebra can be written as 2 complex variables given by
\begin{align} \label{variablesintermsofroots11} 
 \delta(u) = -2 \pi i \tau, \  \tau \in \mathbb{C} \  \ \text{and} \  \ \alpha_{0}(u) = -2 \pi i v, \ v \in \mathbb{C},  
\end{align}
following (\ref{variablesintermsofroots1}) up to the extension of the range of $\tau$ from $\mathbb{H}^{+}$ to $\mathbb{C}$. 

\end{dfn}

Now the highest weight changes in all the possible chambers on both sides of the imaginary wall starting in the fundamental chamber where none of the roots exist. For the other domain of definition one can start in a chamber opposite to the first fundamental chamber. In this chamber all the roots exist. One can then work backwards from this to find the highest weight in other chambers.

\subsection{Change in highest weight}

Here we continue on from sections \ref{subsection1N=4affinewallcrossing}
and \ref{subsection2N=4affinewallcrossing} by looking at the wall crossing of the generating function (\ref{eq:squareofaffineweyldenominator}), this time focusing on the function inside the square, which we call $g'_{\hat{A}_{1}}(u)$. This is done by using the the highest weight of the Verma module associated to the Weyl denominator for $\hat{A}_{1}$. However, this is now generalised to wall crossing beyond the affine wall. First, we recall that the generating function (\ref{eq:squareofaffineweyldenominator}) in the fundamental Weyl chamber can be written as
 \begin{align} \label{eq:N=2generatingfunctionaffinehighestweight}
g'_{\hat{A}_{1}}(u) = \frac{e^{ \lambda(u)} }{\prod^{\infty}_{  m = 1 }(1-e^{ -m \delta(u)})(1-e^{-(m-1) \delta(u)+ \alpha_{0}(u)}) (1-e^{  -m \delta(u)- \alpha_{0}(u) })}.
\end{align} 
This is written in terms of a highest weight $\lambda = \frac{1}{2}(\lambda_{k,h} +\alpha_{0}+ \delta) $ of a Verma module as is also done in \cite{Cheng_2008}. As one crosses into different regions in the moduli space the highest weight jumps, depending on how many Weyl chambers are crossed and what direction one moves in. For example, we can start by moving to \footnote{If one considers the generating function just in this form one also has a factor of -1 for every jump. For example we can include $(-1)^{k}$ in this equation.} 
\begin{align}
& g'_{\hat{A}_{1}}(u) = \\ \nonumber & \frac{e^{ \lambda'(u)} }{\prod^{\infty}_{l =1 }(1-e^{ -l \delta(u)}) (1-e^{  -l \delta(u)- \alpha_{0}(u) })\prod^{\infty}_{m =k+1}(1-e^{-(m-1) \delta(u)+ \alpha_{0}(u)})\prod^{k}_{m = 1 }(1-e^{(m-1) \delta(u)- \alpha_{0}(u)})  }. 
\\ \nonumber
\end{align}
This is now a different representation with a different highest weight given by $\lambda'$. Following (\ref{changeinN=4affinehighestweight1}) the relationship between the highest weights is given by 
\begin{align}
 \lambda' = \lambda + \sum_{m=1}^{k} \Big((m-1) \delta- \alpha_{0}  \Big).   
\end{align}
The remaining jumps to all the different chambers are discussed extensively in appendix section \ref{ChangesinhighestweightaffineA1appendix}. We can keep passing through the remaining chambers until the full BPS spectrum exists

\begin{align}
g'_{\hat{A}_{1}}(u) = \frac{e^{\lambda(u)}  \prod^{\infty}_{l =1 }  e^{  l \delta(u)+ \alpha_{0}(u) }}{ \prod^{\infty}_{l =1 }  (1-e^{  l\delta(u)+ \alpha_{0}(u) }) }   \frac{\prod^{\infty}_{l =1 }e^{l \delta(u)}}{\prod^{\infty}_{l =1 }(1-e^{l \delta(u)})}\frac{\prod^{\infty} _{m = 1} e^{(m-1) \delta(u)- \alpha_{0}(u)}}{\prod^{\infty}_{m = 1 }(1-e^{(m-1) \delta(u)- \alpha_{0}(u)}) }, 
\\ \nonumber
\end{align}
and the highest weight is
\begin{align}
 \lambda' = \lambda + \sum_{m=1}^{\infty} \Big((m-1) \delta- \alpha_{0}  \Big) + \sum_{l=1}^{\infty}l \delta+ \sum_{n=1}^{\infty}\Big(n \delta+ \alpha_{0}  \Big).
\end{align}
It is possible to move in the other direction in which the later BPS states enter the spectrum first. An example was also given by (\ref{changeinN=4affinehighestweight2}). The representations can be ordered in the following way, such that one can write all the possible $\lambda'$ in the table below. This is also shown in Fig.\ref{fig:affinea1walls}
below. Here all the boundaries of the Weyl chambers which are also the walls are shown in blue. These are now written in terms of the inner products with the roots first given in (\ref{variablesintermsofroots1}). The roots that exist within each chamber are also shown.

\begin{table}[h!]

\begin{center}
    
\begin{tabular}{ |l|l|}
	\hline
	\multicolumn{2}{|c|}{ \centering \vphantom{\Huge{HH}} All possible highest weights  } \\[8pt]
	\hline 
	\multicolumn{2}{|c|}{\vphantom{ \Huge{\Huge{\Huge{HH}}}} $ \lambda + \sum_{m=1}^{\infty} \Big((m-1) \delta- \alpha_{0}  \Big) + \sum_{l=1}^{\infty}l \delta+ \sum_{n=1}^{\infty} \Big(n \delta+ \alpha_{0}  \Big) \  \  \  \  \  \  \  \  \ $ \vphantom{\Huge{HH}}} \\[8pt]
	\hline
	& \\
	$ \lambda + \sum_{m=h+1}^{\infty} \Big((m-1) \delta- \alpha_{0}  \Big) +$ & 	$ \lambda + \sum_{m=1}^{\infty} \Big((m-1) \delta- \alpha_{0}  \Big) + $  \\
	$ \sum_{l=1}^{\infty}l \delta+ \sum_{n=1}^{\infty}\Big(n \delta+ \alpha_{0}  \Big)$ & $ \sum_{l=1}^{\infty}l \delta+ \sum_{n=k+1}^{\infty}\Big(n \delta+ \alpha_{0} \Big) $ \\ 
	& \\
	\hline
	& \\
	$\lambda + \sum_{l=1}^{\infty}l \delta+ \sum_{n=1}^{\infty}\Big(n \delta+ \alpha_{0}  \Big)$& $\lambda + \sum_{m=1}^{\infty} \Big((m-1) \delta - \alpha_{0}  \Big) + \sum_{l=1}^{\infty}l \delta$ \\
	& \\
	\hline
	& \\
	$\lambda + \sum_{n=1}^{\infty}\Big(n \delta+ \alpha_{0}  \Big)$ & $\lambda + \sum_{m=1}^{\infty} \Big((m-1) \delta- \alpha_{0}  \Big) $  \\
	& \\
	\hline
	& \\
	$\lambda + \sum_{n=1}^{k}\Big(n \delta+ \alpha_{0}  \Big)$ & $\lambda + \sum_{m=1}^{h} \Big((m-1) \delta- \alpha_{0}  \Big) $ \\
	& \\
	
	\hline
	\multicolumn{2}{|c|}{$\lambda$} \\
	\hline
\end{tabular} 

\end{center}

\begin{flushleft}
\caption{All possible highest weights in the $\mathcal{N}=2$ analog of the $\hat{A}_{1}$ Lie algebra. This is also shown on Fig. \ref{fig:affinea1walls}.}
\label{tab:2}

\end{flushleft}

\end{table}
\begin{figure}

\vspace*{-3cm}%

\begin{center}

\begin{tikzpicture}

\node at (6.5, 0.25) {$\text{Im}[i\alpha_{0}(u)] = 0$};

\node at (0, 9.7) {$\text{Im}[i\delta(u)] = 0$};

\node at (6.3, 6.9) {$\text{Im}[i\alpha_{0}(u)+i\delta(u)] = 0$};

\node at (6.3, -7.2) {$\text{Im}[i\alpha_{0}(u)-i\delta(u)] = 0$};

\node at (6.8, -8.6) {$\text{Im}[i\alpha_{0}(u)-2i\delta(u)] = 0$};

\node at (6.5, 8.3) {$ \  \  \  \ \text{Im}[i\alpha_{0}(u)+2i\delta(u)] = 0$};

\node at (0.68, 8.6) {$\cdot  \cdot  \cdot$};

\node at (0.63, -9.2) {$\cdot  \cdot  \cdot$};

\node at (3.4, 8) {$\alpha_{0}+2\delta$};

\node at (3.06, 7.2) {$\alpha_{0}+\delta$};

\node at (2.6, 6.2) {$\alpha_{0}$};

\node at (3.06, -7.55) {$\alpha_{0}-\delta$};

\node at (3.63, -8.9) {$\alpha_{0}-2\delta$};

\node at (3.5, -5) {$\alpha_{0}-\delta$};

\node at (3.5, 4.6) {$\alpha_{0}+\delta$};

\node at (2.6, 3.3) {$\alpha_{0}$};

\node at (4, 2) {$\alpha_{0}$};

\node at (0.75, 9) {\small$\alpha_{0}+n\delta$};

\node at (-0.75, 9) {\small$\alpha_{0}+n\delta$};

\node at (-0.68, 8.6) {$\cdot  \cdot  \cdot$};

\node at (-0.45, 6.9) {\small$n\delta$};

\node at (-0.46, 6.55) {$\cdot  \cdot  \cdot$};

\node at (-0.7, -9.5) {\small$\alpha_{0}-n\delta$};
\node at (-0.63, -9.2) {$\cdot  \cdot  \cdot$};

\node at (-0.45, -6.83) {\small$n\delta$};

\node at (-0.46, -6.55) {$\cdot  \cdot  \cdot$};

\node at (0.7, -9.5) {\small$\alpha_{0}-n\delta$};

\node at (-2.5, 1.4) {\small$n\delta$};

\node at (-2.5, 1.05) {$\cdot  \cdot  \cdot$};

\node at (-5.5, 3.4) {\small$\alpha_{0}+n\delta$};

\node at (-5.5, 3.05) {$\cdot  \cdot  \cdot$};

\node at (-4, 2.4) {\small$\alpha_{0}-n\delta$};
\node at (-4, 2.05) {$\cdot  \cdot  \cdot$};

\node at (-2.5, 1.4) {\small$n\delta$};

\node at (-2.5, 1.05) {$\cdot  \cdot  \cdot$};

\node at (-2.5, 3.4) {\small$n\delta$};

\node at (-2.5, 3.05) {$\cdot  \cdot  \cdot$};

\node at (-4.5, 6.4) {\small$\alpha_{0}+n\delta$};

\node at (-4.5, 6.05) {$\cdot  \cdot  \cdot$};

\node at (-3.5, 5.0) {\small$\alpha_{0}-n\delta$};
\node at (-3.5, 4.7) {$\cdot  \cdot  \cdot$};
\node at (-3.5, 4.5) {\small$\alpha_{0}-2\delta$};

\node at (-2.5, 1.4) {\small$n\delta$};

\node at (-2.5, 1.05) {$\cdot  \cdot  \cdot$};

\node at (-5.5, -3.6) {\small$\alpha_{0}+n\delta$};

\node at (-5.5, -3.35) {$\cdot  \cdot  \cdot$};

\node at (-5.5, -3.05) {\small$\alpha_{0}+\delta$};

\node at (-4, -2.4) {\small$\alpha_{0}-n\delta$};
\node at (-4, -2.1) {$\cdot  \cdot  \cdot$};

\node at (-2.5, -1.4) {\small$n\delta$};

\node at (-2.5, -1.1) {$\cdot  \cdot  \cdot$};

\node at (3.4, -0.5) {$\text{Fundamental Weyl chamber 1}$};

\node at (-3.4, 0.23) {$\text{Fundamental Weyl chamber 2}$};

\draw[thick,blue,yshift=-4pt,
]
(0,0) -- (6.5,6.5) ;

\draw[thick,blue,yshift=-4pt,
]
(-6.5,-6.5) -- (0,0) ;

\draw[thick,blue,yshift=-4pt,
]
(0,0)  -- (-6.5,6.5);

\draw[thick,blue,yshift=-4pt,
]
(6.5,-6.5)  -- (0,0);

\draw[thick,blue,yshift=-4pt,
]
(0,0)  -- (6.5,0);

\draw[thick,blue,yshift=-4pt,
]
(-6.5,0)  -- (0,0);

\draw[thick,blue,yshift=-4pt,
]
(0,0)  -- (4.5,9);

\draw[thick,blue,yshift=-4pt,
]
(-4.5,-9)  -- (0,0);

\draw[thick,blue,yshift=-4pt,
]
(0,0)  -- (-4.5,9);

\draw[thick,blue,yshift=-4pt,
]
(4.5,-9)  -- (0,0);

\draw[thick,blue,yshift=-4pt,
]
(0,0)  -- (3,9);

\draw[thick,blue,yshift=-4pt,
]
(-3,-9)  -- (0,0);

\draw[thick,blue,yshift=-4pt,
]
(0,0)  -- (-3,9);

\draw[thick,blue,yshift=-4pt,
]
(3,-9)  -- (0,0);

\draw[thick,blue,yshift=-4pt,
]
(0,0)  -- (9/4,9);

\draw[thick,blue,yshift=-4pt,
]
(-9/4,-9)  -- (0,0);

\draw[thick,blue,yshift=-4pt,
]
(0,0)  -- (-9/4,9);

\draw[thick,blue,yshift=-4pt,
]
(9/4,-9)  -- (0,0);

\draw[thick,blue,yshift=-4pt,
]
(0,0)  -- (9/5,9);

\draw[thick,blue,yshift=-4pt,
]
(-9/5,-9)  -- (0,0);

\draw[thick,blue,yshift=-4pt,
]
(0,0)  -- (-9/5,9);

\draw[thick,blue,yshift=-4pt,
]
(9/5,-9)  -- (0,0);

\draw[thick,blue,yshift=-4pt,
]
(0,0)  -- (1.5,9);

\draw[thick,blue,yshift=-4pt,
]
(-1.5,-9)  -- (0,0);

\draw[thick,blue,yshift=-4pt,
]
(0,0)  -- (-1.5,9);

\draw[thick,blue,yshift=-4pt,
]
(1.5,-9)  -- (0,0);

\draw[thick,blue,yshift=-4pt,
]
(0,0)  -- (9/7,9);

\draw[thick,blue,yshift=-4pt,
]
(-9/7,-9)  -- (0,0);

\draw[thick,blue,yshift=-4pt,
]
(0,0)  -- (-9/7,9);

\draw[thick,blue,yshift=-4pt,
]
(9/7,-9)  -- (0,0);

\draw[thick,black,yshift=-4pt,
]
(0,0)  -- (0,9);

\draw[thick,black,yshift=-4pt,
]
(0,-9)  -- (0,0);

\end{tikzpicture}

\end{center}

\caption{This diagram again shows the  standard walls of the Weyl chambers of the affine Lie algebra $\hat{A}_{1}$ but this time extended into the lower half plane after the continuation described above. The highest weights in this diagram are shifted as $\Tilde{\lambda} \rightarrow \lambda - \alpha_{0}$ relative to those in table \ref{tab:2}. The blue lines represent the BPS walls.}

\label{fig:affinea1walls}

\end{figure}
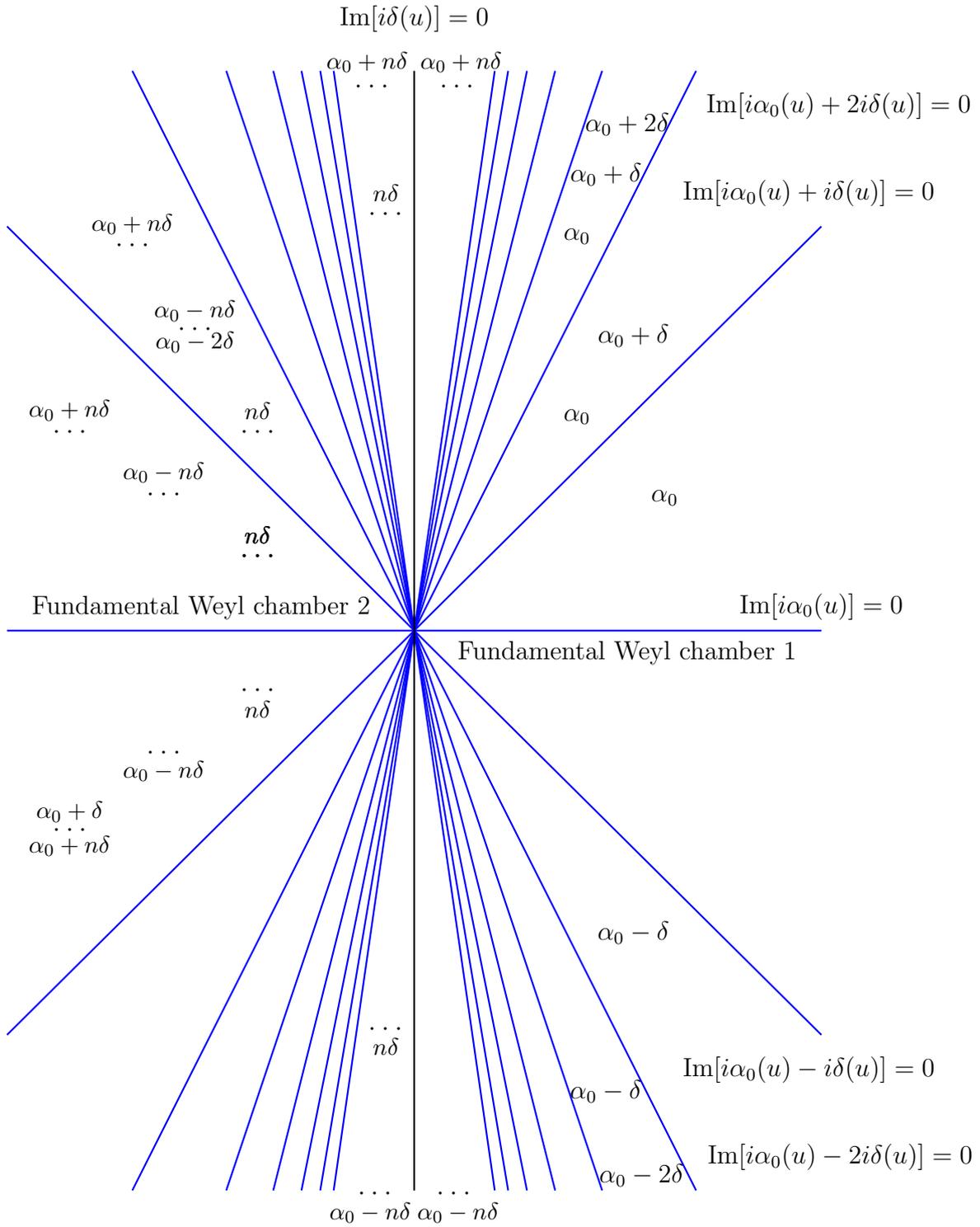

\subsection{Counting roots and weights in Verma modules}

The BPS states with particular charge are represented by the roots of the Lie algebra. To count these roots one uses the weight system of the Verma module as defined in (\ref{eqverma}). The roots representing the charges of the BPS states are found by taking the differences between the weights. One can start in the fundamental Weyl chamber where none of the BPS states exists. One chooses an initial weight from (\ref{eq:N=2generatingfunctionaffinehighestweight}) called $\lambda_{i} \in \Gamma$. One can then move in the 2 possible directions and pick up roots, which are added to the highest weight. For every highest weight $\lambda'_{i}$ all the possible weights under it exist. One can use the indices in the table above $h,k,l$ to label the weights. This means we can write $\lambda = \lambda_{0,0,0}$ and $\lambda' = \lambda_{h,k,l}$ and for a particular representation which can run to infinity for $\lambda_{h,k,l}$ where all possible weights exist. 

\subsubsection{Inclusion of modules}

In each chamber there is then a different highest weight labelling a sub-module of a module in a higher chamber denoted by $M(\lambda_{h,k,l}) \subset M(\lambda_{h',k',l'})$. This is also known as the weak Bruhat ordering noted in \cite{Cheng_2008} (see e.g \cite{hersh2017weak} for definitions) of the highest weights $\lambda_{h,k,l} \rightarrow \lambda_{h',k',l'}$. Here this is extended across the affine wall such that it covers two domains of definition of Weyl chambers and Verma modules $M$ and $\tilde{M}$. One generalises the notion of a Verma module to include an additional module $G(\lambda_{\infty,\infty,\infty l})$ between the 2 domains. One can start in one fundamental Weyl chamber and move in two possible directions. For the first direction one can write the sequence of submodules as:
\begin{align} \label{affineVermamodules}
& M(\lambda_{0,0,0}) \ \subset  \  M(\lambda_{1,0,0})  \   \subset   \    M(\lambda_{2,0,0})  \  \dots  \  M(\lambda_{m-1,0,0})   \  \subset  \   M(\lambda_{m,0,0}) \  \ \dots  \ \\ \nonumber \\ \nonumber 
 & M(\lambda_{\infty-2,0,0}) \ \subset  \   M(\lambda_{\infty-1,0,0})  \ \subset \   M(\lambda_{\infty,0,0}) \ \subset \  G(\lambda_{\infty,\infty,\infty l}) \ \subset  \ \\ \nonumber \\ \nonumber  & \tilde{M}(\lambda_{\infty, \infty-1,\infty l}) \ \subset \ \tilde{M}(\lambda_{\infty, \infty-2,\infty l}) \ \   \dots \ \ \tilde{M}(\lambda_{\infty,k+2,\infty l}) \ \subset  \  \tilde{M}(\lambda_{\infty,k+1, \infty l})  \ \subset  \ \\ \nonumber \\ \nonumber
 & \tilde{M}(\lambda_{\infty,k,\infty l}) \ \dots \ \tilde{M}(\lambda_{\infty,2,\infty l}) \ \subset \ \tilde{M}(\lambda_{\infty,1,\infty l}).  
\end{align}
This is the weight system in 1 direction, and these are all the weights that exist for a particular representation in one direction one can take from the fundamental Weyl chamber to the opposite chamber where all states can exist. 

\subsubsection{Difference in weights}

Now one can define a prescription for determining whether a particular root exists. For this we have a rule that the difference in the weights must be a sum of integer combinations of roots, meaning that  
\begin{align}
\lambda_{h,k,l}- \lambda_{h',k',l'} = \sum_{i} n_{i}\alpha_{i}  \  \in \Gamma.
\end{align}
This means that if we know a set of weights that exist one can deduce the roots that exist by taking the difference of each pair of weights. Then one can determine the spectrum of roots by finding all the roots that must be present to span these differences between the weights. In the setup of the weight system above a new root enters existence as the highest weight jumps. We have the differences as
\begin{align}
\lambda_{m,0,0} -  \lambda_{m-1,0,0} = (m-1) \delta- \alpha_{0},
\end{align}
and 
\begin{align}
\lambda_{\infty,n,\infty l} -  \lambda_{\infty,n+1, \infty l} = n \delta + \alpha_{0}.
\end{align}
When these differences exist in the weight system the roots also enter the spectrum. For the imaginary roots we have
\begin{align}
\lambda_{\infty,\infty,\infty l} -  \lambda_{\infty,0,0} = \sum_{l=1}^{\infty} l \delta \   \in \Gamma.
\end{align}
So the imaginary roots enter the spectrum with this highest weight. Now that we have a prescription for counting roots from highest weight Verma modules in (\ref{affineVermamodules}) we can apply this  to counting BPS states in various $\mathcal{N}=2$ theories. Now, remembering (\ref{variablesintermsofroots11}), we have inner products of the roots with which we can write the complex 2d space of central charges $Z(w)$ parameterised in the form $\alpha_{0}(u) \in \mathbb{C}$ and $\delta(u) \in \mathbb{C}$ for all examples. However, for each specific example one can write these complex numbers in terms of different complex numbers associated to the particular theory, for example the central charges. Therefore, the same root system can give rise to different wall crossing phenomena in different $\mathcal{N}=2$ theories.

\subsection{Example 1: Jafferis-Moore D6-D2-D0 bound state} \label{D6-D2-D0 walls}

Jafferis and Moore \cite{D6D2D0JafferisMoorehttps://doi.org/10.48550/arxiv.0810.4909} define a class of generalised Donaldson-Thomas invariants that count D6-D2-D0 boundstates in type IIA string theory on the resolved conifold which undergo sucessive jumps at walls defined by $\hat{A}_{1}$ Weyl chamber boundaries. In the region with the maximal number of boundstates, also known as the Szendrői region \cite{Szendr_i_2008}, they are defined in terms of non-commutative Donaldson-Thomas invariants \cite{joyce2010theory}. Here the relevant central charges, expressed in terms of the Kähler parameter $t = z \mathcal{P}+ \Lambda e^{i \phi} \mathcal{P'}$, are $Z_{\gamma_{1}}(t) = \Lambda^{3}e^{3 i \phi}$ and $Z_{\gamma_{2}}(t) = -mz-n, \  n,m \in \mathbb{Z}$ \cite{D6D2D0JafferisMoorehttps://doi.org/10.48550/arxiv.0810.4909}. In this case the parameters $z$ and $\phi$ are also those found in the topological string partition function derived by \cite{ASTThttps://doi.org/10.48550/arxiv.2109.06878} which reproduces the wall crossing for these framed BPS states \footnote{This is done by taking the Borel transformation of the topological string free energy along different rays.}. In this case the variables are identified as the argument of the topological string coupling $\arg(\lambda')= 3 \phi$ and the complexified Kähler parameter $z= t'$. The central charges are obtained by taking the positive real parameter $\Lambda \rightarrow \infty$. In this case the stability condition on the central charges takes the form
\begin{align}
\langle \gamma_{1}, \gamma_{2} \rangle \text{Im}[Z_{\gamma_{1}}(t) \bar{Z}_{\gamma_{2}}(t)] = -n  \Lambda^{3} \text{Im}[e^{3 i \phi}(-mz^{*}-n)] > 0.
\end{align}
The physical walls that correspond to D6-D2-D0 bound states can be further be simplified to: $m = \pm 1, 0$ such that the stability conditions become:
\begin{align}
 -n \text{Im}[-z^{*}e^{3 i \phi}-ne^{3 i \phi}] > 0,   \    \    \
   -n \text{Im}[z^{*}e^{3 i \phi}-ne^{3 i \phi}] > 0,  \  \  \   -n \text{Im}[-ne^{3 i \phi}]  > 0.
\end{align}
\begin{ex}
This example involving the linear combinations of moduli constituting the D6-D2-D0 central charges can be matched to the parameters we have defined in (\ref{variablesintermsofroots1}) for the $\mathcal{N} = 4$ theory and can also be used in the $\mathcal{N}=2$ analog for the generating function  in (\ref{eq:N=2generatingfunctionaffinehighestweight}). 
In this example the analog is constructed from the $\hat{A}_{1}$ root system by writing the roots in terms of the central charges
\begin{align}
\pm \alpha_{0}(u)+n\delta(u) = -i Z_{\gamma_{1}}(t) \bar{Z}_{\gamma_{2}}(t) ,  \  \  \  n \in \mathbb{Z}
\end{align}
In this case we have the walls at $\text{Im}[\pm i\alpha_{0}(u)+n i\delta(u)] = 0 $ and can define the imaginary part of the contour as \footnote{We follow a contour prescription analagous to the $\mathcal{N}=4$ case discussed in section \ref{subsection:affineA1contourprescription}.}
\end{ex}

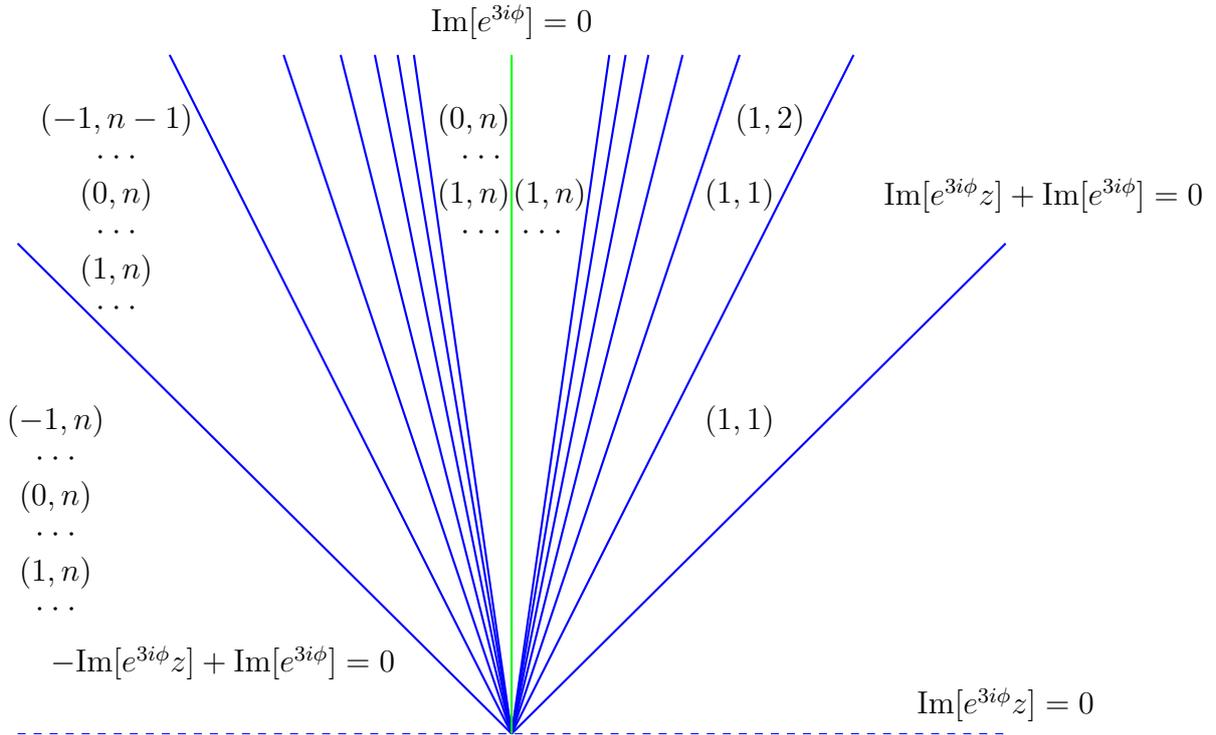
\begin{figure}
\hspace*{0cm}%
 \begin{center}
 \begin{tikzpicture}	
	\node at (-3.8,0.8) {$-\text{Im}[e^{3i \phi}z] + \text{Im}[e^{3i \phi}] = 0$};
	\node at (7, 7) {$\text{Im}[e^{3i \phi}z] + \text{Im}[e^{3i \phi}] = 0$};
 
	\node at (6.5, 0.25) {$\text{Im}[e^{3i \phi}z] = 0$};
	
	\node at (0, 9.3) {$\text{Im}[e^{3i \phi}] = 0$};
	
	\node at (0.4, 6.5) {$\cdot  \cdot  \cdot$};	
	
	\node at (3, 4) {$(1,1)$};
	\node at (3.4, 8) {$(1,2)$};	
	
	\node at (3, 7) {$(1,1)$};

	\node at (0.5, 7) {$(1,n)$};
	
	\node at (-0.5, 7) {$(1,n)$};
	\node at (-0.4, 7.5) {$\cdot  \cdot  \cdot$};
	
	\node at (-0.5, 8) {$(0,n)$};
	\node at (-0.4, 6.5) {$\cdot  \cdot  \cdot$};

	\node at (-6, 4) {$(-1,n)$};
	\node at (-6, 3.5) {$\cdot  \cdot  \cdot$};

	\node at (-6, 3) {$(0,n)$};
	\node at (-6, 2.5) {$\cdot  \cdot  \cdot$};
	
	\node at (-6, 2) {$(1,n)$};
	\node at (-6, 1.5) {$\cdot  \cdot  \cdot$};

	\node at (-5.2, 8) {$(-1,n-1)$};
	\node at (-5.2, 7.5) {$\cdot  \cdot  \cdot$};

	\node at (-5.2, 7) {$(0,n)$};
	\node at (-5.2, 6.5) {$\cdot  \cdot  \cdot$};
	
	\node at (-5.2, 6) {$(1,n)$};
	\node at (-5.2, 5.5) {$\cdot  \cdot  \cdot$};	
	
	\draw[thick,blue,yshift=-4pt,
	]
	(0,0) -- (6.5,6.5) ;
	\draw[thick,blue,yshift=-4pt,
	]
	(0,0)  -- (-6.5,6.5);
	
	\draw[dashed,blue,yshift=-4pt,
	]
	(-6.5,0)  -- (6.5,0);
	
	\draw[thick,blue,yshift=-4pt,
	]
	(0,0)  -- (4.5,9);
	
	\draw[thick,blue,yshift=-4pt,
	]
	(0,0)  -- (-4.5,9);

	\draw[thick,blue,yshift=-4pt,
	]
	(0,0)  -- (3,9);
	
	\draw[thick,blue,yshift=-4pt,
	]
	(0,0)  -- (-3,9);

	\draw[thick,blue,yshift=-4pt,
	]
	(0,0)  -- (9/4,9);
	
	\draw[thick,blue,yshift=-4pt,
	]
	(0,0)  -- (-9/4,9);

	\draw[thick,blue,yshift=-4pt,
	]
	(0,0)  -- (9/5,9);
	
	\draw[thick,blue,yshift=-4pt,
	]
	(0,0)  -- (-9/5,9);
	
	\draw[thick,blue,yshift=-4pt,
	]
	(0,0)  -- (1.5,9);
	
	\draw[thick,blue,yshift=-4pt,
	]
	(0,0)  -- (-1.5,9);

	\draw[thick,blue,yshift=-4pt,
	]
	(0,0)  -- (9/7,9);
	
	\draw[thick,blue,yshift=-4pt,
	]
	(0,0)  -- (-9/7,9);
 
	\draw[thick,green,yshift=-4pt,
	]
	(0,0)  -- (0,9);
	
	\end{tikzpicture}
	
\end{center}
\caption{Walls for the bound states of D2-D0 branes to a large D6 brane.}		
\label{D6D2D0fig2}	
\end{figure}

\vspace{-1.2cm}

\begin{align}
\text{\underline{\textbf{\small{Relation of roots to central charges}}}} \\ \nonumber
\\ \nonumber
\text{Im}[\pm i\alpha_{0}(u)+ni\delta(u)] = \text{Im}[Z_{\gamma_{1}}(t) \bar{Z}_{\gamma_{2}}(t)],   \\ \nonumber
\text{Im}[i\alpha_{0}(u)] = \frac{\text{Im}[z^{*}e^{3 i \phi}]}{\epsilon},  \  \  \\ \nonumber
\text{Im}[i\delta(u)] = -\frac{\text{Im}[e^{3 i \phi}]}{\epsilon},  \  \  \  \text{where} \  \frac{1}{\epsilon} = \Lambda^{3}.
\end{align}
This will generate the wall crossing of the Jafferis-Moore BPS states with splitting given by charges $(\pm 1,n)$ if $\text{Im}[\pm \alpha_{0}(u)+n \delta(u)] $ changes sign. Each Weyl chamber in the lower half of the $\text{Im}[\delta(u)]$ axis then contains a different combination of BPS states. Examples of the possible combinations of the moduli parameters include:
\begin{align}
\text{\underline{\textbf{\small{Inner product of roots with the moduli}}}} \\ \nonumber
\\ \nonumber
\pm \text{Im}[i\alpha_{0}(u)] &= \pm \frac{\text{Im}[z^{*}e^{3 i \phi}]}{\epsilon}, \\ \nonumber
\pm \text{Im}[i\alpha_{0}(u)] -\text{Im}[i\delta(u)]&=  \pm \frac{\text{Im}[z^{*}e^{3 i \phi}]}{\epsilon} + \frac{\text{Im}[e^{3 i \phi}]}{\epsilon},  \\ \nonumber
\pm \text{Im}[i\alpha_{0}(u)] -2 \text{Im}[i\delta(u)] &=  \pm \frac{\text{Im}[z^{*}e^{3 i \phi}]}{\epsilon} + 2 \frac{\text{Im}[e^{3 i \phi}]}{\epsilon},  \\ \nonumber
\pm \text{Im}[i\alpha_{0}(u)]-3\text{Im}[i\delta(u)] &= \pm \frac{\text{Im}[z^{*}e^{3 i \phi}]}{\epsilon} + 3\frac{\text{Im}[e^{3 i \phi}]}{\epsilon},  \\ \nonumber
\dots & = \dots  \\ \nonumber
- n \text{Im}[i\delta(u)] & =  n \Big(\frac{\text{Im}[e^{3 i \phi}]}{\epsilon} \Big).
\end{align}
If these equations are set to 0, such that e.g. $\pm\text{Im}[i\alpha_{0}(u)] +n \text{Im}[i\delta(u)] = 0$, these represent the walls for the bound state in 2 parameters. In this case the lower half plane for the BPS walls for the $\hat{A}_{1}$ root system matches the wall crossing described in \cite{D6D2D0JafferisMoorehttps://doi.org/10.48550/arxiv.0810.4909,Szendr_i_2008,joyce2010theory}. This means the highest weight in the generating function counts boundstates of the charges in the $\hat{A}_{1}$ root system to the large D6 brane core. However, it is a different generating function to the partition function for higher linear combinations of charges outside this $\hat{A}_{1}$ root system \cite{D6D2D0JafferisMoorehttps://doi.org/10.48550/arxiv.0810.4909,ASTThttps://doi.org/10.48550/arxiv.2109.06878}. This is shown in Figure \ref{D6D2D0fig2}.

\subsection{Example 2: Seiberg-Witten theory} \label{SeibergWittenexamplesubsection}

Seiberg-Witten theory is represented by affine SU(2) Lie algebra or $\hat{A}_{1}$, see e.g. Cecotti and Vafa \cite{cecottivafacompleteclassificationhttps://doi.org/10.48550/arxiv.1103.5832}. It was introduced by Seiberg and Witten in \cite{Seiberg_1994} and reviewed extensively in e.g. \cite{1997Lerche}.
The theory has also been described by a 2 node quiver with 2 arrows $\bullet \rightrightarrows \bullet$ (see \cite{quivershallhaloDenef_2002,quivergaugetheoryFiol:2006jz,cecottivafacompleteclassificationhttps://doi.org/10.48550/arxiv.1103.5832,N=2BPSquivershttps://doi.org/10.48550/arxiv.1112.3984}). As described in section \ref{section:rootsystemBPSliealgebras} one can construct the Cartan matrix for $\hat{A}_{1}$ from the undirected adjacency matrix of the quiver.  
Another way of looking at this is through the Seiberg-Witten curve, an elliptic curve that defines the central charges and monodromies of the theory and is parameterised by a complex one-dimensional modulus $w \in \mathbbm{P}^1\setminus  \{-1,1,\infty \}$. The number and type of BPS states that exist within the theory depends on the region within this one-dimensional moduli space - there exist two chambers. In one chamber this theory has infinitely many BPS states given by the hypermultiplet dyons $-m(\alpha_{0}+\alpha_{1})-\alpha_{1}$,  $-(m-1) (\alpha_{0}+\alpha_{1})+\alpha_{1}$ and the W-boson with a charge corresponding to the roots $\alpha_{0}+\alpha_{1}$. In the other chamber only the 2 states (a monopole and a dyon) exist as  a basis $\alpha_{1}$ and $\alpha_{0}$. These results have been previously derived through other methods including quiver mutations and attractor flow but can also be derived via the generating function. 
	
As with the previous examples, to derive the BPS state counts through the generating function one must read off the highest weight using the Weyl denominator of the $\hat{A}_{1}$ Lie algebra. 

\begin{dfn}

Again, we recall from (\ref{variablesintermsofroots1}-\ref{variablesintermsofroots11}) that the roots act on the Cartan subalgebra, such that we have 2 complex variables: 
\begin{align}
\alpha_{1}(u) = \tau \in \mathbb{C}, \ \ \alpha_{0}(u) = z \in \mathbb{C} \ \  \text{and} \ \ \delta(u)= \alpha_{0}(u)+ \alpha_{1}(u) = \tau+z.
\end{align} 
Here we have redefined our basis of roots given in (\ref{variablesintermsofroots1}-\ref{variablesintermsofroots11}) by exchanging $\alpha_{0}(u)$ and $\alpha_{1}(u)$ to better compare to the charges existing in Seiberg-Witten theory. 

\end{dfn}

One can also integrate to find Fourier coefficients in different chambers as in the $\mathcal{N}=4$ example in section \ref{subsection:affineA1contourprescription}. The counting function itself can be written as \footnote{The highest weight can now be written as  $\lambda = n \alpha_{0}(u)+l\alpha_{1}(u)$.}
	\begin{align} \label{eq:seibergwittenspecificgeneratingfunction}
g'_{\hat{A}_{1}}(u) = \  & \frac{e^{n \alpha_{0}(u)+l\alpha_{1}(u)}}{ \prod_{m=1}^{\infty}(1- e^{-m (\alpha_{0}(u)+\alpha_{1}(u))})(1-e^{-m(\alpha_{0}(u)+\alpha_{1}(u))-\alpha_{1}(u) })(1- e^{-(m-1)(\alpha_{0}(u)+\alpha_{1}(u))+ \alpha_{1}(u)})  }, \\ \nonumber
 & = \frac{e^{n\tau+m z}}{ \prod_{m=1}^{\infty}(1- e^{-m (z+\tau)})(1-e^{-m(z+\tau)- \tau })(1- e^{-(m-1)(z+\tau)+ \tau})}, 
	\end{align} 	
where we follow (\ref{eq:N=2generatingfunctionaffinehighestweight}) but with the roots exchanged and the introduction of a new labelling of the highest weight using $n,l \in \mathbb{Z}$. 

\subsubsection{Seiberg-Witten walls} \label{seibergwittenframedwalls}

In Seiberg-Witten theory there is only 1 parameter $w$ describing the moduli space of the Seiberg-Witten curve that enters the stability condition. The equation for the wall of marginal stability can then be written as: $\text{Im}[Z_{\alpha_{0}}(w) \bar{Z}_{\alpha_{1}}(w)] = 0$. We know that in Seiberg-Witten theory there is one region of the moduli space with just 2 basis BPS states existing in $\mathcal{S}_w$ and 1 other region on the other side of the wall with infinitely many BPS states existing - the full spectrum of roots of the $\hat{A}_{1}$ Lie algebra. This means, if we are to match this with the chambers of the generating function, we must jump from one chamber to another and thus obtain a change in BPS degeneracies that describes this. For this we must choose a moduli prescription under which this jumping occurs. 

For this we must be careful to distinguish the framed BPS states described in GMN \cite{FramedBPSGMNhttps://doi.org/10.48550/arxiv.1006.0146} and reviewed here in section \ref{framedbpsstates} from the original vanilla BPS states in the theory. 

Recalling from section \ref{framedbpsstates} that the BPS walls, now detoned as $W_{\alpha_{i}}$, intersect on the wall of marginal stability $MS_{\alpha_{0}, \alpha_{1}}$. One can now consider that the walls for the framed halo BPS states $W_{\alpha_{i}}$ form a cone for which in the upper region there exists the full infinite Seiberg-Witten BPS spectrum and in the lower just the 2 basis states. Which region is which can be determined by the attractor flow existence conditions on the central charges from section \ref{attractorflowexistence} discussed previously in the work on attractor flow \cite{Alim:2023doi} following the literature \cite{MooreArithmeticAttractorshttps://doi.org/10.48550/arxiv.hep-th/9807087,DenefSeibergWittenemptyhole_2000, DenefGreeneRaugas_2001}. 

\begin{ex}

The identification between the roots and the central charges is as follows:
\begin{align}
\alpha_{0}(u) = \frac{Z_{\alpha_{0}}(w)}{\mu}, \ \  
\alpha_{1}(u) =  \frac{Z_{\alpha_{1}}(w)}{\mu}, \ \ \delta(u) = \frac{1}{\mu}(Z_{\alpha_{0}}(w)+Z_{\alpha_{1}}(w)).
\end{align}
Here we can take $\mu = \epsilon \zeta$ where $\zeta:= e^{i \theta}, \ \theta \in \{0,2\pi \}$ is again a phase, and we recall from (\ref{eq:rootscharges}) that $\epsilon \in \mathbb{R}$ is a small parameter that can be used to define the contour (see also sec. \ref{subsection:affineA1contourprescription}) when extracting Fourier coefficients. This is analogous to the $\mathcal{N} =4$ examples in \cite{Cheng_2007,Cheng_2008} where one takes contours at infinity $\epsilon \rightarrow 0$ to generate a consistent jumping of dyon degeneracies.
In $\mathcal{N}=2$ language the phase $\zeta$ is a complex parameter representing the phase of an infinitely heavy dyon to which the framed particles are bound.

\end{ex}

Now we will briefly review the wall crossing for the framed BPS states: On the upper half plane the jumping occurs at the walls $\pm \text{Im}[i\alpha_{0}(u)]+m \text{Im}[i\alpha_{1}(u)+i\alpha_{0}(u)] = 0, \ m \in \mathbb{Z}$. We can choose the moduli that give the matched jumping for the framed halo BPS states. The walls are given by the vanishing locus of the imaginary parts of the central charges:\footnote{Here we are letting $-i \mu \rightarrow \mu$.}
\begin{center}  
\begin{align*}
\text{\underline{ \textbf{Imaginary parts of central charges on one side of affine wall} $W_{\alpha_{0}+\alpha_{1}}$ }}   
\end{align*}
\begin{align}
\text{Im}[i\alpha_{0}(u)] & = \text{Im}\left[\frac{Z_{\alpha_{0}}(w)}{\mu} \right], 
\\ \nonumber 
\\ \nonumber 
\text{Im}[i\alpha_{0}(u)]+\text{Im}[i\alpha_{1}(u)+i\alpha_{0}(u)] & =  \text{Im}\left[\frac{2Z_{\alpha_{0}}(w)+Z_{\alpha_{1}}(w)}{\mu}\right],
\\ \nonumber
\\ \nonumber
\text{Im}[i\alpha_{0}(u)]+2\text{Im}[i\alpha_{1}(u)+i\alpha_{0}(u)] & =  \text{Im}\left[\frac{3Z_{\alpha_{0}}(w)+2Z_{\alpha_{1}}(w)}{\mu}\right], \\ \nonumber
\\ \nonumber
\dots & = \dots .
\end{align}
\end{center}
where the dots represent an infinite continuation of this pattern of BPS walls. Now we can look at the combinations on the other side of the affine wall and determine what happens as these BPS walls are crossed:  
\begin{align}
\text{\underline{ \textbf{\small{On other side, and including}} $W_{\alpha_{0}+\alpha_{1}}$ }} \\ \nonumber
\\ \nonumber
-\text{Im}[i\alpha_{0}(u)] &= -\text{Im}\left[\frac{Z_{\alpha_{0}}(w)}{\mu}\right],  \\ \nonumber
\\ \nonumber
-\text{Im}[i\alpha_{0}(u)] +\text{Im}\left[i\alpha_{1}(u)+i\alpha_{0}(u)\right]&=  \text{Im}\left[\frac{Z_{\alpha_{1}}(w)}{\mu}\right],  \\ \nonumber \\ \nonumber
-\text{Im}[i\alpha_{0}(u)] +2 \text{Im}[i\alpha_{1}(u)+i\alpha_{0}(u)] &=  \text{Im}\left[\frac{Z_{\alpha_{0}}(w)+2Z_{\alpha_{1}}(w)}{\mu}\right],  \\ \nonumber \\ \nonumber
-\text{Im}[i\alpha_{0}(u)]+3\text{Im}[i\alpha_{1}(u)+i\alpha_{0}(u)] &= \text{Im}\left[\frac{2Z_{\alpha_{0}}(w)+3Z_{\alpha_{1}}(w)}{\mu}\right], \\ \nonumber \\ \nonumber
\dots & = \dots  \\ \nonumber
\\ \nonumber
n \text{Im}[i\alpha_{1}(u)+i\alpha_{0}(u)] & =  n \text{Im}\left[\frac{Z_{\alpha_{0}}(w)+Z_{\alpha_{1}}(w)}{\mu}\right].
\end{align}

All these walls intersect on the wall of marginal stability $MS_{\alpha_{0}, \alpha_{1}}$ for the vanilla states. This produces a cone with the full $\hat{A}_{1}$ spectrum on one side and the basis states on the other. The diagram below Fig. \ref{fig:SWwalls} shows the existence conditions for the framed halo BPS states of Seiberg-Witten theory. The walls of the framed halo BPS states all intersect on the wall of the vanilla BPS degeneracies. The inside of the wall represents the region where only 2 vanilla BPS states exist.

\subsubsection{Existence conditions from generating function}

Now we recall the attractor flow existence conditions in section \ref{attractorflowexistence} so that flow lines (or BPS walls) terminating at regular points in the moduli space, where the central charge vanishes, are excluded. In terms of the generating function, this condition can be stated such that the generating function must not have any poles at regular points in the moduli space. To do this one can redefine the factor in the denominator that contains the pole, at the wall of marginal stability $MS_{\alpha_{0}, \alpha_{1}}$, where the central charges align. This can be done by writing the central charge that causes the pole at a regular point in terms of a real function multiplying one of the other central charges that vanishes at a singular point.
Then one can choose such a function that can be continued across the wall of marginal stability $MS_{\alpha_{0}, \alpha_{1}}$. This is then no longer a root and can no longer be written as the linear combination of the other roots. Hence this BPS state doesn't exist or contribute to any count of BPS states in a highest weight of a representation or module. One can then just consider the factor in the generating function for the non-existing BPS state as just a normalisation.

We start by explicitly writing the generating function (\ref{eq:seibergwittenspecificgeneratingfunction}) in terms of central charges
\begin{align} \label{eq:seibergwittenspecificgeneratingfunction2}
& g_{\hat{A}_{1}}'(w) = \\ \nonumber  & \frac{e^{\lambda(w)}}{ \prod_{m=1}^{\infty}(1- e^{-\frac{m}{\mu} (Z_{\alpha_{0}}(w)+Z_{\alpha_{1}}(w))})(1-e^{-\frac{m}{\mu} (Z_{\alpha_{0}}(w)+Z_{\alpha_{1}}(w))-\frac{Z_{\alpha_{1}}(w)}{\mu} })(1- e^{-\frac{(m-1)}{\mu} (Z_{\alpha_{0}}(w)+Z_{\alpha_{1}}(w))+\frac{Z_{\alpha_{1}}(w)}{\mu}})  }.
\end{align} 
Now we can write the factors representing central charges that vanish at regular points in terms of ratios of the central charges such that: 
\begin{align*}
\text{\underline{\textbf{\small{Central charges in terms of ratio}}}}
\end{align*}
\begin{align}
-\frac{Z_{\alpha_{1}}(w)}{\mu} \Big(1+\frac{Z_{\alpha_{0}}(w)}{Z_{\alpha_{1}}(w)} \Big)m \ \ & = \  \ \frac{Z_{\alpha_{1}}(w)}{\mu}r_{m}(w),  \\ \nonumber
\frac{Z_{\alpha_{1}}(w)}{\mu} \Big(1-(m-1) \big(1+\frac{Z_{\alpha_{0}}(w)}{Z_{\alpha_{1}}(w)} \big) \Big) \  & =  \ \frac{Z_{\alpha_{1}}(w)}{\mu}r_{m+1,m}(w) ,  \\ \nonumber
\frac{Z_{\alpha_{1}}(w)}{\mu} \Big(-1-m \big(1+\frac{Z_{\alpha_{0}}(w)}{Z_{\alpha_{1}}(w)} \big) \Big) \  & =  \  \frac{Z_{\alpha_{1}}(w)}{\mu}r_{m,m+1}(w),
\end{align} 
\begin{align}
& r_{m}(w) =   -\Big(1+\frac{Z_{\alpha_{0}}(w)}{Z_{\alpha_{1}}(w)}\Big)m, \  \   \  
r_{m+1,m}(w) = 1-(m-1) \Big(1+\frac{Z_{\alpha_{0}}(w)}{Z_{\alpha_{1}}(w)}\Big), \  m >2, \\ \nonumber
& r_{m,m+1}(w) = -1-m \Big(1+\frac{Z_{\alpha_{0}}(w)}{Z_{\alpha_{1}}(w)}\Big).
\end{align} 
On the wall of marginal stability, the ratio of the central charges is real. Therefore, we have on the wall of marginal stability: $r_{m}(w), \ r_{m+1,m}(w), \ r_{m,m+1}(w) \in \mathbb{R}$.
\begin{align*}
\text{\underline{\textbf{\small{Example of continuation through wall of marginal stability}} $MS_{\alpha_{0}, \alpha_{1}}$}}     
\end{align*}
Now we find a continuation of the generating function through the wall of marginal stability $MS_{\alpha_{0}, \alpha_{1}}$ with a function that avoids the poles at a regular point in the moduli space. This is done for the central charges in the exponents that would otherwise vanish at a regular point on the other side of the wall by choosing real continuations of $r_{m}(w), \ r_{m+1,m}(w), $ and $r_{m,m+1}(w)$, such that the central charges now behave as those that flow to singular points. 

\begin{dfn}
For example, one can choose \footnote{This is only an example and there are potentially many other possible continuations that work.}
\begin{align}
  & r_{m}(w) =   - \Big(1+ \Big| \frac{Z_{\alpha_{0}}(w)}{Z_{\alpha_{1}}(w)} \Big| \Big)m, \  \  \
 r_{m+1,m}(w) = 1-(m-1) \Big(1+\Big| \frac{Z_{\alpha_{0}}(w)}{Z_{\alpha_{1}}(w)} \Big| \Big) , \  \ m >2,  \\ \nonumber  \ \
 & r_{m,m+1}(w) = -1-m \Big(1+\Big|\frac{Z_{\alpha_{0}}(w)}{Z_{\alpha_{1}}(w)} \Big| \Big),
\end{align} 
on the other side of the wall of marginal stability $MS_{\alpha_{0}, \alpha_{1}}$. 
\end{dfn}

Now the generating function avoids all poles at regular points. Essentially the BPS walls $W_{\alpha_{i}}$ have collapsed into 2 walls representing the basis states. This means there are 2 BPS states that exist in this region as the generating function can maximally count only 2 BPS states in its highest weight. 

\begin{ex}[$\mathbf{Excluded \ BPS \ states}$]  \ \newline
The exponents of the non-existing BPS states are no longer in the lattice of positive roots  
$ Z_{\alpha_{1}}(w) / \mu \ \ r_{m}(w), $ 
$ \  Z_{\alpha_{1}}(w) / \mu \ \ r_{m,m+1}(w), \ Z_{\alpha_{1}}(w) / \mu \ \ r_{m+1,m}(w), \ \ m >2 \ \ \notin \Delta^{+}$. This is because 
\begin{align}
 \frac{Z_{\alpha_{1}}(w)}{\mu}r_{m}(w) \neq & \ m \delta(u), \ \ \frac{Z_{\alpha_{1}}(w)}{\mu}r_{m,m+1}(w) \neq \alpha_{1}(u) - (m-1) \delta(u) \ \ \text{and} \\ \nonumber \ \ \frac{Z_{\alpha_{1}}(w)}{\mu}r_{m+1,m}(w) \neq & -\alpha_{1}(u)-m\delta(u).
\end{align} 
\end{ex}

\begin{ex}[$\mathbf{Existing \ BPS \ states}$] \ \newline
Now the only BPS states that can still be written as roots include $\alpha_{0}(u) = Z_{\alpha_{0}}(w) / \mu$
and $\alpha_{0}(u)-\delta(u) =-\alpha_{1}(u) = -Z_{\alpha_{1}}(w) / \mu$ as these were not modified at the wall.
The attractor flow for these states terminates at singular points.
\end{ex}
\begin{align*}
\text{\underline{ \textbf{\small{Continuation of generating function through}} $MS_{\alpha_{0}, \alpha_{1}}$}}
\end{align*}
On this side of the wall of marginal stability $MS_{\alpha_{0}, \alpha_{1}}$ the generating function (\ref{eq:seibergwittenspecificgeneratingfunction2}) can be re-written as
\begin{align*}
& \tilde{g}'_{\hat{A}_{1}}(w) = \\ \nonumber & \frac{e^{\lambda(w)}}{(1-e^{-\frac{Z_{\alpha_{0}}(w)}{\mu}})(1-e^{\frac{Z_{\alpha_{1}}(w)}{\mu}})\prod_{m=1}^{\infty}(1- e^{\frac{Z_{\alpha_{1}}(w)}{\mu}r_{m}(w)})(1-e^{\frac{Z_{\alpha_{1}}(w)}{\mu}r_{m,m+1}(w)})\prod_{m=3}^{\infty}(1- e^{\frac{Z_{\alpha_{1}}(w)}{\mu}r_{m+1,m}(w)}) },
	\end{align*} 
such that it now becomes
\begin{align} \label{eq:seibergwittenspecificgeneratingfunction3}
& \frac{e^{\lambda(w)}}{(1-e^{-\frac{Z_{\alpha_{0}}(w)}{\mu}})(1-e^{\frac{Z_{\alpha_{1}}(w)}{\mu}}) }f\Big(\frac{Z_{\alpha_{1}}(w)}{\mu}r_{m}(w),\frac{Z_{\alpha_{1}}(w)}{\mu}r_{m,m+1}(w),\frac{Z_{\alpha_{1}}(w)}{\mu}r_{m+1,m}(w)\Big),
	\end{align} 
where this is now a generating function of 2 BPS states only. All BPS walls $W_{\alpha_{i}}$ have collapsed onto two and the function $f\big(\frac{Z_{\alpha_{1}}(w)}{\mu}r_{m}(w),\frac{Z_{\alpha_{1}}(w)}{\mu}r_{m,m+1}(w),\frac{Z_{\alpha_{1}}(w)}{\mu}r_{m+1,m}(w)\big)$ in (\ref{eq:seibergwittenspecificgeneratingfunction3}) can be treated now just as a normalisation. The framed wall crossing on this side of the wall of marginal stability can now be tabulated.
\begin{table}[h!]
    
  \begin{center}

\begin{tabular}{ |l|}
	\hline
	   \vphantom{\Huge{H}}	Highest weight on side with 2 BPS states \\[8pt]
	\hline
	 \vphantom{\Huge{H}}	
  $\lambda + (\alpha_{0}-\delta) + \alpha_{0}$
		
		 \\[8pt]
	\hline
\vphantom{\Huge{H}} $\lambda +\alpha_{0}-\delta $ 
 \\[8pt]	
	\hline
\vphantom{\Huge{H}} $\lambda$ \\[8pt]
	\hline
\end{tabular}

\end{center}

    \caption{Highest weights of the modules when only 2 BPS states exist. Shown on lower half of Fig. \ref{fig:SWwalls}. }
    \label{tab:SWhighestweightsonsidewith2}
\end{table}
A complete tabulation (\ref{tab:completetabulationforaffinea1}) of crossing of both the BPS walls $W_{\alpha_{i}}$ and the wall of marginal stability $MS_{\alpha_{0}, \alpha_{1}}$ can also be written down.
\begin{table}[h!]

\begin{center}
\begin{tabular}{ |l|l|}
	\hline
	\multicolumn{2}{|c|}{ \vphantom{\Huge{H}} Tabulation on both sides of the wall of marginal stability $MS_{\alpha_{0}, \alpha_{1}}$ $\ \ \ \ \ \ \ \ \ \ \ \ \ \ \ \ \ \ \ \ \ \ \ \ \ \ \ $}
	
	\\[8pt]
	\hline
	 \vphantom{\Huge{H}}	Side with full $\hat{A}_{1}$ spectrum & 
\vphantom{\Huge{H}} Side with 2 basis states \\[8pt]
	
	\hline
	& \\
	$\lambda + \sum_{m=2}^{\infty} \Big((m-1) \delta- \alpha_{0}  \Big) + \sum_{l=1}^{\infty}l \delta+ \sum_{n=0}^{\infty} \Big(n \delta+ \alpha_{0}  \Big) $ & $ \lambda + (\delta- \alpha_{0})+\alpha_{0} $ \\
	& \\
	\hline
	& \\
	$ \lambda + \sum_{m=h+1}^{\infty} \Big((m-1) \delta- \alpha_{0}  \Big) +\sum_{l=1}^{\infty}l \delta+ \sum_{n=0}^{\infty}\Big(n \delta+ \alpha_{0}  \Big)$ & 	$ \lambda + \delta- \alpha_{0} $   \\ 
	& \\
	\hline
	& \\
	$\lambda + \sum_{l=1}^{\infty}l \delta+ \sum_{n=0}^{\infty}\Big(n \delta+ \alpha_{0}  \Big)$& $\lambda + \delta- \alpha_{0}$ \\
	& \\
	\hline
	& \\
	$\lambda + \sum_{n=0}^{\infty}\Big(n \delta+ \alpha_{0}  \Big)$ & $\lambda + \delta- \alpha_{0} $  \\
	& \\
	\hline
	& \\
	$\lambda + \sum_{n=0}^{k}\Big(n \delta+ \alpha_{0}  \Big)$ & $\lambda + \delta- \alpha_{0} $ \\
	& \\
	
	\hline
	\multicolumn{2}{|c|}{$\lambda$} \\
	\hline
\end{tabular}

\end{center}

    \caption{Complete tabulation of all highest weights in the original basis for the Seiberg-Witten analog. First column is on the side of the wall with the full infinite spectrum. Second column is on the side with just the 2 basis states. This is shown on Fig. \ref{fig:SWwalls}.}
    \label{tab:completetabulationforaffinea1}
\end{table}
\begin{figure}

\vspace*{-3cm}%

\begin{center}

\begin{tikzpicture}

\fill[yellow] (0,-0.115)
-- (-7,-0.115) 
-- (-7,6.85) 
-- cycle;

\node at (6.5, 0.25) {$\text{Im}[Z_{\alpha_{0}}(w)]= 0$};

\node at (0.13, 9.8) {$\text{Im}[Z_{\alpha_{0}}(w)+Z_{\alpha_{1}}(w)] = 0$};

\node at (6.7, 6.7) {$\text{Im}[2Z_{\alpha_{0}}(w)+Z_{\alpha_{1}}(w)] = 0$};

\node at (6.7, -7.1) {$\text{Im}[Z_{\alpha_{1}}(w)] = 0$};

\node at (6.8, -8.3) {$\text{Im}[-Z_{\alpha_{0}}(w)+2Z_{\alpha_{1}}(w)] = 0$};

\node at (6.8, 8.3) {$ \  \  \  \ \text{Im}[3Z_{\alpha_{0}}(w)+2Z_{\alpha_{1}}(w)] = 0$};

\node at (0.65, 8.8) {$\cdot  \cdot  \cdot$};

\node at (-0.65, -0.4) {$w_{1}$};

\node at (3.4, 8) {$\alpha_{0}+2\delta$};

\node at (3.06, 7.2) {$\alpha_{0}+\delta$};

\node at (2.6, 6.2) {$\alpha_{0}$};

\node at (3.5, -5) {$\alpha_{0}-\delta$};

\node at (3.5, 4.6) {$\alpha_{0}+\delta$};

\node at (2.6, 3.3) {$\alpha_{0}$};

\node at (4, 2) {$\alpha_{0}$};

\node at (0.7, 9.15) {\small$\alpha_{0}+n\delta$};

\node at (-0.7, 9.15) {\small$\alpha_{0}+n\delta$};

\node at (-0.63, 8.8) {$\cdot  \cdot  \cdot$};

\node at (-0.45, 6.9) {\small$n\delta$};

\node at (-0.46, 6.55) {$\cdot  \cdot  \cdot$};

\node at (-5.5, 3.4) {\small$\alpha_{0}+n\delta$};

\node at (-5.5, 3.05) {$\cdot  \cdot  \cdot$};

\node at (-4, 2.4) {\small$\alpha_{0}-n\delta$};
\node at (-4, 2.05) {$\cdot  \cdot  \cdot$};

\node at (-2.5, 3.4) {\small$n\delta$};

\node at (-2.5, 3.05) {$\cdot  \cdot  \cdot$};

\node at (-4.5, 6.4) {\small$\alpha_{0}+n\delta$};

\node at (-4.5, 6.05) {$\cdot  \cdot  \cdot$};

\node at (-3.5, 5.0) {\small$\alpha_{0}-n\delta$};
\node at (-3.5, 4.7) {$\cdot  \cdot  \cdot$};
\node at (-3.5, 4.5) {\small$\alpha_{0}-2\delta$};

\node at (-2.5, 1.7) {\small$n\delta$};

\node at (-2.5, 1.35) {$\cdot  \cdot  \cdot$};

\node at (-4, 1.0) {\small$\alpha_{0}-\delta$};

\node at (-6, 1.5) {\small$\alpha_{0}$};

\node at (-4, -2.4) {\small$\alpha_{0}-\delta$};

\node at (4.1, -0.5) {$\text{Fundamental Weyl chamber 1}$};

\node at (-4.1, 0.23) {$\text{Fundamental Weyl chamber 2}$};

\draw[thick,red] plot [smooth] coordinates{(-7,3) (0,-0.15) (7,-4)};

\node at (6.4, -1.7) {$MS_{\alpha_{0}, \alpha_{1}}: \text{Im}[Z_{\alpha_{0}}(w) \bar{Z}_{\alpha_{1}}(w)] = 0$};

\node at (6.5, 9.8) {$MS: \text{wall of marginal stability}$};

\node at (-6, 7.9) {$\text{BPS walls}$};

\draw[thick,blue,yshift=-4pt,
]
(0,0) -- (6.5,6.5) ;

\draw[dashed,blue,yshift=-4pt,
]
(-6.5,-6.5) -- (0,0);

\node at (-6.5,-6.64)  {$\times$};

\draw[thick,blue,yshift=-4pt,
]
(0,0)  -- (-6.5,6.5);

\draw[thick,blue,yshift=-4pt,
]
(6.5,-6.5)  -- (0,0);

\draw[thick,blue,yshift=-4pt,
]
(0,0)  -- (6.5,0);

\draw[thick,blue,yshift=-4pt,
]
(-6.5,0)  -- (0,0);

\draw[thick,blue,yshift=-4pt,
]
(0,0)  -- (4.5,9);

\draw[dashed,blue,yshift=-4pt,
]
(-4.5,-9)  -- (0,0);

\node at (-4.5,-9.14)  {$\times$};

\draw[thick,blue,yshift=-4pt,
]
(0,0)  -- (-4.5,9);

\draw[dashed,blue,yshift=-4pt,
]
(4.5,-9)  -- (0,0);

\node at (4.5,-9.14)  {$\times$};

\draw[thick,blue,yshift=-4pt,
]
(0,0)  -- (3,9);

\draw[dashed,blue,yshift=-4pt,
]
(-3,-9)  -- (0,0);

\node at (-3,-9.14)  {$\times$};

\draw[thick,blue,yshift=-4pt,
]
(0,0)  -- (-3,9);

\draw[dashed,blue,yshift=-4pt,
]
(3,-9)  -- (0,0);

\node at (3,-9.14)  {$\times$};

\draw[thick,blue,yshift=-4pt,
]
(0,0)  -- (9/4,9);

\draw[dashed,blue,yshift=-4pt,
]
(-9/4,-9)  -- (0,0);

\node at (-9/4,-9.14)  {$\times$};

\draw[thick,blue,yshift=-4pt,
]
(0,0)  -- (-9/4,9);

\draw[dashed,blue,yshift=-4pt,
]
(9/4,-9)  -- (0,0);

\node at (9/4,-9.14)  {$\times$};

\draw[thick,blue,yshift=-4pt,
]
(0,0)  -- (9/5,9);

\draw[dashed,blue,yshift=-4pt,
]
(-9/5,-9)  -- (0,0);

\node at (-9/5,-9.14)  {$\times$};

\draw[thick,blue,yshift=-4pt,
]
(0,0)  -- (-9/5,9);

\draw[dashed,blue,yshift=-4pt,
]
(9/5,-9)  -- (0,0);

\node at (9/5,-9.14)  {$\times$};

\draw[thick,blue,yshift=-4pt,
]
(0,0)  -- (1.5,9);

\draw[dashed,blue,yshift=-4pt,
]
(-1.5,-9)  -- (0,0);

\node at (-1.5,-9.14)  {$\times$};

\draw[thick,blue,yshift=-4pt,
]
(0,0)  -- (-1.5,9);

\draw[dashed,blue,yshift=-4pt,
]
(1.5,-9)  -- (0,0);

\node at (1.5,-9.14)  {$\times$};

\draw[thick,blue,yshift=-4pt,
]
(0,0)  -- (9/7,9);

\draw[dashed,blue,yshift=-4pt,
]
(-9/7,-9)  -- (0,0);

\node at (-9/7,-9.14)  {$\times$};

\draw[thick,blue,yshift=-4pt,
]
(0,0)  -- (-9/7,9);

\draw[dashed,blue,yshift=-4pt,
]
(9/7,-9)  -- (0,0);

\node at (9/7,-9.14)  {$\times$};

\draw[thick,black,yshift=-4pt,
]
(0,0)  -- (0,9);

\draw[dashed,black,yshift=-4pt,
]
(0,-9)  -- (0,0);

\node at (0,-9.14)  {$\times$};

\end{tikzpicture}

\end{center}

\caption{This diagram again shows the  walls of the Weyl chambers of the affine $A_{1}$ Lie algebra, but with the wall of marginal stability for the vanilla BPS states in red. This time we show the intersection of the BPS walls at the point $w_{1}$ on the wall of marginal stability. Here the existing BPS walls for Seiberg-Witten theory are shown in dark blue. The dashed blue lines are those excluded by the existence condition of vanishing central charges at regular points represented by the crosses on the ends of the line.}

\label{fig:SWwalls}

\end{figure}

The wall crossing on both sides of $MS_{\alpha_{0}, \alpha_{1}}$ is depicted in Fig. \ref{fig:SWwalls}. The crossing of $MS_{\alpha_{0}, \alpha_{1}}$ can clearly be seen by looking at just the 2 outer BPS walls. These can be used to construct a cone with upper and lower region. The upper region contains the full spectrum and the lower just the 2 basis states. The parameter $\zeta$ can be varied such that the cone can be taken at any point along the wall. On Fig. \ref{fig:SWwalls} we directly show the intersection of the BPS walls $W_{\alpha_{i}}$ with the wall of marginal stability $MS_{\alpha_{0}, \alpha_{1}}$. We choose a parameter $\zeta_{1}=1$. One can change the parameter from $\zeta_{1}$ to $\zeta_{2}$. This changes the intersection point of the BPS walls with the wall of marginal stability, and gives a different cone, with the same wall crossing behaviour. Such cones can be used to sweep all regions inside and outside the wall of marginal stability. Indeed we have actually reproduced the scattering diagram for Seiberg-Witten theory developed by Bridgeland  \cite{BridgelandOriginalscatteringdiagram,Bousseau:2022snmnewattractorflowscatteringdiagram}.

This therefore reproduces the spectrum $\mathcal{S}_w$ of Seiberg-Witten theory (see table (\ref{tab:finalswspectrumweylchambers})) with 2 basis states on one side of the wall and infinitely many on the other. 

\begin{table}[h!]
    
\begin{center}
	\begin{tabular}{ |l|l|}
		\hline
		\multicolumn{2}{|c|}{BPS states in chambers} \\[5pt]
		\hline
		
	         Chamber 1 
		& \vphantom{\Huge{H}} $\pm(-\alpha_{0}+\delta), \pm \alpha_{0}$\\[5pt]
		\hline
		Chamber 2
		& \vphantom{\Huge{H}} $\pm\alpha_{0}+n\delta, \pm \delta$\\[5pt]
		\hline

\hline

\end{tabular}
\end{center}

    \caption{Final existing BPS states in Seiberg-Witten theory on both sides of the wall of marginal stability represented by the red line in Fig. \ref{fig:SWwalls}.}
    \label{tab:finalswspectrumweylchambers}
\end{table}

\newpage

\section{Argyres-Douglas $A_{2}$} \label{A2subsection}

In this case we look at Argyres-Douglas theory represented by a 2 node quiver with 1 arrow $\bullet \longrightarrow \bullet$. This theory has BPS states described by the root system of the SU(3) Lie algebra \cite{Argyres_1995}. As with Seiberg-Witten theory, this theory is parameterised by a complex 1d moduli space of an elliptic curve consisting of a single parameter $w \in \mathbbm{P}^1\setminus  \{-1,1,\infty \}$. The number of BPS states that exist in the theory depends on the region of the moduli space. It has been found that in one chamber this theory has 2 basis BPS states given by electric and magnetic monopoles $\alpha_{1}$ and $\alpha_{2}$. In the other chamber 3 BPS states exist including the 2 monopoles $\alpha_{1}, \alpha_{2}$ and a dyon $\alpha_{3} = \alpha_{1}+ \alpha_{2}$. 

\subsection{Introduction to counting function for Argyres-Douglas $A_{2} \  \  \ $ theory}

One can conjecture that a similar generating function exists for other ADE Lie Algebras that are not affine. This is part of a general relation between BPS quivers and these counting functions. Here we look at the simplest ADE example with a wall of marginal stability in addition to BPS walls. This is the Argyres-Douglas $A_{2}$ theory \cite{Argyres_1995}. Here we find a similar generating function and wall crossing behaviour for this Argyres-Douglas $A_{2}$ theory  following the methods detailed in \cite{Cheng_2008}. In this case we again look for a match between the boundaries of the Weyl chambers and the walls. If such a match is found one can use an analogous formulation to that in \cite{Cheng_2008} to count the existence of certain BPS states by looking at how many additional roots are added to the highest weight in each Weyl chamber. In this case only the Weyl denominator is needed to determine the change in the highest weight of the Verma module. However, one can also show that the full character also transforms in such a way under wall crossing. This means it remains invariant (after change of basis of roots) up to an additional root being added or subtracted to the highest weight every time a boundary of a Weyl chamber is crossed. When in the full character the highest weight is of a representation.  

\subsubsection{$A_{2}$ Weyl character}

 The Weyl character formula can be generalised from the characters of Borcherds-Kac-Moody algebras to those for ADE type Lie algebras. In this way one can compare the prescription for highest weights in Bocherds-Kac-Moody algebras to those in models with the $A_{1}, A_{2}$ and $\hat{A}_{1}$ root systems. Again, following from sec. \ref{subsectionADEWeyldenominator} for these roots systems the analog counting function is written as the Weyl denominator that is simple to expand as a geometric series. This being the denominator of the character (recall eq. \ref{eq:firstfullcharacter}) \footnote{A good overview of these Weyl character formulae is given in \cite{WeylCharacterWalton_2013}.}
 \begin{equation} \label{eq:secondfullcharacter}
ch_{\lambda}(u) = \frac{\sum_{\mathbf{w} \in W} (\det{\mathbf{w}}) e^{\mathbf{w}(\lambda+\rho)(u)}}{e^{\rho(u)}\prod_{\alpha \in \Delta^{+}}(1-e^{-\alpha(u)})},
\end{equation}
where the vector $\rho = \frac{1}{2} \sum_{\alpha_{i} \in \Delta^{+}} \alpha_{i}$ is again the Weyl vector - written as the half sum of the positive roots. As mentioned before in section \ref{subsectionADEWeyldenominator} $\mathbf{w}$ are the Weyl group elements and one chooses a representation to start with that has a highest weight  $\lambda$.

The number and charge of the roots representing the BPS states should be distinct for every possible Weyl chamber. As with Seiberg-Witten theory in \ref{seibergwittenframedwalls} such a match again exists for the framed BPS states from \cite{FramedBPSGMNhttps://doi.org/10.48550/arxiv.1006.0146, BPSgalaxiesAndriyash_2012}. In \cite{Cheng_2008} the generating function corresponds to the Weyl denominator of the Borcherds-Kac-Moody algebra. In this case we just look at the simple case of the $SU(3)$ Weyl denominator. One can also look at the full character of this Lie algebra. This is the Weyl character for the $SU(3)$ Lie Algebra. One can choose a  2 dimensional moduli parameter for example $u=(u_{1},u_{2})$ to define in which chamber one is in. This can then be substituted into the character formula.

\begin{dfn}

The character itself reads:
\begin{align}
& ch_{\lambda}(u) = \\ \nonumber
& \scriptsize\text{$	
	-\frac{e^{\lambda(u) - \frac{2}{(\alpha_{1},\alpha_{1})}((\lambda + \frac{1}{2}(\alpha_{1}+ \alpha_{2}+ \alpha_{3})),\alpha_{1})\alpha_{1}(u)}+e^{\lambda(u) - \frac{2}{(\alpha_{2},\alpha_{2})}((\lambda + \frac{1}{2}(\alpha_{1}+ \alpha_{2}+ \alpha_{3})),\alpha_{2})\alpha_{2}(u)}
		+e^{\lambda(u) - \frac{2}{(\alpha_{3},\alpha_{3})}((\lambda + \frac{1}{2}(\alpha_{1}+ \alpha_{2}+ \alpha_{3})),\alpha_{3})(\alpha_{3}(u))}}{(1-e^{-\alpha_{1}(u)})(1-e^{-\alpha_{2}(u)}(1-e^{-\alpha_{3}(u)})}
	$}.
\end{align}	
The highest weight of the representation is denoted as  $\lambda$ and the roots system of the algebra contains $\alpha_{1}, \  \alpha_{2}, \  \alpha_{3} = \alpha_{1}+ \alpha_{2}$. 

\end{dfn}

We know from the quiver representation theory \cite{N=2BPSquivershttps://doi.org/10.48550/arxiv.1112.3984} that these roots correspond to BPS states and that we should observe at least 2 chambers with 2 and 3 BPS states respectively. These walls should however not match the Weyl chambers directly as they (as in the Seiberg-Witten example in sec. \ref{SeibergWittenexamplesubsection}) correspond to the BPS walls  $W_{\alpha_{i}}$ for the framed halo BPS states \cite{FramedBPSGMNhttps://doi.org/10.48550/arxiv.1006.0146} which are different from the wall of marginal stability $MS_{\alpha_{1},\alpha_{2}}$. Here we will first look at the wall crossing for the BPS walls $W_{\alpha_{i}}$ in which there is one chamber with all the BPS states existing and another (which we can label the fundamental Weyl chamber) with none of the BPS states existing. 

Therefore, as with the $\mathcal{N} =4$ black hole example in \cite{Cheng_2008} and the analogs of the $\hat{A}_{1}$ Lie algebra, we again find wall crossing behavior with the BPS walls corresponding to the boundaries of Weyl chambers $\text{Im}[i\alpha_{i}(u)]=0$. Furthermore, this again manifests itself either as a shift in the highest weight or in the Fourier expansion coefficients jumping between distinct Weyl chambers. For example in the transition from the chamber  $\text{Im}[i\alpha_{i}(u)] > 0 \ \ \ \forall{i \in {1,2,3}}$ to $\text{Im}[i\alpha_{1}(u)] < 0, \ \ \text{Im}[i\alpha_{2}(u)] > 0, \ \ \text{Im}[i\alpha_{3}(u)] > 0$, we determined that the highest weight changes such that $ch_{\lambda_{2}} = ch_{\lambda_{1}-\alpha'_{2}}$ after a suitable basis transformation. In the following section \ref{sec:WeyldenominatorA2} and in the appendix \ref{Weyl denominator expansions apendix} we calculate this change for all the remaining chambers just from the denominator.

\subsection{Wall crossing and change of basis for denominator} \label{sec:WeyldenominatorA2}

As with the affine Lie algebra we look at the denominator and try to expand it in the different Weyl chambers. This alone should encode the wall crossing for the BPS walls in the same way that the Verma modules do in the $\hat{A}_{1}$ case. The Weyl denominator of a general Lie algebra, introduced in (\ref{eq:originalweyldenominatordefinition}), is independent of the representation chosen and is what can be expanded in the different Weyl chambers 
\begin{equation} \label{eq:weyldenominator2}
\frac{e^{\Lambda(u)}}{e^{\rho(u)}\prod_{\alpha \in \Delta^{+}}(1-e^{-\alpha(u)})} = \frac{e^{\lambda(u)}}{\prod_{\alpha \in \Delta^{+}}(1-e^{-\alpha(u)})},
\end{equation}
where as with the case of degeneracies in $\mathcal{N}=4$ examples we multiply the denominator by an additional charge factor $e^{\Lambda(u)}$. The product is over all positive roots and the exponentials must be negative to allow a convergent geometric series expansion \footnote{
This formula can be expanded in this form \cite{WeylCharacterWalton_2013} and is known as the Kostant partition function.}
\begin{equation}
\frac{1}{\prod_{\alpha \in \Delta^{+}}(1-e^{-\alpha(u)})} = \sum_{\beta \in \mathbb{N} \Delta^{+} } K(\beta) e^{-\beta(u)},
\end{equation}
where $K(\beta)$ is the count of all the possible combinations in which $\beta$ can be written as a linear combination of positive roots \cite{WeylCharacterWalton_2013}. Now we look at the specific case of (\ref{eq:weyldenominator2}) for the $A_{2}$ Weyl denominator
\begin{align} \label{eq:a2weyldenominator}
\frac{e^{\lambda(u)}}{(1-e^{-\alpha_{1}(u)})(1-e^{-\alpha_{2}(u)})(1-e^{-\alpha_{3}(u)})},
\end{align}	
where $\alpha_{3}(u)=\alpha_{1}(u)+\alpha_{2}(u)$. 

\subsubsection{Weyl denominator expansion in different chambers}\label{sec:expansiondifferentchambers}

Now for each Weyl chamber in which we do the expansion we will change the basis of positive roots in such a way that the exponents in the expansion remain negative for all $n, m, l$, where $n,m$ and $l$ are the term numbers of this expansion. As discussed in \cite{N=2BPSquivershttps://doi.org/10.48550/arxiv.1112.3984} this corresponds to taking successive Weyl reflections and can also can be thought of as quiver mutations or changes of basis. Physically this means choosing a basis of particles. Evaluated at a particular modulus outside the fundamental Weyl chamber (after BPS walls have been crossed) one must therefore take Weyl reflections of the form 
\begin{equation} \label{weylreflectioneq}
\mathbf{w}_{\alpha_{i}}(\alpha_{j}) = \alpha_{j} - \frac{2}{(\alpha_{i},\alpha_{i})}(\alpha_{j},\alpha_{i})\alpha_{i}.
\end{equation}

This transformation is done to distinguish whether the expansions in the different Weyl chambers just correspond to a different basis or quiver mutation (see \cite{N=2BPSquivershttps://doi.org/10.48550/arxiv.1112.3984}) by successive Weyl reflection, or if they  correspond to walls in the moduli space and actually distinguish whether there are different numbers of BPS states present in the model. \footnote{In addition to the quiver mutation in the form of a Weyl reflection we have also exchanged $\alpha'_{1}$ and $\alpha'_{2}$ to better distinguish the roots before and after the transformation.} 

\begin{align}
 \text{Im}[i\alpha_{i}(u)] > 0 \ \ \ \forall{i \in {1,2,3}}\\ \nonumber
 \sum^{ \infty }_{n,m,n = 0} e^{-(n\alpha_{1}(u)+m\alpha_{2}(u)+l\alpha_{3}(u))} \\ \nonumber \text{the exponents are} \  \  
-n\alpha_{1}(u)-m\alpha_{2}(u)-l\alpha_{3}(u),
\end{align}

The expansion in the fundamental Weyl chamber is already in the required form with negative exponents so there is no change of basis required.
\begin{align}
\ \\ \nonumber
 \text{Im}[i\alpha_{1}(u)] < 0, \  \text{Im}[i\alpha_{2}(u)] > 0, \ \text{Im}[i\alpha_{3}(u)] > 0 \\ \nonumber
- \sum^{ \infty }_{n,m,l = 0} e^{(n+1)\alpha_{1}(u)-m\alpha_{2}(u)-l\alpha_{3}(u)},\\ \nonumber
-(n+1)(-\alpha_{1}(u))-m\alpha_{2}(u)-l(\alpha_{1}+\alpha_{2})(u), \\ \nonumber
\text{change basis} \ \  \alpha'_{1} = \alpha_{1}+\alpha_{2} = \alpha_{3}, \  \alpha'_{2} = -\alpha_{1}, \  \alpha'_{3} = \alpha_{2},\\ \nonumber
\text{expression becomes} -((n+1)\alpha'_{2}(u)+m\alpha'_{3}(u)+l\alpha'_{1}(u)).
\end{align}
In this Weyl chamber the exponent changes sign. We therefore perform a change of basis to restore the negative exponents. This results in a shift in the highest weight by a single root.

So now the denominator in the new basis reads
\begin{align} \label{jumpequation}
-\frac{e^{(\lambda'-\alpha'_{2})(u)}}{(1-e^{-\alpha'_{1}(u)})(1-e^{-\alpha'_{2}(u)})(1-e^{-\alpha'_{3}(u)})},
\end{align}	
such that the module jumps to one with highest weight $\lambda'-\alpha'_{2}$ relative to that in (\ref{eq:a2weyldenominator}).
We have continued the computation for the expansion of the Weyl denominator in the remaining Weyl chambers. All the changes in the possible expansions are listed in appendix \ref{Weyl denominator expansions apendix}. For the remaining expansions it was always possible to write the expansion in terms of a set of negative exponents. When this was done we always obtained a shift in one of the coefficients.

Therefore it is always possible to change the basis in such a way that all the exponents in the expansion are again negative in $n,m$ and $l$ for all $n, m, l$. These basis changes also correspond to rotations of the positive plane. We have not found any obstructions to this. So the expansions are distinguished after changing the basis by which and how many roots are shifted by one in the exponent and the negative pre-factor in front of the sum. This means that this shift does in fact represent a jump in the number of existing BPS states and is not just a basis change. Physically this should have an interpretation of framed halo BPS states binding to a line operator or core charge, or in terms of quivers as a mutation of a framed quiver. 

\subsubsection{Inclusion of modules and structure of chambers}

In this way 6 chambers are found with different highest weight modules $M(\lambda_{n,m})$ which can be defined in the same way as for the $\hat{A}_{1}$ root system in (\ref{affineVermamodules}).
One can again write this in terms of inclusions if we write $\lambda_{n,m} = \lambda+ n\alpha_{1}+m\alpha_{2}$. \footnote{Here if we follow on from (\ref{jumpequation}) we must switch convention $\alpha'_{i}\rightarrow \alpha_{i}$ to have the chambers in order of increasing highest weight.} In one direction we have
\begin{align}
M(\lambda_{0,0}) \subset M(\lambda_{1,0}) \subset M(\lambda_{2,1}) \subset M(\lambda_{2,2}),
\end{align}	
and in the other
\begin{align}
M(\lambda_{0,0}) \subset M(\lambda_{0,1}) \subset M(\lambda_{1,2}) \subset M(\lambda_{2,2}).
\end{align}
This means that there is indeed a non-trivial jump in the number of BPS states when the series is expanded in different Weyl chambers. This can be seen by distinct expansion coefficients and can be encoded in the distinct highest weight module.

Therefore, we have found 6 chambers with a different combination of roots (representing the BPS charges) existing in each chamber. This should then correspond to 1 chamber with no roots existing another with just $\alpha_{1}$ existing but with no other roots, another with $\alpha_{2}$ and no other roots, another with $\alpha_{2}, \alpha_{3}$, one with $\alpha_{1}, \alpha_{3}$ and finally one with all 3 roots existing as $\alpha_{1},\alpha_{2}, \alpha_{3}$. This corresponds to the wall crossing for the framed BPS states. This can be shown in a table.

\begin{table}[h!]
    
  \begin{center}

\begin{tabular}{ |l|l|}
	\hline
	\multicolumn{2}{|c|}{\vphantom{\Huge{H}} $\ \ $ \centering All possible highest weights $\ \ $}
	\\[7pt]
	\hline
	\multicolumn{2}{|c|}{\centering \vphantom{\Huge{H}} $\  \  \  \lambda + \alpha_{1}+\alpha_{2}+ \alpha_{3} \  \  \ $}  
	\\[8pt]
	\hline
	 \vphantom{\Huge{H}}	$\lambda + \alpha_{1}+\alpha_{3}  \ \  \  $& \vphantom{\Huge{H}} $\lambda + \alpha_{2}+\alpha_{3}$ \\[7pt]
		
	\hline
\vphantom{\Huge{H}} $\lambda +\alpha_{1}$ & $\lambda + \alpha_{2} $  \\[7pt]	
	\hline
 \multicolumn{2}{|c|}{\vphantom{\Huge{H}} $\lambda$} \\[7pt]
	\hline
\end{tabular}

\end{center}

    \caption{Highest weight states representing framed BPS boundstates in 6 Weyl chambers.}
    \label{tab:A2statesinweylchambers}
\end{table}

\subsubsection{Central charge representation of the roots}

If one wants to connect this wall crossing to BPS walls for framed halo states one must again write the complex inner products of the roots in terms of central charges. These walls can then be mapped into the moduli space to see the wall crossing at the wall of marginal stability $MS_{\alpha_{1},\alpha_{2}}$ in addition to that at the BPS walls $W_{\alpha_{i}}$. 

\begin{ex}

The roots can thus be written as follows:
\begin{align}
\alpha_{1}(u) = \frac{Z_{\alpha_{1}}(w)}{\mu}, \ \  
\alpha_{2}(u) =  \frac{Z_{\alpha_{2}}(w)}{\mu}.
\end{align}
Here, as with the Seiberg-Witten example, we can let $\mu = \epsilon \zeta \in \mathbb{C}$ where $\epsilon \in \mathbb{R}$ is a small parameter defining the contour and $\zeta:= e^{i \theta}, \ \theta \in [0,2 \pi ]$ is the phase that is chosen. 

\end{ex}

The BPS walls can be written as the vanishing locus of the imaginary parts of the central charges which read:\footnote{Here we again let $-i \mu \rightarrow \mu$.}
\begin{align*}
& \text{Im}[i\alpha_{1}(u)] = \text{Im}\left[\frac{Z_{\alpha_{1}}(w)}{\mu}\right], \ \  
\text{Im}[i\alpha_{2}(u)] =  \text{Im}\left[\frac{Z_{\alpha_{2}}(w)}{\mu}\right], \ \ \\ \nonumber \nonumber \\ & \text{Im}[i\alpha_{1}(u)+i\alpha_{2}(u)] =  \text{Im}\left[\frac{Z_{\alpha_{1}}(w)+Z_{\alpha_{2}}(w)}{\mu}\right].
\end{align*}

\subsubsection{Crossing the wall of marginal stability $MS_{\alpha_{1},\alpha_{2}}$ }

At the wall of marginal stability $MS_{\alpha_{1},\alpha_{2}}$  , the central charges align (as is the case for Seiberg-Witten theory in section \ref{SeibergWittenexamplesubsection})
such that their ratio is some real function $r_{i}(w) \in \mathbb{R}$ along the wall. This means that on the wall of marginal stability 
$MS_{\alpha_{1},\alpha_{2}}$
one can write the sum of the central charges as 
\begin{align} \label{rewrittencomposite}
&\frac{Z_{\alpha_{1}}(w)+Z_{\alpha_{2}}(w)}{\mu} = \frac{Z_{\alpha_{1}}(w)}{\mu} \Big(1+\frac{Z_{\alpha_{2}}(w)}{Z_{\alpha_{1}}(w)} \Big),  \\ \nonumber
\\ \nonumber
&\text{and on the wall of marginal stability we have}
\\ \nonumber
\\ \nonumber
&r_{3}(w)\frac{Z_{\alpha_{1}}(w)}{\mu},
\\ \nonumber
\\ \nonumber
&\text{where $r_{3}(w) = 1+\frac{Z_{\alpha_{2}}(w)}{Z_{\alpha_{1}}(w)} \in \mathbb{R}$.}
\end{align}
Now as in the example of Seiberg-Witten theory we follow the attractor flow existence conditions from \cite{DenefSeibergWittenemptyhole_2000,DenefGreeneRaugas_2001,Bousseau:2022snmnewattractorflowscatteringdiagram}. We again choose a continuation of the generating function across the wall of marginal stability that avoids a pole at a regular point. The 2 basis charges only have allowed poles at singular points so we must only change the continuation for the composite state (\ref{rewrittencomposite}) above. For example a possible continuation one can choose is:
\begin{align}
r_{3}(w) = 1+ \Bigg| \frac{Z_{\alpha_{2}}(w)}{Z_{\alpha_{1}}(w)} \Bigg| \in \mathbb{R}.
\end{align}
\begin{ex}[$\bold{Exclusion \ of \ composite \ state}$]

The function $r_{3}(w) Z_{\alpha_{1}}(w)/\mu \neq \alpha_{1}(u)+\alpha_{2}(u)$. It is no longer a positive root in the $A_{2}$ Lie algebra $r_{3}(w)\frac{Z_{\alpha_{1}}(w)}{\mu} \notin  \Delta^{+}$. Therefore, it does not exist as a BPS state. To illustrate this further one can explicitly write the generating function (\ref{eq:a2weyldenominator}) in terms of central charges on both sides of the wall.

\end{ex}
\begin{align}
\text{One one side}: \ \ &\frac{e^{\lambda(u)}}{(1-e^{-\frac{Z_{\alpha_{1}}(w)}{\mu}})(1-e^{-\frac{Z_{\alpha_{2}}(w)}{\mu}})(1-e^{-\frac{Z_{\alpha_{1}}(w)+Z_{\alpha_{2}}(w)}{\mu}})}. \\ \nonumber
\\ \nonumber
\text{On the other}: \ \ &\frac{e^{\lambda(u)}}{(1-e^{-\frac{Z_{\alpha_{1}}(w)}{\mu}})(1-e^{-\frac{Z_{\alpha_{2}}(w)}{\mu}})(1-e^{-r_{3}(w)\frac{Z_{\alpha_{1}}(w)}{\mu}})} = \\ \nonumber
\\ \nonumber
& \frac{e^{\lambda(u)}}{(1-e^{-\frac{Z_{\alpha_{1}}(w)}{\mu}})(1-e^{-\frac{Z_{\alpha_{2}}(w)}{\mu}})} f\Big(r_{3}(w)Z_{\alpha_{1}}(w)\Big),
\end{align}

where the function $f\big(r_{3}(w)Z_{\alpha_{1}}(w)\big)$ representing the non-existing BPS state can be treated just as a normalisation factor. What is left for the 2 existing basis BPS states is then just a product over 2 factor representing the 2 vanilla BPS states still existing in the spectrum. In this case the 2 remaining BPS states can also be tabulated in terms of framed wall crossing.
\begin{table}[h!]
    
\begin{center}

\begin{tabular}{ |l|}
	\hline
	\vphantom{\Huge{H}}	
 Highest weight on one side of $MS_{\alpha_{1},\alpha_{2}}$
	\\[7pt]
	\hline
	 \vphantom{\Huge{H}}	
  $\lambda + \alpha_{1}+\alpha_{2} $
	\\[7pt]
	\hline
	 \vphantom{\Huge{H}}	
  $\lambda +\alpha_{2}$ 
 \\[7pt]	
	\hline
	\vphantom{\Huge{H}}	
 $\lambda$ \\ [7pt]
	\hline
\end{tabular}

\end{center}

    \caption{Highest weights representing framed BPS states in all the chambers: this is on the side of the wall of marginal stability where only the 2 basis BPS states in $A_{2}$ exist.}
    \label{tab:A2allframed2basis}
\end{table}
We can now tabulate all the wall crossing for both the BPS walls $W_{\alpha_{i}}$ and the walls of marginal stability $MS_{\alpha_{1},\alpha_{2}}$ together.
\begin{table}[h!]
    
\begin{center}

\begin{tabular}{ |l|l|}
	\hline
	 \vphantom{\Huge{H}}  
 On side with 2  &  On side with 3 
	 \\[7pt]
	\hline
	\vphantom{\Huge{H}}	
 $\lambda + \alpha_{1}+\alpha_{2}$ & $\lambda + \alpha_{1}+\alpha_{2}+ \alpha_{3}$  \\[7pt]	
	\hline
	\vphantom{\Huge{H}}	
 $\lambda + \alpha_{2} $& $\lambda + \alpha_{1}+\alpha_{3}$ \\ [7pt]
	\hline 
\vphantom{\Huge{H}}	$\lambda +\alpha_{2}$ & $\lambda + \alpha_{1} $ \\[7pt]	
	\hline
	\multicolumn{2}{|c|}{\vphantom{\Huge{H}} $\lambda \  \  \  $} \\ [7pt]
	\hline
\end{tabular}

\end{center}

    \caption{This is a complete tabulation for framed BPS states in all chambers present on both sides of the wall of marginal stability for primitive wall crossing in the $A_{2}$ example. See Fig. \ref{fig:A2allwallcrossing}.}
    \label{tab:completetabulationA2}
\end{table}
This is plotted on Fig. \ref{fig:A2allwallcrossing} below.
\begin{center}
\begin{figure}[h!]

\begin{tikzpicture}

\fill[lime] (0,-0.123)
-- (-6.51,-0.123) 
-- (-3.83,5.87) 
-- cycle;

\node at (-4.3,6.3) {$W_{\alpha_{2}}: \text{Im}[Z_{\alpha_{2}}(w)] = 0$};
\node at (4.5, 6.3) {$ \   \  W_{\alpha_{1}+\alpha_{2}}:  \text{Im}[ Z_{\alpha_{2}}(w)+Z_{\alpha_{1}}(w)] = 0$};

\node at (7.0, 0.2) {$W_{\alpha_{1}}: \text{Im}[Z_{\alpha_{1}}(w)] = 0$};

\node at (-0.5, 4) {$\alpha_{3}$};
\node at (0.5, 4) {$\alpha_{1}$};

\node at (-4.5, 2.6) {$\alpha_{1}$};
\node at (-2.5, 2.6) {$\alpha_{3}$};
\node at (-3.5, 2.6) {$\alpha_{2}$};

\node at (0, -3) {$\alpha_{2}$};

\node at (-5, 1) {$\alpha_{1}$};
\node at (-4, 1) {$\alpha_{2}$};

\node at (-4, -2) {$\alpha_{2}$};

\node at (3.6, 2) {$\alpha_{1}$};

\node at (1.0, -0.4) {$w_{1}$};

\node at (6, 5) {$W: \text{BPS wall}$};
\node at (7, 4.2) {$MS: \text{Wall of marginal stability}$};

\node at (6.5, -4.5) {$MS_{\alpha_{1},\alpha_{2}}: \text{Im}[Z_{\alpha_{1}}(w) \bar{Z}_{\alpha_{2}}(w)] = 0$};

\draw[thick,red] plot [smooth] coordinates{(-7,3) (0,-0.15) (7,-4)};

\draw[thick,blue,yshift=-4pt,
]
(0,0) -- (3.8,6) ;

\draw[dashed,blue,yshift=-4pt,
]
(-3.8,-6) -- (0,0) ;

\node at (-3.8, -6.14) {$\times$};

\draw[thick,blue,yshift=-4pt,
]
(3.8,-6)  -- (-3.8,6);

\draw[thick,blue,yshift=-4pt,
]
(-6.5,0)  -- (6.5,0);

(0,-7) arc (-90:90:7) -- (0,7);

\end{tikzpicture}

	\caption{The BPS walls are shown by the lines in blue. The wall of marginal stability is shown by the line in red. The dashed blue line is the BPS wall excluded by the existence condition of vanishing central charge at a regular point.}
\label{fig:A2allwallcrossing}	
	
\end{figure}
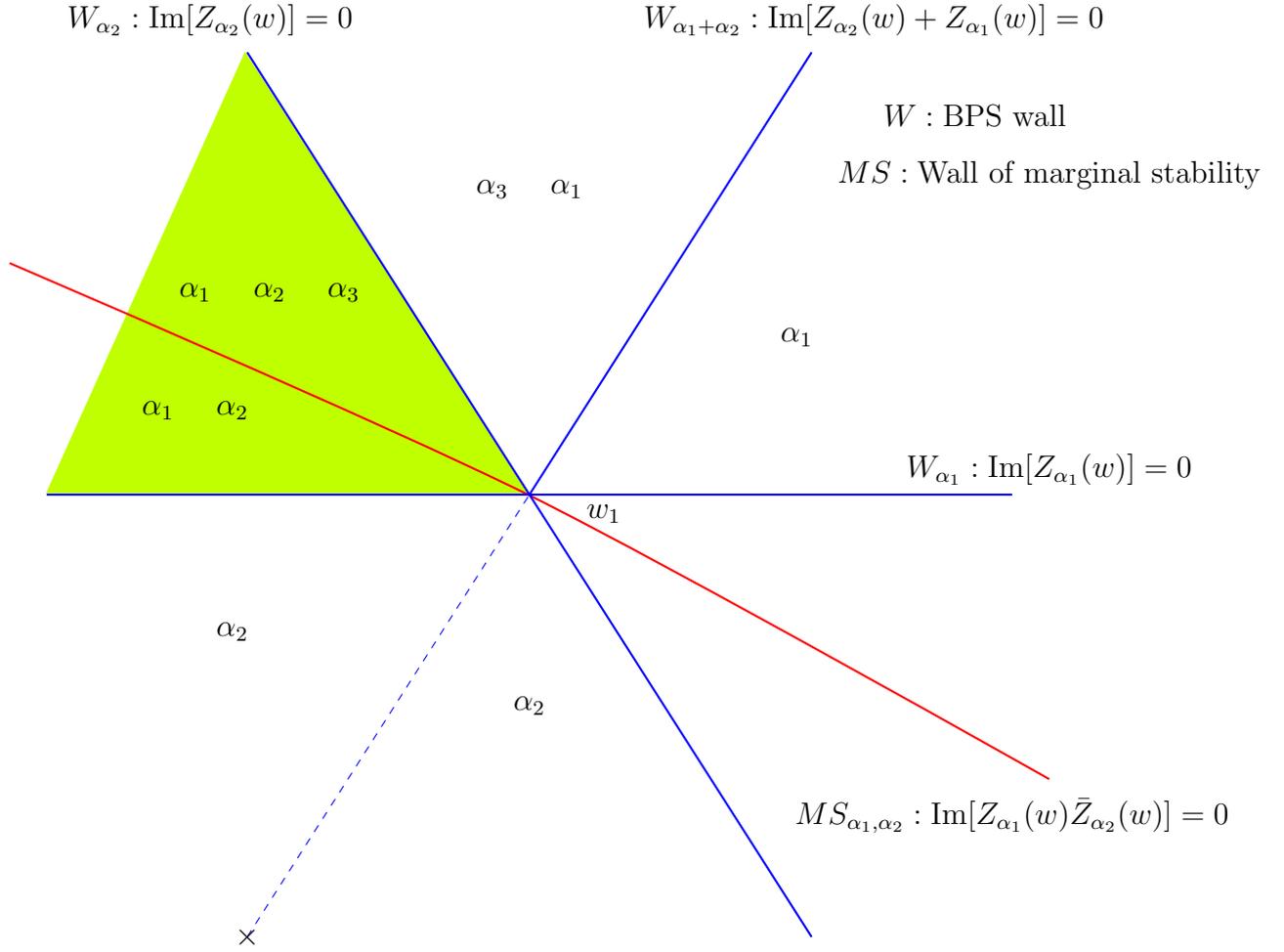
\end{center}

We can now see from Fig. \ref{fig:A2allwallcrossing} (where we have again chosen $\zeta =1$) that the wall crossing for the regular BPS states for the Argyres-Douglas $A_{2}$ BPS structure just has 2 regions - that is one region containing all 3 BPS states and their antiparticles $\pm \alpha_{1},\pm \alpha_{2}, \pm \alpha_{3}$ and another region with just the 2 basis states $\pm \alpha_{1},\pm \alpha_{2}$.  In this case the wall of marginal stability for the regular (vanilla) BPS states is again the usual locus in the moduli space at which the ratio of the central charges is real. We have also again recovered the scattering diagram for the $A_{2}$ quiver from \cite{BridgelandOriginalscatteringdiagram,Bousseau:2022snmnewattractorflowscatteringdiagram}. In this case we have the following final count of the roots in the different chambers.

\subsubsection{Interpretation in terms of attractor flow lines} \label{subsec:A2attractorflow}

\begin{figure}[h!]
	\centering
	{\includegraphics[width=0.9\textwidth]{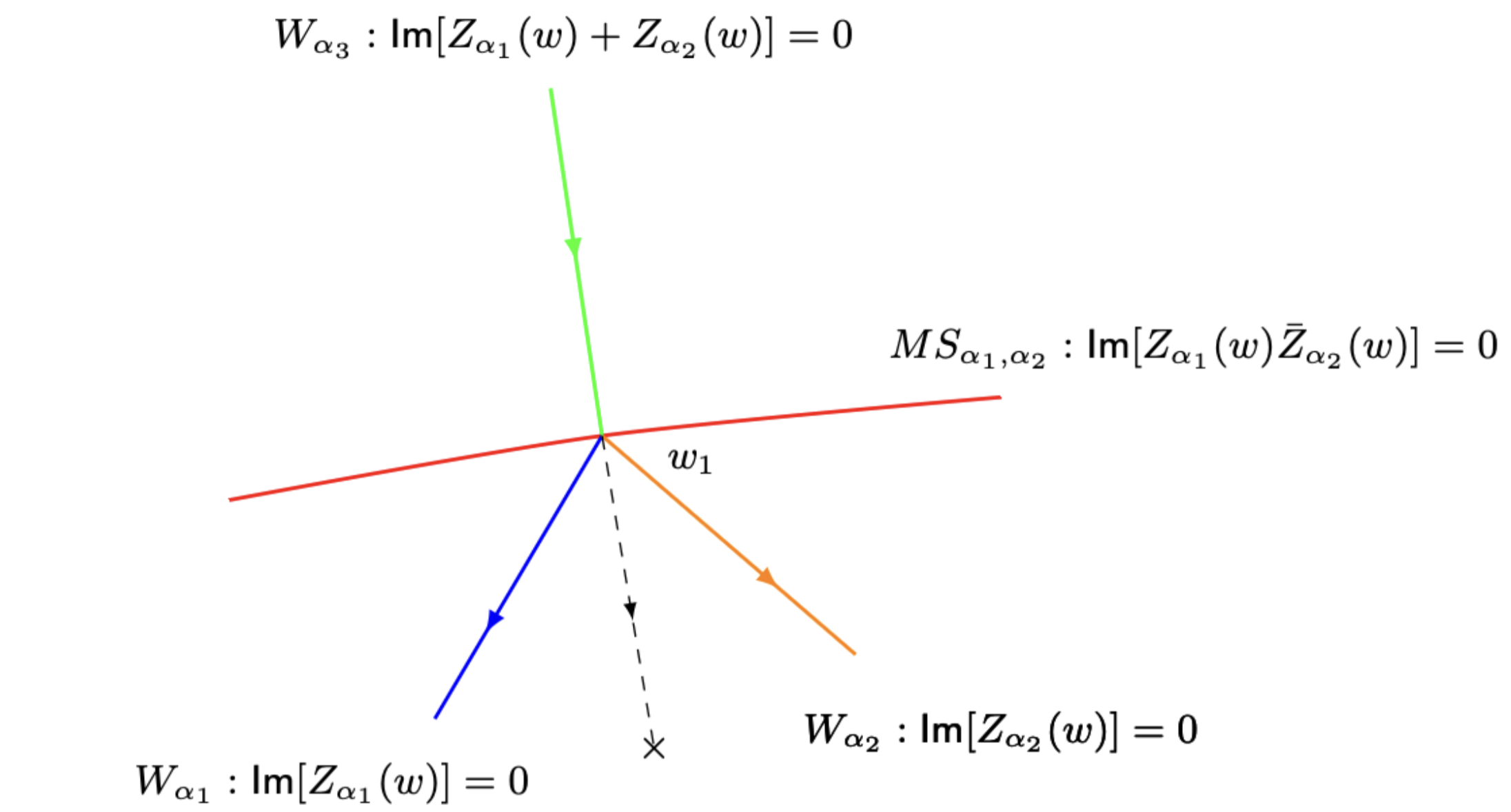}}
	\caption{Split attractor flow for the Argyres-Douglas $A_{2}$ model.}
	\label{A2attractorlinechamber1}
\end{figure}

This picture of the wall crossing in terms of both the BPS walls and the walls of marginal stability can be shown on a split attractor flow diagram. These have recently been derived for the $A_{2}$ model in \cite{Alim:2023doi}. As the attractor flow lines for a central charge are lines of constant phase (see e.g.
\cite{Bousseau:2022snmnewattractorflowscatteringdiagram}) our BPS walls can be mapped onto these flow lines. The excluded BPS wall corresponds to the flow line excluded by its termination/end point where the central charge vanishes at a regular point in the moduli space. Here we show the split flow line of the composite BPS state (shown in green), with root $\alpha_{1}+\alpha_{2}$, splitting into its constituents with roots $\alpha_{1}$ and $\alpha_{2}$, on Fig. \ref{A2attractorlinechamber1}, represented by the blue and orange lines respectively. The grey line represents the flow line of the non-existing composite on the other side of the wall of marginal stability $MS_{\alpha_{1},\alpha_{2}}$.

\begin{figure}[h!]
	\centering
	{\includegraphics[width=1.1\textwidth]{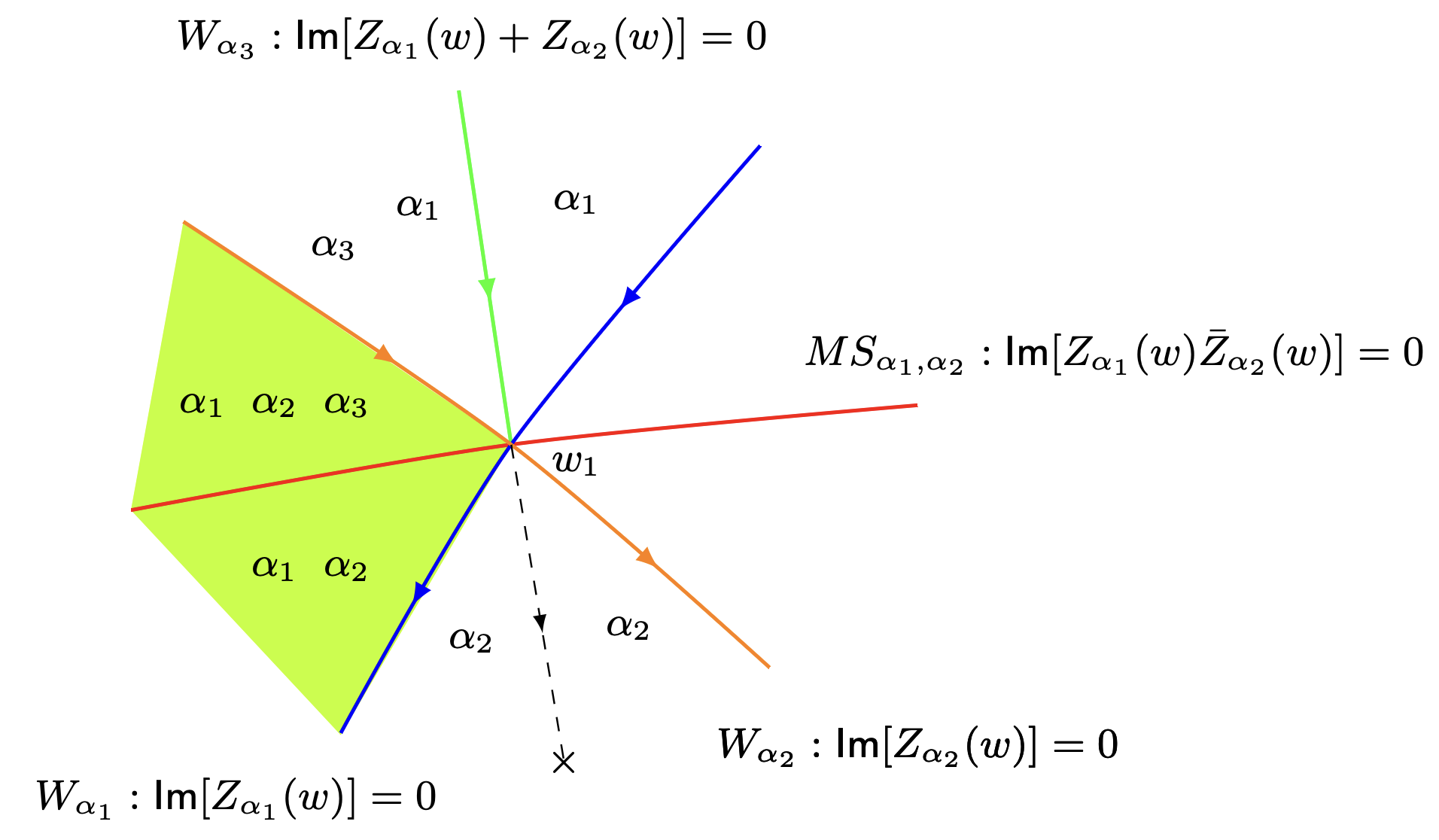}}
	\caption{We zoom in and match the chambers bounded by the flow lines with the combination of roots that are counted in the highest weight.}
 \label{A2attractorlinechamber2}
\end{figure}

On Fig. \ref{A2attractorlinechamber2} we continue the flow of the constituent (basis) BPS states back across $MS_{\alpha_{1},\alpha_{2}}$ into the region where the composite still exists. Then we zoom in and match all the existing flow lines of all the BPS states with the BPS walls. Then we can again see which roots (counted in the highest weight) exist in which chambers as various BPS walls are crossed. As before the regions shaded in green represent those regions where the count of the framed halo states in the highest weight of the generating function matches the total number of vanilla BPS states in the theory and we once again see the jump between 2 and 3 roots depending on which side of the wall of marginal stability $MS_{\alpha_{1},\alpha_{2}}$ we perform the count.

\begin{table}[h!]
    
 \begin{center}
	\begin{tabular}{ |l|l|}
		\hline
		\multicolumn{2}{|c|}{ BPS states in chambers} \\
		\hline
        \vphantom{\Huge{H}}
         Chamber 1 
		& $\pm\alpha_{1}, \pm \alpha_{2}, \pm (\alpha_{1}+\alpha_{2})$\\[7pt]		\hline  \vphantom{\Huge{H}}		
  Chamber 2
		&  $\pm\alpha_{1}, \pm \alpha_{2}$\\[7pt]
		\hline

\hline

\end{tabular}
\end{center}

    \caption{Final existing BPS states in $A_{2}$ Argyres-Douglas on both sides of wall of marginal stability. That is on both sides of red line in Fig. \ref{fig:A2allwallcrossing}.}
    \label{tab:finalA2existingstatesmarginalstability}
\end{table}

\begin{align*}
\\
\\
\end{align*}

\section{Conclusions}

We have developed a new type of generating function describing the BPS states in special 4d $\mathcal{N}=2$ theories using a Lie algebraic formulation. The main conjecture is that this takes the form of the Weyl denominator for ADE type and affine Lie algebras constructed from ADE type quivers. The wall crossing for the allowed set of framed halo BPS states \cite{FramedBPSGMNhttps://doi.org/10.48550/arxiv.1006.0146} is encoded in the Weyl chambers with the boundaries corresponding to the BPS walls such that the highest weight of a Verma module jumps when a BPS wall is crossed. Therefore the total number of vanilla BPS states in the theory corresponds to the maximal highest weight on one side of the wall of marginal stability (distinct from the BPS walls) for vanilla BPS states. The number of BPS states jumps when one crosses the wall of marginal stability because some of the BPS walls are excluded on one side by mapping them to attractor flow lines that would terminate at regular points in the moduli space \cite{DenefSeibergWittenemptyhole_2000,DenefGreeneRaugas_2001,Alim:2023doi}. We also conjecture that this can be generalised to other quivers with the framed wall crossing being generated by a generalised mutation sequence rather than just the Weyl group with a corresponding denominator function in which the Fourier coefficients also jump
when a BPS wall of this type is crossed.

We have derived these generating functions by seeing that one can take the $\mathcal{N}=4$ generating function and restrict it to subalgebras that retain some of the wall crossing behavior of the full algebra. Hence, one can also use the highest weight modules to count the BPS states that exist in the subalgebras. What is particularly interesting, is that the affine Lie algebra $\hat{A}_{1}$ is a subalgebra. This is immediately generalisable to BPS structures in $\mathcal{N}=2$ theories which also have root systems described by $\hat{A}_{1}$. Firstly, we have found that one can write the argument of the affine Weyl denominator in terms of the central charges of the D6-D2-D0 bound states in  \cite{D6D2D0JafferisMoorehttps://doi.org/10.48550/arxiv.0810.4909}, such that this directly matches the walls as the Weyl chamber boundaries. Therefore, this generating function counts a particular set of BPS boundstates that have been previously used to compute non-commutative Donaldson-Thomas invariants by \cite{Szendr_i_2008,joyce2010theory} and more recently in \cite{ASTThttps://doi.org/10.48550/arxiv.2109.06878}. In this case the jumps are also comparable to the Stokes jumps obtained using the Borel transform of the Gromov-Witten potential on along different rays. It would be interesting in future to directly relate the full partition function for the framed BPS states derived in \cite{D6D2D0JafferisMoorehttps://doi.org/10.48550/arxiv.0810.4909,ASTThttps://doi.org/10.48550/arxiv.2109.06878,Szendr_i_2008}, which contain a higher linear combination of charges outside the $\hat{A}_{1}$ root system, to our generating function.

We have also used this idea in Seiberg-Witten theory (which is also described by the $\hat{A}_{1}$ root system) and have matched BPS walls to Weyl chamber boundaries. Again, it would be interesting in future to exactly match our BPS boundstates in this example with the framed halo states counted by GMN \cite{FramedBPSGMNhttps://doi.org/10.48550/arxiv.1006.0146} in their generating function for framed degeneracies. In the case of Seiberg-Witten theory we have an extra wall of marginal stability. As this wall is crossed, the attractor flow existence conditions from \cite{DenefSeibergWittenemptyhole_2000,DenefGreeneRaugas_2001,Alim:2023doi} are applied such that no flow can end at a regular point where the central charge vanishes. From this we find that one must re-write the central charges of the non-existing BPS states in terms of those of existing states with a prefactor. This means that within the wall the non-existing BPS states are no longer in the positive root lattice and hence cannot be counted in any highest weight module. For Seiberg-Witten theory this then reproduces the known spectrum $\mathcal{S}_w$ on both sides of the wall \cite{Seiberg_1994,1997Lerche,N=2BPSquivershttps://doi.org/10.48550/arxiv.1112.3984}. Therefore there exists a monopole and dyon on one side and the full spectrum including infinitely many dyons, plus the W-boson on the other. The diagram representing this wall crossing recovers the scattering diagram of Bridgeland \cite{BridgelandOriginalscatteringdiagram}. We then generalised this principle to the Argyres-Douglas $A_{2}$ theory, again recovering the wall crossing and scattering diagram. This is no longer a subalgebra of the original $\mathcal{N}=4$ generating function, but this principle still holds and reproduces the primitive wall crossing process of the decay of a single dyon. This therefore has given us a testing case that should in future be generalisable to ADE type Argyres-Douglas theories and ultimately to more general quivers. The results of this work are also presented in the thesis \cite{thesis}.

\subsection*{Acknowledgements}

We would like to thank Arpan Saha for discussions about the project. 
This work is supported by the DFG Emmy Noether grant AL 1407/2-1. Daniel Bryan was supported by the Friedrich-Naumann Foundation for Freedom (FNF) scholarship programme from the Ministry for Education and Research of Germany. The work of Daniel Bryan has been further supported by the OPUS grant, number 2022/47/B/ST2/03313, titled “Quantum Geometry and BPS states”, funded by the National Science Center, Poland.

\appendix

\section{All changes in highest weight for $\hat{A}_{1}$} \label{ChangesinhighestweightaffineA1appendix}

We write the generating function in terms of a highest weight. This highest weight then shifts in all the different chambers in the moduli space. 

\subsection{Before affine wall}

The original generating function in the fundamental Weyl chamber is

\begin{align}
f(k,l) =  \oint_{\gamma} d\alpha_{0}(u) d \delta(u) e^{\lambda_{k,l}(u)}    \frac{e^{ \alpha_{0}(u)+ \delta(u)} }{\prod^{\infty}_{  m = 1 }(1-e^{ -m \delta(u)})^{2} (1-e^{-(m-1) \delta(u)+ \alpha_{0}(u)})^{2} (1-e^{  -m \delta(u)- \alpha_{0}(u) })^{2} }, 
\end{align}

which we write as

\begin{align}
f(k,l) =  \oint_{\gamma} d\alpha_{0}(u) d \delta(u) \Bigg(    \frac{e^{ \lambda(u)} }{\prod^{\infty}_{  m = 1 }(1-e^{ -m \delta(u)})(1-e^{-(m-1) \delta(u)+ \alpha_{0}(u)}) (1-e^{  -m \delta(u)- \alpha_{0}(u) }) } \Bigg)^{2}.
\end{align} 

In terms of the highest weight $\lambda$. Now we see how the generating function changes as we cross $k$ walls in one direction

\begin{align}
f(k,l)  =  \oint_{\gamma} d\alpha_{0}(u) d \delta(u) \Bigg(   \frac{e^{ \lambda(u)} }{\prod^{\infty}_{l =1 }(1-e^{ -l \delta(u)}) (1-e^{  -l \delta(u)- \alpha_{0}(u) })\prod^{\infty}_{m =k+1}(1-e^{-(m-1) \delta(u)+ \alpha_{0}(u)}) } \\ \nonumber \frac{\prod^{k} _{m = 1} e^{(m-1) \delta(u)- \alpha_{0}(u)}}{\prod^{k}_{m = 1 }(1-e^{(m-1) \delta(u)- \alpha_{0}(u)}) } \Bigg)^{2}.   
\end{align}

We can keep moving in this direction such that we let $k \rightarrow \infty$. In this case the generating function becomes

\begin{align}
f(k,l)  =  \oint_{\gamma} d\alpha_{0}(u) d \delta(u)  \Bigg(     \frac{e^{\lambda(u)}}{\prod^{\infty}_{l =1 }(1-e^{ -l \delta(u)})  (1-e^{  -l \delta(u)- \alpha_{0}(u) }) }   \frac{\prod^{\infty} _{m = 1} e^{(m-1) \delta(u)- \alpha_{0}(u)}}{\prod^{\infty}_{m = 1 }(1-e^{(m-1) \delta(u)- \alpha_{0}(u)}) }   \Bigg)^{2}.   
\end{align}

The new highest weight becomes: $\lambda' = \lambda + \sum_{m=1}^{\infty} \Big((m-1) \delta- \alpha_{0}  \Big)$. 

\subsection{After crossing affine wall}

If one proceeds to cross the affine wall then you get an infinite number of affine roots added to the highest weight

\begin{align}
f(k,l)  =  \oint_{\gamma} d\alpha_{0}(u) d \delta(u)    \Bigg( \frac{e^{\lambda(u)}}{\prod^{\infty}_{l =1 }  (1-e^{  -l \delta(u)- \alpha_{0}(u) }) }   \frac{\prod^{\infty}_{l =1 }e^{l \delta(u)}}{\prod^{\infty}_{l =1 }(1-e^{l \delta(u)})}\frac{\prod^{\infty} _{m = 1} e^{(m-1) \delta(u)- \alpha_{0}(u)}}{\prod^{\infty}_{m = 1 }(1-e^{(m-1) \delta(u)- \alpha_{0}(u)}) }  \Bigg)^{2}.     
\end{align}

$\lambda' = \lambda + \sum_{m=1}^{\infty} \Big((m-1) \delta- \alpha_{0}  \Big) + \sum_{l=1}^{\infty}l \delta$

One can continue to cross walls in this direction with the integer $p$ decreasing

\begin{align}
& f(k,l)  = \\ & \nonumber \oint_{\gamma} d\alpha_{0}(u) d \delta(u) \\ \nonumber & \Bigg(   \frac{e^{\lambda(u)} \prod^{\infty}_{n =k+1 }  e^{  n \delta(u)+ \alpha_{0}(u) }}{\prod^{\infty}_{n =p+1 }  (1-e^{  l \delta(u)+ \alpha_{0}(u) }) \prod^{k}_{n =1 }  (1-e^{  -n\delta(u)- \alpha_{0}(u) }) }  \frac{\prod^{\infty}_{l =1 }e^{l \delta(u)}}{\prod^{\infty}_{l =1 }(1-e^{l \delta(u)})}\frac{\prod^{\infty} _{m = 1} e^{(m-1) \delta(u)- \alpha_{0}(u)}}{\prod^{\infty}_{m = 1 }(1-e^{(m-1) \delta(u)- \alpha_{0}(u)})}  \Bigg)^{2}. 
\end{align}

Now the final highest weight becomes 

\begin{align}
\lambda' =  \lambda + \sum_{m=1}^{\infty} \Big((m-1) \delta- \alpha_{0}  \Big) + \sum_{l=1}^{\infty}l \delta+ \sum_{n=p+1}^{\infty}\Big(n \delta+ \alpha_{0}  \Big),
\end{align}

in the chamber where all the BPS states exist.

\section{$A_{2}$ Weyl character changes in representation} 

The full list of basis changes and computations for the change in the highest weight (described in sec. \ref{A2subsection}) of the Weyl character/denominator of $SU(3)$ is shown here. This is distinct in the different Weyl chambers. This then reveals which BPS boundstates exist within each chamber. This gives us the 6 chambers with the combinations of BPS states existing within each one.

\subsection{Weyl denominator expansions} \label{Weyl denominator expansions apendix}

We look at the possible expansions of the Weyl denominator and find that we can write it in terms of negative exponents after a suitable change of basis of positive roots. However, we obtain a shift in the exponents in one of the roots which generates different coefficients. This indicates a wall has been crossed. For clarity, we here write the inner product as $\alpha(u)= (\alpha,u)$:  

\begin{align}
\text{Im}[i(\alpha_{1}, u)] > 0, \ \text{Im}[i(\alpha_{2}, u)] < 0, \ \text{Im}[i(\alpha_{3}, u)] > 0, \\ \nonumber
-\sum^{ \infty }_{n,m,l = 0} e^{-n(\alpha_{1}, u)+(m+1)(\alpha_{2}, u)-l(\alpha_{3}, u)}\\ \nonumber
-(n(\alpha_{1}, u)+(m+1)(-\alpha_{2}, u)+l(\alpha_{1}+\alpha_{2}, u)), \\ \nonumber
\text{change basis} \ \  \alpha'_{1} = -\alpha_{2}, \  \alpha'_{2} = \alpha_{1}+\alpha_{2}=\alpha_{3}, \  \alpha'_{3} = \alpha_{1},\\ \nonumber
\text{expression becomes} -(n(\alpha'_{3}, u)+(m+1)(\alpha'_{1}, u)+l(\alpha'_{2}, u)),\\ \nonumber
\end{align}

\begin{align}
\text{Im}[i(\alpha_{1}, u)] > 0, \ \text{Im}[i(\alpha_{2}, u)] < 0, \ \text{Im}[i(\alpha_{3}, u)] < 0,\\ \nonumber
\sum^{ \infty }_{n,m,l = 0} e^{-n(\alpha_{1}, u)+(m+1)(\alpha_{2}, u)+(l+1)(\alpha_{3}, u)}\\ \nonumber
-(n(\alpha_{1}, u)+(m+1)(-\alpha_{2}, u)+(l+1)(-\alpha_{1}-\alpha_{2}, u), \\ \nonumber
\text{change basis} \ \  \alpha'_{1} = -\alpha_{1}-\alpha_{2} = -\alpha_{3}, \  \alpha'_{2} = \alpha_{1}, \  \alpha'_{3} = -\alpha_{2},\\ \nonumber
\text{expression becomes} -(n(\alpha'_{2}, u)+(m+1)(\alpha'_{3}, u)+(l+1)(\alpha'_{1}, u)),
\end{align}

\begin{align}
\text{Im}[i(\alpha_{1}, u)] < 0, \ \text{Im}[i(\alpha_{2}, u)] > 0, \ \text{Im}[i(\alpha_{3}, u)] < 0,\\ \nonumber
\sum^{ \infty }_{n,m,l = 0} e^{(n+1)(\alpha_{1}, u)-m(\alpha_{2}, u)+(l+1)(\alpha_{3}, u)}\\ \nonumber
-((n+1)(-\alpha_{1}, u)+m(\alpha_{2}, u)+(l+1)(-\alpha_{1}-\alpha_{2}, u)), \\ \nonumber
\text{change basis} \ \  \alpha'_{1} = \alpha_{2}, \  \alpha'_{2} = -\alpha_{1}-\alpha_{2} = -\alpha_{3}, \  \alpha'_{3} = -\alpha_{1},\\ \nonumber
\text{expression becomes} -((n+1)(\alpha'_{3}, u)+m(\alpha'_{1}, u)+(l+1)(\alpha'_{2}, u)),
\end{align}

\begin{align}
\text{Im}[i(\alpha_{i}, u)] < 0, \ \ \ \forall{i \in {1,2,3}},\\ \nonumber
\sum^{\infty }_{n,m,l = 0} e^{(n+1)(\alpha_{1}, u)+(m+1)(\alpha_{2}, u)+(l+1)(\alpha_{3}, u)}\\ \nonumber
-((n+1)(-\alpha_{1}, u)+(m+1)(-\alpha_{2}, u)+(l+1)(-\alpha_{1}-\alpha_{2}, u)), \\ \nonumber
\text{change basis} \ \  \alpha'_{1} = -\alpha_{1}, \  \alpha'_{2} = -\alpha_{2}, \  \alpha'_{3} = -\alpha_{3} = -\alpha_{1}-\alpha_{2}, \\ \nonumber
-((n+1)(\alpha'_{1}, u)+(m+1)(\alpha'_{2}, u)+(l+1)(\alpha'_{3}, u)).
\end{align}


\providecommand{\href}[2]{#2}\begingroup\raggedright\endgroup

\end{document}